\documentclass[aps,pra,reprint,superscriptaddress]{revtex4-2}

\usepackage{graphicx}
\usepackage[caption=false,subrefformat=parens]{subfig}
\usepackage{dcolumn}
\usepackage{bm}
\usepackage{amsmath,amssymb}
\usepackage{hyperref}
\usepackage{braket}
\usepackage[T1]{fontenc}
\usepackage{color}

\DeclareMathOperator{\Tr}{Tr}
\DeclareMathOperator{\Real}{Re}
\DeclareMathOperator{\Imag}{Im}

\makeatletter
\newcommand\exchange[2]{\let\@tempa#1\let#1#2\let#2\@tempa}
\makeatother

\exchange{\Gamma}{\varGamma}
\exchange{\Delta}{\varDelta}
\exchange{\epsilon}{\varepsilon}
\exchange{\Theta}{\varTheta}
\exchange{\Lambda}{\varLambda}
\exchange{\Xi}{\varXi}
\exchange{\Pi}{\varPi}
\exchange{\Phi}{\varPhi}
\exchange{\Psi}{\varPsi}
\exchange{\Omega}{\varOmega}

\newcommand{\add}[1]{#1}

\begin{document}

\title{Designing nontrivial one-dimensional Floquet topological phases using a spin-1/2 double-kicked rotor}

\author{Yusuke Koyama}
\affiliation{Department of Applied Physics, Nagoya University, Nagoya 464-8603, Japan}
\author{Kazuya Fujimoto}
\affiliation{Department of Physics, Tokyo Institute of Technology, 2-12-1 Ookayama, Meguro-ku, Tokyo 152-8551, Japan}
\author{Shuta Nakajima}
\affiliation{Center for Quantum Information and Quantum Biology, Osaka University, Toyonaka, Osaka 560-0043, Japan}
\author{Yuki Kawaguchi}
\affiliation{Department of Applied Physics, Nagoya University, Nagoya 464-8603, Japan}
\affiliation{Research Center for Crystalline Materials Engineering, Nagoya University, Nagoya 464-8603, Japan}

\date{\today}

\begin{abstract}
A quantum kicked rotor model is one of the promising systems to realize various Floquet topological phases.
We consider a double-kicked rotor model for a one-dimensional quasi-spin-1/2 Bose-Einstein condensate with spin-dependent and spin-independent kicks which are implementable for cold atomic experiments.
We theoretically show that the model can realize all the Altland-Zirnbauer classes with nontrivial topology in one dimension.
In the case of class CII, we show that a pair of winding numbers $(w_0,w_\pi)\in 2\mathbb{Z}\times 2\mathbb{Z}$ featuring the edge states at zero and $\pi$ quasienergy, respectively, takes various values depending on the strengths of the kicks.
We also find that the winding numbers change to $\mathbb{Z}$ when we break the time-reversal and particle-hole symmetries by changing the phase of a kicking lattice.
We numerically confirm that the winding numbers can be obtained by measuring the mean chiral displacement in the long-time limit in the present case with four internal degrees of freedom.
We further propose two feasible methods to experimentally realize the spin-dependent and spin-independent kicks required for various topological phases.
\end{abstract}

\maketitle

\section{\label{sec:introduction}Introduction}
Periodic driving to design topological phases has become one of the fundamental techniques in recent studies of condensed matter physics.
A time-periodic field, such as laser fields radiating to electrons in solids and shaking external potentials for ultracold atoms, introduces periodicity in the frequency space, enabling us to manipulate the band topology~\cite{okaPhotovoltaicHallEffect2009,lindnerFloquetTopologicalInsulator2011,goldmanLightinducedGaugeFields2014,goldmanTopologicalQuantumMatter2016,eckardtColloquiumAtomicQuantum2017,zhangTopologicalQuantumMatter2018}.
It is also possible to construct discrete time evolution by a series of unitary operators, such as quantum walks~\cite{kitagawaExploringTopologicalPhases2010,kitagawaObservationTopologicallyProtected2012,kitagawaTopologicalPhenomenaQuantum2012} and micro motions~\cite{kitagawaTopologicalCharacterizationPeriodically2010,rudnerAnomalousEdgeStates2013,carpentierTopologicalIndexPeriodically2015,nathanTopologicalSingularitiesGeneral2015}, where one can design the effective Hamiltonian, defined as a logarithm of a single-period time-evolution operator, to be topologically nontrivial.
Moreover, since the topology of time-periodic systems, so-called Floquet systems, are distinct from the static ones~\cite{eckardtColloquiumAtomicQuantum2017,cooperTopologicalBandsUltracold2019,royPeriodicTableFloquet2017,yaoTopologicalInvariantsFloquet2017,ozawaTopologicalPhotonics2019,harperTopologyBrokenSymmetry2020,rudnerBandStructureEngineering2020}, one can access exotic phenomena that are absent in static systems, e.g., the $\pi$ modes observed in ultracold atoms~\cite{xieTopologicalQuantumWalks2020,winterspergerRealizationAnomalousFloquet2020} and photonic crystals~\cite{kitagawaObservationTopologicallyProtected2012,gaoProbingTopologicalProtection2016,mukherjeeExperimentalObservationAnomalous2017,maczewskyObservationPhotonicAnomalous2017,chengObservationAnomalousEnsuremath2019}.

A kicked rotor is a typical Floquet system~\cite{casatiStochasticBehaviorQuantum1979,izrailevSimpleModelsQuantum1990,raizenQuantumChaosCold1999,sadgrovePseudoclassicalMethodAtomOptics2011} and can host various Floquet topological phases.
A quantum version of the kicked rotor was first considered in the context of quantum chaos and Anderson localization~\cite{casatiStochasticBehaviorQuantum1979,izrailevSimpleModelsQuantum1990}, which were experimentally demonstrated using a Bose-Einstein condensate (BEC) periodically subjected to optical lattice pulses~\cite{chabeExperimentalObservationAnderson2008,lemarieCriticalStateAnderson2010}.
\add{The periodic time evolution of the cold atoms can be precisely controlled by hyperfine-state-dependent optical lattice pulses~\cite{karskiQuantumWalkPosition2009,summyQuantumRandomWalk2016,dadrasQuantumWalkMomentum2018,dadrasExperimentalRealizationMomentumspace2019}.}
A particular interest is in the case when the period of the kicking pulses is on resonance~\add{\cite{izrailevSimpleModelsQuantum1990,raizenQuantumChaosCold1999,sadgrovePseudoclassicalMethodAtomOptics2011}}, for which the effective Hamiltonian reduces to a tight-binding model in the momentum space.
Due to this momentum space lattice structure, topological phases can appear in the kicked rotor systems, and internal degrees of freedom play essential roles in emergence of topological phases.
For an on-resonance kicked rotor, spin internal degrees of freedom have been experimentally introduced by using a spin-dependent optical lattice, and a discrete-time quantum walk in the momentum space was demonstrated~\cite{summyQuantumRandomWalk2016,dadrasQuantumWalkMomentum2018,dadrasExperimentalRealizationMomentumspace2019}.
Furthermore, such a kicked rotor model is theoretically shown to have the sublattice degrees of freedom when we apply two kicks in a single period setting the free evolution time between kicks to meet a high-order quantum resonance condition~\add{\cite{izrailevSimpleModelsQuantum1990,wangExponentialQuantumSpreading2013}}.
With these methods, exotic Floquet topological phases were theoretically studied~\cite{zhouFloquetTopologicalPhases2018,zhouNonHermitianFloquetTopological2019,zhouFloquetSecondOrderTopological2021,bolikDetectingTopologicalPhase2022}.

In this paper, we propose a kicked rotor model with both the spin and sublattice degrees of freedom that can realize all the Altland-Zirnbauer (AZ) classes with nontrivial topology in one dimension~\cite{schnyderClassificationTopologicalInsulators2008,kitaevPeriodicTableTopological2009,ryuTopologicalInsulatorsSuperconductors2010,royPeriodicTableFloquet2017,yaoTopologicalInvariantsFloquet2017}.
The AZ classification characterizes a system according to the time-reversal symmetry, the particle-hole symmetry, and the chiral symmetry.
The chiral operator $\hat{\Gamma}$ is a unitary operator given by $\hat{\Gamma}=\hat{T}\hat{C}$, which always satisfies $\hat{\Gamma}^2=1$.
The time-reversal operator $\hat{T}$ and the particle-hole operator $\hat{C}$ are antiunitary operators and their squares take $+1$ or $-1$.
Here, we would like to emphasize that an antiunitary operator whose square is $-1$ is achieved only in a system more than two internal degrees of freedom.
Thus, to realize all the symmetry classes in the AZ classification, we need four internal degrees of freedom.
Among them, we have a particular interest in class CII, which has both the time-reversal and particle-hole symmetries with $\hat{T}^2=\hat{C}^2=-1$, and hence four internal degrees of freedom are required for emergence of class CII.
Although the class CII models have been theoretically proposed~\cite{gentileEdgeStatesTopological2015,liuFractionalChargedEdge2019,zhouFloquetTopologicalPhases2020,zhouNonHermitianFloquetPhases2020,malardMulticriticalityOnedimensionalTopological2020}, the corresponding experiments have never been reported.
Besides class CII, classes AIII, BDI, D, and DIII host nontrivial topological phases in one dimension.
Classes AIII and BDI are experimentally realized in cold atomic systems~\cite{meierObservationTopologicalSoliton2016,meierObservationTopologicalAnderson2018,xieTopologicalCharacterizationsExtended2019,xieTopologicalQuantumWalks2020}.
In this paper, we show that the single kicked rotor model can access all the above classes by changing the parameters of the kicking lattices.

We consider a double-kicked rotor (DKR) model for a one-dimensional quasi-spin-1/2 BEC.
Here, the key ingredient is to simultaneously apply spin-dependent and spin-independent kicks.
We examine the symmetry of the Floquet operator and find parameter sets for realizing each AZ class.
We further investigate the phase diagram for the case of class CII in the parameter space of the kick strengths.
We confirm that a pair of winding numbers $(w_0, w_\pi)\in 2\mathbb{Z}\times 2\mathbb{Z}$ featuring the edge states at zero and $\pi$ quasienergy, respectively, take various values depending on the kick strengths.
We also find that the winding numbers change to $\mathbb{Z}$ when we break the time-reversal and particle-hole symmetries by changing the phase of a kicking lattice.
The winding numbers can be experimentally measured from the mean chiral displacement (MCD) for chiral symmetric systems~\cite{cardanoDetectionZakPhases2017,meierObservationTopologicalAnderson2018,maffeiTopologicalCharacterizationChiral2018,xieTopologicalCharacterizationsExtended2019,xieTopologicalQuantumWalks2020,derricoBulkDetectionTimedependent2020,st-jeanMeasuringTopologicalInvariants2021,xiaoObservationTopologicalPhase2021}.
We numerically confirm that in the present case with the four internal degrees of freedom, the time average of the MCD converges to the winding number in the long-time limit.

The rest of this paper is organized as follows.
In Sec.~\ref{sec:model}, we introduce the DKR model for a quasi-spin-1/2 BEC and calculate the time evolution operator over a single period.
In Sec.~\ref{sec:symmetry}, we discuss the symmetry properties of our system and derive the condition for realizing Floquet topological phases.
In Sec.~\ref{sec:topological_phase}, we study Floquet topological phases especially in class CII.
In Sec.~\ref{sec:experiment}, we discuss how to experimentally realize our model.
In Sec.~\ref{sec:conclusion}, we conclude this work.

\section{\label{sec:model}On-resonance spin-1/2 double-kicked rotor model}
We consider the DKR model for a one-dimensional quasi-spin-1/2 BEC, which has two spin degrees of freedom and two sublattice degrees of freedom.
We start with a general form of the spin-1/2 DKR (SDKR) model that includes the spin-dependent and spin-independent kicks.
The Hamiltonian is given by
\begin{align}
    \hat{H}(t) &= \frac{\hat{p}^2}{2M} \otimes \hat{\sigma}_0 + \hat{H}_1 \sum_{m=-\infty}^{\infty} \delta(t - mT) \notag \\
    &\quad + \hat{H}_2 \sum_{m=-\infty}^{\infty} \delta(t - T_2 - mT),\label{eq:SDKR_Hamiltonian} \\
    \hat{H}_j &= \lambda_j^0\cos\left(\frac{2\pi}{a}\nu_j^0\hat{x} + \alpha_j^0\right) \otimes \hat{\sigma}_0 \notag \\
    &\quad + \lambda_j\cos\left(\frac{2\pi}{a}\nu_j\hat{x} + \alpha_j\right) \otimes \bm{n}_j\cdot\hat{\bm{\sigma}} \quad (j=1, 2), \label{eq:SDKR_kick_Hamiltonian}
\end{align}
where $M$ is the atomic mass, $\hat{x}$ and $\hat{p}$ are the position and momentum of the atom, and $\hat{\bm{\sigma}}=(\hat{\sigma}_x,\hat{\sigma}_y,\hat{\sigma}_z)$ and $\hat{\sigma}_0$ are the vector of Pauli matrices and the identity matrix, respectively, in the spin space.
Here, we assume that the interatomic interactions are sufficiently weak and thus negligible.
The BEC is simultaneously kicked by the spin-independent and spin-dependent optical lattices twice in the single period $T$.
The duration between the first ($j=1$) and second ($j=2$) kicks is $T_2$.
The two optical lattices in each kick ($j=1, 2$) have the strengths $\lambda_j^0, \lambda_j$, the wave lengths $a/\nu_j^0, a/\nu_j$ ($\nu_j^0, \nu_j\in\mathbb{Z}$), and the phases $\alpha_j^0,\alpha_j$, respectively.
We require that $\nu_{1,2}^0$ and $\nu_{1,2}$ have no common divisor in order to ensure that the spatial period of Eq.~(\ref{eq:SDKR_Hamiltonian}) is $a$.
The unit vector $\bm{n}_j$ specifies the spin dependence of the optical lattice.
We choose $\bm{n}_1\nparallel\bm{n}_2$ so that the Hamiltonian $\hat{H}(t)$ is not block-diagonalized.
Without loss of generality, we can choose $\alpha_1^0=0$.
We will show below that this model belongs to the classes CII, AIII, BDI, D, or DIII depending on the choices of the parity of $\nu_{1,2}^0, \nu_{1,2}$ and the other phases $\alpha_2^0, \alpha_{1,2}$.

Because the Hamiltonian (\ref{eq:SDKR_Hamiltonian}) is time periodic, we investigate the symmetry property of the system by calculating the time evolution operator over a single period, i.e., the Floquet operator, $\hat{\mathcal{U}}(T)=\mathcal{T} \exp[-i\int_{-0}^{T-0}\hat{H}(t)\,dt/\hbar]$, where $\mathcal{T}$ is the time ordering operator.
By integrating Eq.~(\ref{eq:SDKR_Hamiltonian}) from $t=-0$ to $t=T-0$, we obtain
\begin{align}
    \hat{\mathcal{U}}(T) &= e^{-\frac{i}{\hbar}\frac{\hat{p}^2}{2M} \otimes \hat{\sigma}_0 (T-T_2)} e^{-\frac{i}{\hbar}\hat{H}_2} e^{-\frac{i}{\hbar}\frac{\hat{p}^2}{2M} \otimes \hat{\sigma}_0 T_2} e^{-\frac{i}{\hbar}\hat{H}_1}. \label{eq:SDKR_Floquet_op}
\end{align}
Because the Hamiltonian (\ref{eq:SDKR_Hamiltonian}) is also spatially periodic with the period $a$, the quasimomentum $\hbar(2\pi/a) \beta$ ($0\leq\beta<1$) becomes a good quantum number.
In other words, the atoms with a given $\beta$ have momentum restricted to
\begin{align}
    p &= \hbar\frac{2\pi}{a}(l+\beta) \quad (l\in\mathbb{Z}), \label{eq:momentum}
\end{align}
during the time evolution.
It follows that when we start from a BEC with atoms in the $\beta=0$ state, we can rewrite the momentum operator $\hat{p}$ as
\begin{align}
    \hat{p} &= \hbar\frac{2\pi}{a}\hat{l},
\end{align}
where $\hat{l}$ is the discretized momentum operator whose eigenvalues are all integers.
Here, we choose $T$ and $T_2$ as~\cite{mooreAtomOpticsRealization1995,wangExponentialQuantumSpreading2013,wangProposalColdatomRealization2008}
\begin{align}
    T &= 4\pi \frac{M}{\hbar}\left(\frac{a}{2\pi}\right)^2 = \frac{h}{4E_\mathrm{R}}, \\
    T_2 &= \frac{T}{4} = \pi \frac{M}{\hbar}\left(\frac{a}{2\pi}\right)^2 = \frac{h}{16E_\mathrm{R}}, \label{eq:quantum_resonance_condition}
\end{align}
where $E_\mathrm{R}=\hbar^2 (\pi/a)^2/2M$ is the recoil energy of the optical lattice in Eq.~(\ref{eq:SDKR_kick_Hamiltonian}) with $\nu=1$.
Then, we obtain the Floquet operator for the on-resonance SDKR (ORSDKR) model:
\begin{align}
    \hat{\mathcal{U}}(T) &= e^{+i\frac{\pi}{2}\hat{l}^2 \otimes \hat{\sigma}_0} e^{-\frac{i}{\hbar}\hat{H}_2} e^{-i\frac{\pi}{2}\hat{l}^2 \otimes \hat{\sigma}_0} e^{-\frac{i}{\hbar}\hat{H}_1}. \label{eq:2_ORSDKR_Floquet_op}
\end{align}
The point here is that due to the choice of $T_2$ in Eq.~(\ref{eq:quantum_resonance_condition}), the first and the third factors in Eq.~(\ref{eq:SDKR_Floquet_op}) become the identity ($i$ and $-i$, respectively) for even (odd) eigenvalues of $\hat{l}$, introducing sublattice structure in the momentum space.

Due to the appearance of the sublattice structure, we decompose the momentum lattice basis $\{\ket{l}\}$ into even $l$ sites (A sites) and odd $l$ sites (B sites) and rewrite $\ket{l=2n}=\ket{n}\otimes\ket{\mathrm{A}}$ and $\ket{l=2n+1}=\ket{n}\otimes\ket{\mathrm{B}}$~\cite{wangExponentialQuantumSpreading2013,zhouNonHermitianFloquetTopological2019}.
We further move to the quasiposition basis $\{\ket{\theta}\}$ defined by
\begin{align}
    \ket{\theta} &= \sum_{n=-\infty}^{\infty} \ket{n}\frac{1}{\sqrt{2\pi}}e^{i\theta n} \quad (-\pi<\theta\leq\pi). \label{eq:Fourier_quasiposition}
\end{align}
Here, $\theta$ is regarded as the ``Bloch wave number'' for the tight-binding model in the momentum space.
Indeed, the Floquet operator is block diagonal in terms of $\theta$ as $\hat{\mathcal{U}}(T) = \int_{-\pi}^{\pi} d\theta \ket{\theta}\bra{\theta} \otimes \hat{U}(\theta)$ where
\begin{gather}
    \hat{U}(\theta) = e^{-i \hat{h}_2(\theta)} e^{-i \hat{h}_1(\theta)}, \label{eq:ORSDKR_Floquet_op_quasiposition} \\
    \hat{h}_j(\theta) = \Lambda_j^0(\theta) \hat{h}_{\tau j}(\theta, \nu_j^0) \otimes \hat{\sigma}_0 + \Lambda_j(\theta) \hat{h}_{\tau j}(\theta, \nu_j) \otimes \bm{n}_j\cdot\hat{\bm{\sigma}}, \label{eq:ORSDKR_Hamiltonian} \\
    \hat{h}_{\tau j}(\theta, \nu) =
    \begin{cases}
        \hat{\tau}_0 & (\text{even }\nu) \\
        \bm{m}_j(\theta)\cdot\hat{\bm{\tau}} & (\text{odd }\nu)
    \end{cases} \quad (j=1, 2). \label{eq:factor_sublattice}
\end{gather}
Here, $\hat{\bm{\tau}}=(\hat{\tau}_x,\hat{\tau}_y,\hat{\tau}_z)$ and $\hat{\tau}_0$ are the vector of Pauli matrices and the identity matrix, respectively, acting on the sublattice space, and we define
\begin{align}
    \Lambda_j^0(\theta) &= \frac{\lambda_j^0}{\hbar} \cos\left(\frac{\nu_j^0}{2}\theta - \alpha_j^0\right), \\
    \Lambda_j(\theta) &= \frac{\lambda_j}{\hbar} \cos\left(\frac{\nu_j}{2}\theta - \alpha_j\right), \\
    \bm{m}_1 &= \left(\cos\frac{\theta}{2}, \sin\frac{\theta}{2}, 0\right), \\
    \bm{m}_2 &= \left(-\sin\frac{\theta}{2}, \cos\frac{\theta}{2}, 0\right).
\end{align}
The detailed derivation of Eq.~(\ref{eq:ORSDKR_Floquet_op_quasiposition}) is given in Appendix~\ref{appx:basis}.

\section{\label{sec:symmetry}Symmetry properties}
\begin{table}
    \caption{\label{tab:AZ_class}Symmetry properties and topological invariants of the AZ classes that host nontrivial static and Floquet topological phases in one dimension~\cite{schnyderClassificationTopologicalInsulators2008,kitaevPeriodicTableTopological2009,ryuTopologicalInsulatorsSuperconductors2010,royPeriodicTableFloquet2017,yaoTopologicalInvariantsFloquet2017}.}
    \begin{ruledtabular}
        \begin{tabular}{cccccc}
            Class & $\hat{T}^2$ & $\hat{C}^2$ & $\hat{\Gamma}^2$ & Static & Floquet \\
            \hline
            AIII & $0$ & $0$ & $1$ & $\mathbb{Z}$ & $\mathbb{Z}\times\mathbb{Z}$ \\
            BDI & $+1$ & $+1$ & $1$ & $\mathbb{Z}$ & $\mathbb{Z}\times\mathbb{Z}$ \\
            D & $0$ & $+1$ & $0$ & $\mathbb{Z}_2$ & $\mathbb{Z}_2\times\mathbb{Z}_2$ \\
            DIII & $-1$ & $+1$ & $1$ & $\mathbb{Z}_2$ & $\mathbb{Z}_2\times\mathbb{Z}_2$ \\
            CII & $-1$ & $-1$ & $1$ & $2\mathbb{Z}$ & $2\mathbb{Z}\times 2\mathbb{Z}$
        \end{tabular}
    \end{ruledtabular}
\end{table}

\begin{table*}
    \caption{\label{tab:class_ORSDKR}Optical lattice parameters and the corresponding symmetry operators for realizing symmetry classes in the ORSDKR.
    We list parameter sets with which the Floquet operator cannot be block-diagonalized.
    Without loss of generality, we choose $\alpha_1^0=0$.
    For the parameter sets with $\lambda_1^0=0$, we choose $\alpha_1=0$.
    We describe $\alpha_{1,2}^0,\alpha_{1,2} \in [0,\pi)$ because adding $\pi$ to them preserves the symmetry of the system.
    The symbol ``-'' in the columns $\nu_{1,2}^0,\nu_{1,2},\alpha_{1,2}^0,\alpha_{1,2}$ means that the corresponding kick must not be included, i.e., the corresponding $\lambda_{1,2}^0,\lambda_{1,2}$ must be zero.
    The symbol ``$*$'' in the columns $\nu_{1,2}^0,\nu_{1,2},\alpha_{1,2}^0,\alpha_{1,2}$ means no restriction except not belonging to the other classes.
    $\bm{n}_\perp$ is an unit vector perpendicular to both $\bm{n}_1$ and $\bm{n}_2$.}
    \begin{ruledtabular}
        \begin{tabular}{cccccccccccc}
            Class & $\nu_1^0$ & $\nu_1$ & $\nu_2^0$ & $\nu_2$ & $\alpha_1^0$ & $\alpha_1$ & $\alpha_2^0$ & $\alpha_2$ & $\hat{T}$ & $\hat{C}$ & $\hat{\Gamma}$ \\
            \hline
            AIII & odd & odd & odd & odd & $*$ & $*$ & $*$ & $*$ & - & - & $\hat{\tau}_z \otimes \hat{\sigma}_0$ \\
            AIII & - & odd & - & even & - & $*$ & - & $*$ & - & - & $\hat{\tau}_0 \otimes \bm{n}_\perp\cdot\hat{\bm{\sigma}}$ \\
            AIII & - & even & - & odd & - & $*$ & - & $*$ & - & - & $\hat{\tau}_0 \otimes \bm{n}_\perp\cdot\hat{\bm{\sigma}}$ \\
            AIII & odd & even & odd & even & $*$ & $*$ & $*$ & $*$ & - & - & $\hat{\tau}_z \otimes \bm{n}_\perp\cdot\hat{\bm{\sigma}}$ \\
            BDI & odd & odd & odd & odd & $0$ & $0$ & $\pi/2$ & $\pi/2$ & $\hat{\tau}_0 \otimes \bm{n}_\perp\cdot\hat{\bm{\sigma}} \hat{\sigma}_y \hat{K}$ & $\hat{\tau}_z \otimes \bm{n}_\perp\cdot\hat{\bm{\sigma}} \hat{\sigma}_y \hat{K}$ & $\hat{\tau}_z \otimes \hat{\sigma}_0$ \\
            D & odd & odd & $*$ & $*$ & $0$ & $0$ & $\pi/2$ & $\pi/2$ & - & $\hat{\tau}_z \otimes \bm{n}_\perp\cdot\hat{\bm{\sigma}} \hat{\sigma}_y \hat{K}$ & - \\
            D & odd & even & $*$ & $*$ & $0$ & $\pi/2$ & $\pi/2$ & $\pi/2$ & - & $\hat{\tau}_z \otimes \bm{n}_\perp\cdot\hat{\bm{\sigma}} \hat{\sigma}_y \hat{K}$ & - \\
            DIII & - & odd & - & even & - & $0$ & - & $\pi/2$ & $\hat{\tau}_z \otimes \hat{\sigma}_y \hat{K}$ & $\hat{\tau}_z \otimes \bm{n}_\perp\cdot\hat{\bm{\sigma}} \hat{\sigma}_y \hat{K}$ & $\hat{\tau}_0 \otimes \bm{n}_\perp\cdot\hat{\bm{\sigma}}$ \\
            DIII & odd & even & odd & even & $0$ & $\pi/2$ & $\pi/2$ & $\pi/2$ & $\hat{\tau}_0 \otimes \hat{\sigma}_y \hat{K}$ & $\hat{\tau}_z \otimes \bm{n}_\perp\cdot\hat{\bm{\sigma}} \hat{\sigma}_y \hat{K}$ & $\hat{\tau}_z \otimes \bm{n}_\perp\cdot\hat{\bm{\sigma}}$ \\
            CII & odd & odd & odd & odd & $0$ & $\pi/2$ & $\pi/2$ & $0$ & $\hat{\tau}_0\otimes\hat{\sigma}_y \hat{K}$ & $\hat{\tau}_z\otimes\hat{\sigma}_y \hat{K}$ & $\hat{\tau}_z\otimes\hat{\sigma}_0$
        \end{tabular}
    \end{ruledtabular}
\end{table*}

We derive the conditions for the system to belong to each AZ class.
For Floquet topological systems, it is convenient to describe the symmetry property in terms of the Floquet operator $\hat{U}(\theta)$.
Since $\theta$ corresponds to the ``Bloch wave number'', the time-reversal symmetry, the particle-hole symmetry, and the chiral symmetry for the Floquet operator are respectively given by~\cite{royPeriodicTableFloquet2017,yaoTopologicalInvariantsFloquet2017}
\begin{subequations}\label{eq:symmetries_Floquet}
    \begin{align}
        \hat{T}\hat{U}(\theta)\hat{T}^{-1} &= \hat{U}^{-1}(-\theta), \label{eq:time_reversal_Floquet} \\
        \hat{C}\hat{U}(\theta)\hat{C}^{-1} &= \hat{U}(-\theta), \label{eq:particle_hole_Floquet} \\
        \hat{\Gamma}\hat{U}(\theta)\hat{\Gamma}^{-1} &= \hat{U}^{-1}(\theta), \label{eq:chiral_Floquet}
    \end{align}
\end{subequations}
where $\hat{T}$, $\hat{C}$, and $\hat{\Gamma}$ are the time-reversal, particle-hole, and chiral operators, respectively.
The former two are antiunitary operators, whereas the last one, $\hat{\Gamma}$, is a unitary operator equal to $\hat{T}\hat{C}$ up to a phase factor.
In Table~\ref{tab:AZ_class}, we summarize the symmetry properties and topological invariants of the AZ classes that have nontrivial topological phases in one dimension.

Because we are considering the time-dependent Hamiltonian, the time-reversal and chiral symmetries exist only under proper choices of the origin of the time axis.
For the case of $\hat{U}(\theta)$ in Eq.~(\ref{eq:ORSDKR_Floquet_op_quasiposition}) composed of two successive unitary operators, we can define the Floquet operators in the two symmetric time frames as~\cite{asbothBulkboundaryCorrespondenceChiral2013}
\begin{align}
    \hat{U}_1(\theta) &= e^{-i \hat{h}_1(\theta) / 2} e^{-i \hat{h}_2(\theta)} e^{-i \hat{h}_1(\theta) / 2}, \label{eq:symmetric_time_frame_1} \\
    \hat{U}_2(\theta) &= e^{-i \hat{h}_2(\theta) / 2} e^{-i \hat{h}_1(\theta)} e^{-i \hat{h}_2(\theta) / 2}. \label{eq:symmetric_time_frame_2}
\end{align}
It follows that both $\hat{U}_1(\theta)$ and $\hat{U}_2(\theta)$ satisfy Eq.~(\ref{eq:symmetries_Floquet}) when both $e^{-i \hat{h}_1(\theta)}$ and $e^{-i \hat{h}_2(\theta)}$ satisfy Eq.~(\ref{eq:symmetries_Floquet}), which in turn is rewritten as
\begin{subequations}
    \begin{align}
        \hat{T} \hat{h}_j(\theta) \hat{T}^{-1} &= \hat{h}_j(-\theta), \label{eq:time_reversal_Hamiltonian} \\
        \hat{C} \hat{h}_j(\theta) \hat{C}^{-1} &= -\hat{h}_j(-\theta), \label{eq:particle_hole_Hamiltonian} \\
        \hat{\Gamma} \hat{h}_j(\theta) \hat{\Gamma}^{-1} &= -\hat{h}_j(\theta). \label{eq:chiral_Hamiltonian}
    \end{align}
    \label{eq:symmetries_Hamiltonian}
\end{subequations}

We first consider the case of class CII.
What we have to do is to find the operators $\hat{T}$, $\hat{C}$, and $\hat{\Gamma}$ that satisfy Eq.~(\ref{eq:symmetries_Hamiltonian}) with $\hat{h}_j$ given in Eq.~(\ref{eq:ORSDKR_Hamiltonian}).
Here, $\hat{T}$ and $\hat{C}$ should satisfy $\hat{T}^2=\hat{C}^2=-1$.
First, for the system to have the chiral symmetry, $\hat{h}_j(\theta)$ should not include the identity matrix, $\hat{\tau}_0\otimes\hat{\sigma}_0$, which leads to that both $\nu_1^0$ and $\nu_2^0$ are odd integers.
The chiral operator that anticommutes with the first term of Eq.~(\ref{eq:ORSDKR_Hamiltonian}) is $\hat{\Gamma}=\hat{\tau}_z\otimes\hat{\sigma}_0$, which also anticommutes with the second term of Eq.~(\ref{eq:ORSDKR_Hamiltonian}) only when $\nu_1$ and $\nu_2$ are also odd integers.
For this $\hat{\Gamma}$, a possible pair of the time-reversal and particle-hole operators are $\hat{\tau}_0\otimes\hat{\sigma}_y \hat{K}$ and $\hat{\tau}_z\otimes\hat{\sigma}_y \hat{K}$, where $\hat{K}$ is the complex conjugate operator:
The changes in $\hat{h}_j(\theta)$ under the other choices of $\hat{T}$ and $\hat{C}$ are not written in simple sign changes as in Eqs.~(\ref{eq:time_reversal_Hamiltonian}) and (\ref{eq:particle_hole_Hamiltonian}).
Because we have chosen $\alpha_1^0=0$ for which $\Lambda_1^0(\theta)$ is an even function of $\theta$, we choose $\hat{T}=\hat{\tau}_0\otimes \hat{\sigma}_y \hat{K}$ and $\hat{C}=\hat{\tau}_z\otimes \hat{\sigma}_y \hat{K}$ such that they satisfy Eqs.~(\ref{eq:time_reversal_Floquet}) and (\ref{eq:particle_hole_Floquet}).
It follows that $\Lambda_2(\theta)$  [$\Lambda_1(\theta)$ and $\Lambda_2^0(\theta)$] should be an even function [odd functions] of $\theta$, resulting in $\alpha_2=0$ and $\alpha_1=\alpha_2^0=\pm\pi/2$.
This is the condition for the system to belong class CII.

We stress here that the existence of the spin-independent lattice is crucial for the realization of class CII system.
Otherwise, $\hat{h}_1(\theta)$ and $\hat{h}_2(\theta)$ commute with the same operator $\hat{\tau}_z\otimes\bm{n}_\perp\cdot\hat{\bm{\sigma}}$ with $\bm{n}_\perp$ being a unit vector perpendicular to both $\bm{n}_1$ and $\bm{n}_2$, which means that the system is merely a combination of the systems written by $2\times 2$ matrices.

Similarly to the class CII case, we obtain the conditions for the ORSDKR to belong to the symmetry classes that possess nontrivial topological phases in one dimension.
The detailed derivation is given in Appendix~\ref{appx:symmetry}, and we summarize the result in Table~\ref{tab:class_ORSDKR}.

\section{\label{sec:topological_phase}Topological phases in class CII}
\subsection{\label{subsec:winding_number}Winding number}
Since the ORSDKR has the chiral symmetry, we can characterize the topological phases with the winding numbers calculated as follows.
Defining the effective Hamiltonian $\hat{H}_{j,\mathrm{eff}}(\theta)$ from the relation $\hat{U}_j(\theta)=e^{-i\hat{H}_{j,\mathrm{eff}}(\theta) T/\hbar}$ ($j=1,2$), we introduce $Q$ matrix as~\cite{asbothBulkboundaryCorrespondenceChiral2013}
\begin{align}
    \hat{Q}_j(\theta) &= \sin\frac{\hat{H}_{j,\mathrm{eff}}(\theta) T}{\hbar} = \frac{\hat{U}_j^\dagger(\theta) - \hat{U}_j(\theta)}{2i}.
\end{align}
In our model, the chiral operator $\hat{\Gamma}=\hat{\tau}_z\otimes\hat{\sigma}_0$ is diagonal, and hence $Q$ matrix is block off-diagonal,
\begin{align}
    \hat{Q}_j(\theta) &= \begin{pmatrix}
        0 & \hat{q}_j(\theta) \\
        \hat{q}_j^\dagger(\theta) & 0
    \end{pmatrix}.
\end{align}
By using the off-diagonal element $\hat{q}_j(\theta)$, we can calculate the winding number as
\begin{align}
    w_j &= \int_{-\pi}^{\pi} \frac{i}{2\pi} \Tr \left[\hat{q}_j^{-1}(\theta) \frac{\partial}{\partial \theta} \hat{q}_j(\theta)\right] \,d\theta. \label{eq:winding_number}
\end{align}
Here, $w_1$ and $w_2$ are the winding numbers calculated for the Floquet operators $\hat{U}_1$ and $\hat{U}_2$, respectively, in the symmetric time frames.
On the other hand, the winding numbers related to the number of edge states at quasienergies $\epsilon=0$ and $\pi\hbar/T$, if the system has edges, are given by~\cite{asbothBulkboundaryCorrespondenceChiral2013}
\begin{align}
    w_0 = \frac{w_1+w_2}{2},\quad w_\pi = \frac{w_1-w_2}{2}. \label{eq:winding_number_Floquet}
\end{align}

\subsection{\label{subsec:phase_diagram}Phase diagram}
\begin{figure}
    \includegraphics[width=\linewidth]{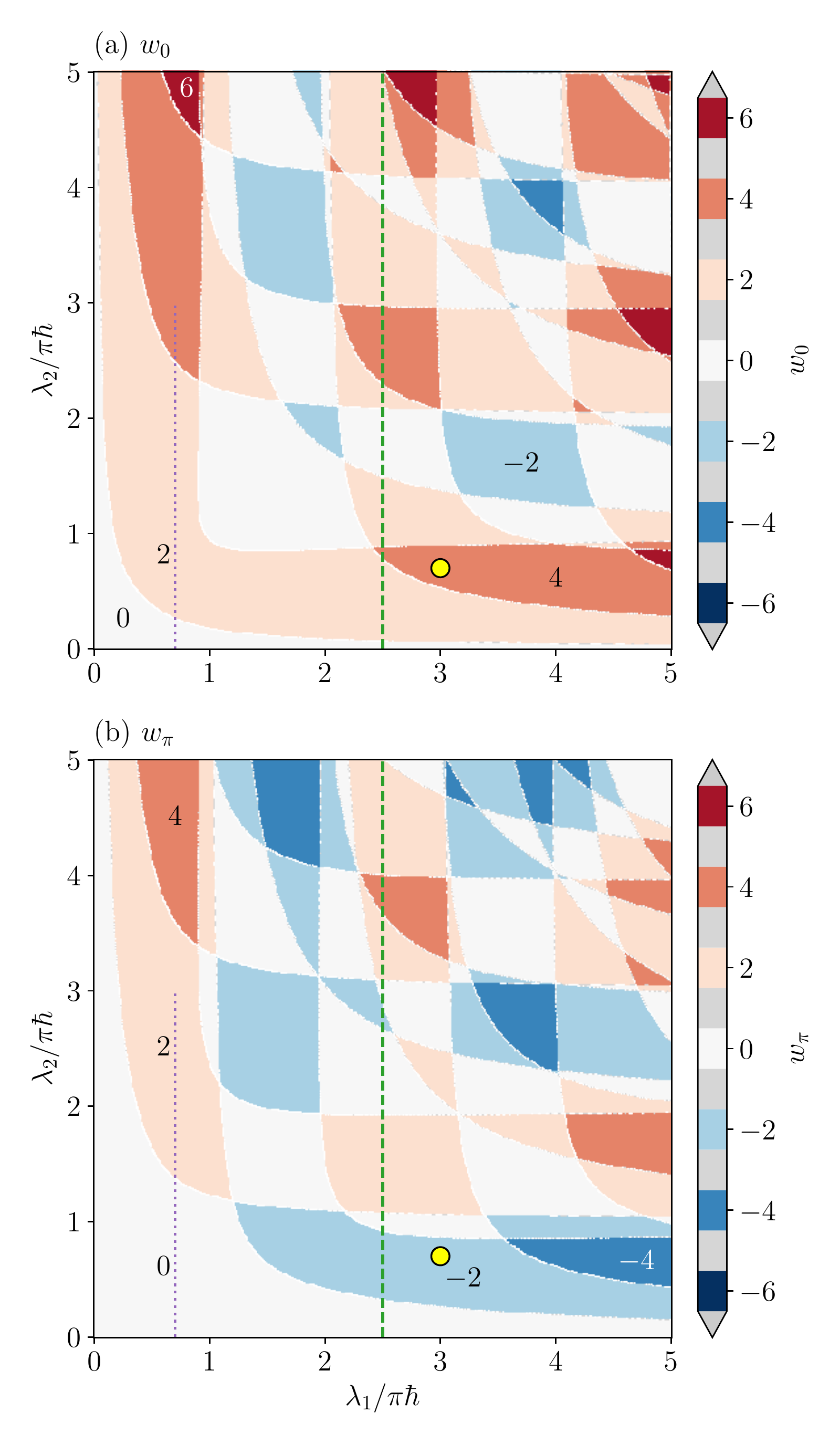}
    \caption{\label{fig:winding_number}Topological phase diagram of the ORSDKR.
    (a) Winding number $w_0$ at quasienergy $\epsilon=0$ and (b) winding number $w_\pi$ at quasienergy $\epsilon=\pi\hbar/T$.
    The parameters are fixed as $\lambda_1^0=0.3\pi\hbar$, $\lambda_2^0=0.4\pi\hbar$, $\bm{n}_1\cdot\bm{n}_2=1/\sqrt{2}(\neq\pm 1)$, $\nu_{1,2}^0=1$, $\nu_{1,2}=1$, $\alpha_1^0=\alpha_2=0$, and $\alpha_1=\alpha_2^0=-\pi/2$.
    The green dashed lines, the yellow filled circles, and the purple dotted lines indicate the values of $\lambda_{1,2}$ used for Figs.~\ref{fig:winding_number_CII_AIII} and \ref{fig:spectrum}, Fig.~\ref{fig:MCD}, and Fig.~\ref{fig:MCD_winding_number}, respectively.}
\end{figure}

\begin{figure}
    \includegraphics[width=\linewidth]{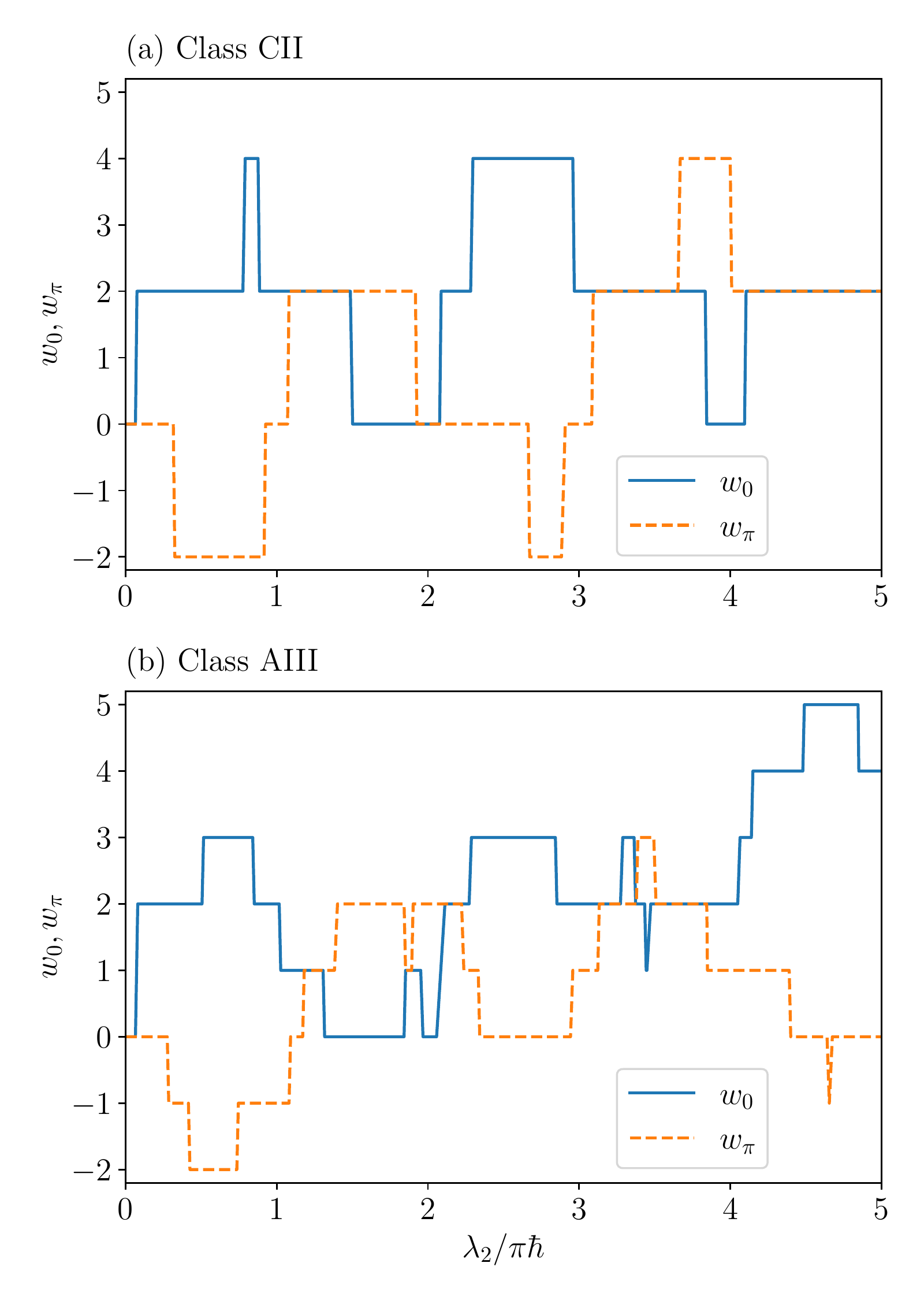}
    \subfloat{\label{fig:winding_number_CII}}
    \subfloat{\label{fig:winding_number_AIII}}
    \caption{\label{fig:winding_number_CII_AIII}Winding numbers $w_0$ (blue solid line) and $w_\pi$ (orange dashed line) of the ORSDKR in (a) class CII and (b) class AIII as functions of the kick strength $\lambda_2$.
    (a) The parameters are the same as those in Fig.~\ref{fig:winding_number} and we choose $\lambda_1=2.5\pi\hbar$ (green dashed lines in Fig.~\ref{fig:winding_number}).
    (b) The parameters are the same as those of (a) except for $\alpha_2$:
    We choose $\alpha_2=0.1\pi$ so that the system does not have the time-reversal and particle-hole symmetries.
    At the point where one of the winding numbers in (a) jumps by 2, a new phase with an odd winding number appears in (b).}
\end{figure}

\begin{figure*}
    \add{
    \includegraphics[width=0.9\linewidth]{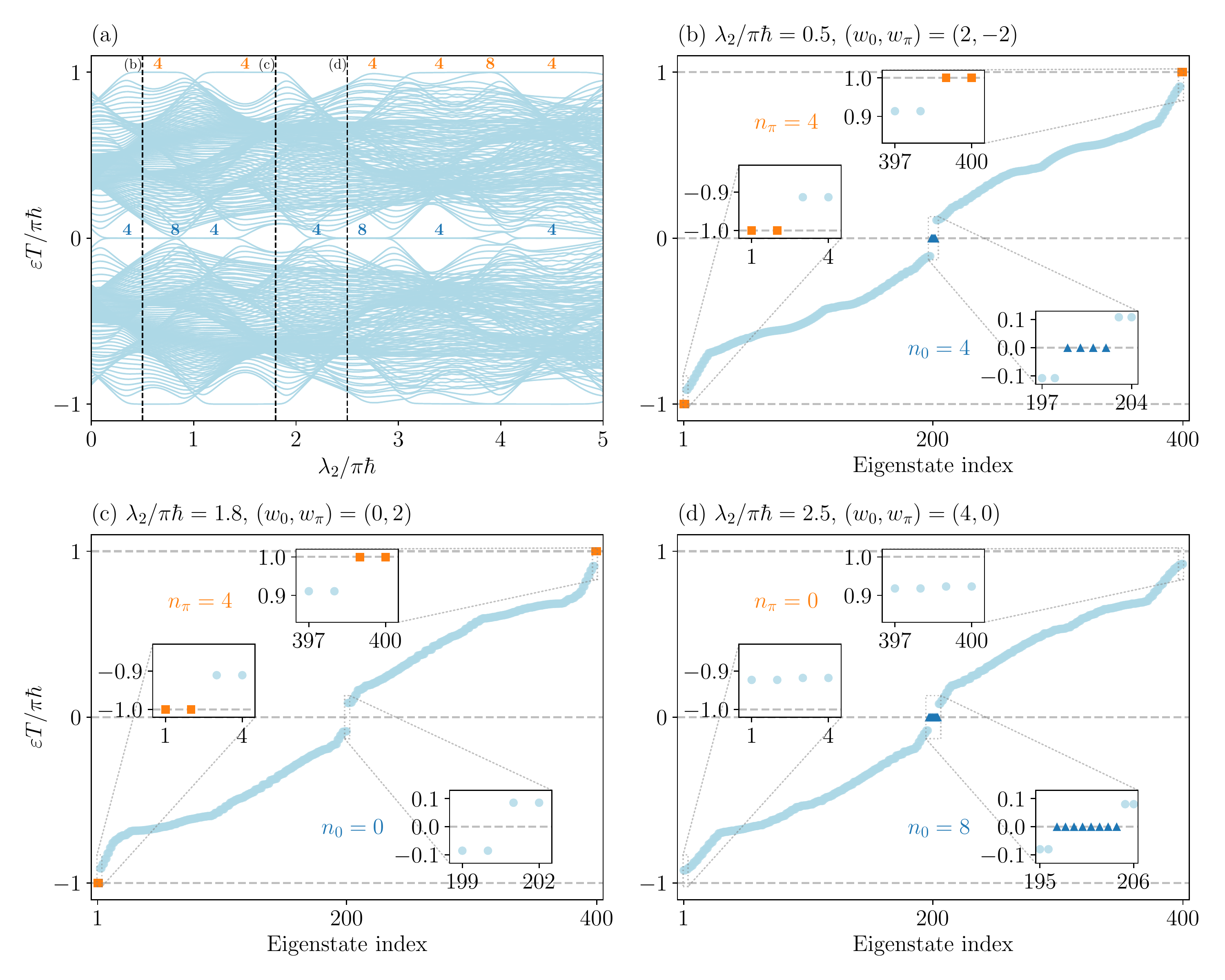}
    \subfloat{\label{fig:spectrum_both}}
    \subfloat{\label{fig:spectrum_pi}}
    \subfloat{\label{fig:spectrum_zero}}
    \caption{\label{fig:spectrum}Spectra of the Floquet operator $\hat{\mathcal{U}}(T)$ under the open boundary condition.
    (a) Spectrum as a function of the kick strength $\lambda_2$.
    The parameters are the same as those in Fig.~\protect\subref*{fig:winding_number_CII} and we set the number of unit cells in the momentum lattice to 100.
    The numbers at $\epsilon=0$ (dark blue) and $\pi\hbar/T$ (orange) denote the numbers of edge states at zero quasienergy ($n_0$) and $\pi$ quasienergy ($n_\pi$), respectively.
    (b)--(d) Spectra for fixed $\lambda_2/\pi\hbar$ of 0.5~(b), 1.8~(c), and 2.5~(d).
    (b) Four eigenstates at zero quasienergy and four eigenstates at $\pi$ quasienergy exist.
    (c) Four eigenstates at $\pi$ quasienergy exist and (d) eight eigenstates at zero quasienergy exist.
    Considering a pair of winding numbers $(w_0,w_\pi)$ shown in Fig.~\protect\subref*{fig:winding_number_CII}, we see the relation $(n_0,n_\pi)=(2|w_0|,2|w_\pi|)$.}
    }
\end{figure*}

\begin{figure}
    \includegraphics[width=\linewidth]{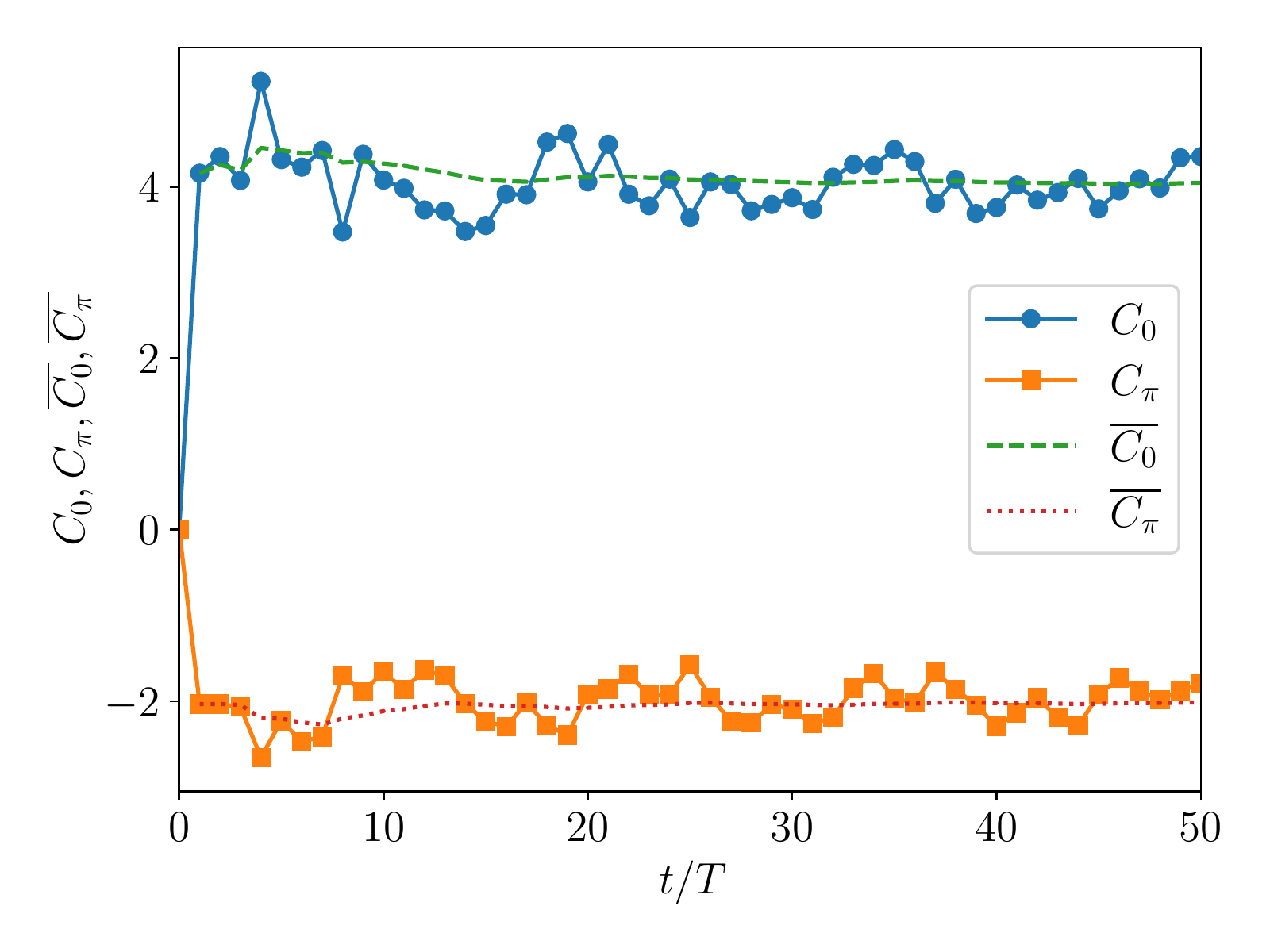}
    \caption{\label{fig:MCD}MCDs $C_0=(C_1+C_2)/2$ (blue solid curve) and $C_\pi=(C_1-C_2)/2$ (orenge solid curve), and their time averages $\overline{C_0}$ (green dashed curve) and $\overline{C_\pi}$ (red dotted curve).
    The parameters are the same as those in Fig.~\ref{fig:winding_number} and we choose $\lambda_1=3.0\pi\hbar$ and $\lambda_2=0.7\pi\hbar$ (yellow filled circles in Fig.~\ref{fig:winding_number}), for which a pair of the winding numbers is $(w_0,w_\pi)=(4,-2)$.
    $C_0$ and $C_\pi$ oscillate and converge to $w_0$ and $w_\pi$, whereas $\overline{C}_0$ and $\overline{C}_\pi$ converge more rapidly.}
\end{figure}

To satisfy the class CII symmetries, $\bm{n}_1\nparallel\bm{n}_2$ is required.
Otherwise, both $\hat{h}_1(\theta)$ and $\hat{h}_2(\theta)$ commute with the same operator $\hat{\tau}_0\otimes\bm{n}_1\cdot\hat{\bm{\sigma}}$.
We remember that all $\nu_{1,2}^0$ and $\nu_{1,2}$ must be odd numbers.
The cases $(\nu_{1,2}^0,\nu_{1,2})=(1,1)$ and $(1,3)$ would be experimentally implementable (see Sec.~\ref{sec:experiment}).
Here, we show the results for $(\nu_{1,2}^0,\nu_{1,2})=(1,1)$.
We have not seen significant difference for other choices of $(\nu_{1,2}^0,\nu_{1,2})$.

Figure~\ref{fig:winding_number} shows the winding numbers $w_0$ and $w_\pi$ as functions of the kick strengths $\lambda_1$ and $\lambda_2$.
Here, we set the parameters as $\lambda_1^0=0.3\pi\hbar$, $\lambda_2^0=0.4\pi\hbar$, $\bm{n}_1\cdot\bm{n}_2=1/\sqrt{2}(\neq\pm 1)$, $\nu_{1,2}^0=1$, $\nu_{1,2}=1$, $\alpha_1^0=\alpha_2=0$, and $\alpha_1=\alpha_2^0=-\pi/2$.
We confirm that both $w_0$ and $w_\pi$ are even integers, being consistent with $2\mathbb{Z}\times 2\mathbb{Z}$ in Table~\ref{tab:AZ_class}.
Figure~\ref{fig:winding_number} shows that we can access various phases by changing the kick strengths.

It is interesting to see how the phase diagram changes as the symmetry changes.
In the ORSDKR model, it is possible to change the system from class CII to class AIII by slightly changing one of the phases $\alpha$ from the fixed values for class CII.
(See the first and last rows in Table~\ref{tab:class_ORSDKR}.)

We show the winding numbers $w_0$ and $w_\pi$ as functions of the kick strength $\lambda_2$ in class CII [Fig.~\subref*{fig:winding_number_CII}] and AIII [Fig.~\subref*{fig:winding_number_AIII}].
Here, we fix $\lambda_1=2.5\pi\hbar$ and choose $\alpha_2=0$ for class CII [Fig.~\subref*{fig:winding_number_CII}] and $0.1\pi$ for class AIII [Fig.~\subref*{fig:winding_number_AIII}].
The other parameters are the same as those in Fig.~\ref{fig:winding_number}.
At small $\lambda_2$, the winding numbers for class CII and AIII have similar structures.
However, there is a significant difference that at the point where one of the winding numbers in class CII changes by 2, a new phase with an odd winding number arises in class AIII (see, e.g., at $\lambda_2/\pi\hbar=0.3$).
The deviation of the winding numbers between the two classes enlarges as $\lambda_2$ becomes large.
We note that the structures of $w_0$ and $w_\pi$ in class AIII approaches the one in class CII as $\alpha_2$ becomes closer to zero.

It is also possible to change the system from class CII to class BDI using the same lattice potential but with changing $\alpha_1$ and $\alpha_2$ by $\pi/2$.
In this case, the winding numbers change drastically because the symmetry operators undergo discrete change (see Table~\ref{tab:class_ORSDKR}).

\add{\subsection{\label{subsec:bulk_edge_correspondence}Bulk-edge correspondence}}
\add{Here, we comment on the bulk-edge correspondence for the ORSDKR.}
We can theoretically consider a finite system in the momentum space by artificially restricting the possible momentum value to a finite range~\cite{zhouFloquetTopologicalPhases2018,zhouNonHermitianFloquetTopological2019}.
By calculating the eigenenergies $\epsilon$ and eigenstates $\ket{\psi}$ of $\hat{\mathcal{U}}(T)$ through $\hat{\mathcal{U}}(T) \ket{\psi} = e^{-i\epsilon T/\hbar} \ket{\psi}$, we can obtain the edge states in a finite system.
\add{The spectra of $\hat{\mathcal{U}}(T)$ under the open boundary condition are shown in Fig.~\ref{fig:spectrum}.
The parameters are the same as those in Fig.~\subref*{fig:winding_number_CII} and we set the number of unit cells in the momentum lattice to 100.
The numbers at $\epsilon=0$ and $\pi\hbar/T$ in Fig.~\subref*{fig:spectrum_both} denote the numbers of edge states at zero quasienergy ($n_0$) and $\pi$ quasienergy ($n_\pi$), respectively.
Comparing the number of edge states with the winding numbers $(w_0,w_\pi)$ in Fig.~\subref*{fig:winding_number_CII}, we see the relation $(n_0,n_\pi)=(2|w_0|,2|w_\pi|)$.
We have confirmed that this relation is also true for other chiral symmetric classes, i.e., classes AIII and BDI.
This relation can be understand by generalizing the relation of the bulk-edge correspondence proven for a chiral symmetric time-independent lattice Hamiltonian~\cite{xiongAnisotropicMagneticResponses2017} to Floquet systems.
The detailed discussion is given in Appendix~\ref{appx:bulk_edge_correspondence}.
}

\subsection{\label{sec:MCD}Mean chiral displacement}
The winding numbers can be experimentally measured from the MCD for chiral symmetric systems~\cite{cardanoDetectionZakPhases2017,meierObservationTopologicalAnderson2018,maffeiTopologicalCharacterizationChiral2018,xieTopologicalCharacterizationsExtended2019,xieTopologicalQuantumWalks2020,derricoBulkDetectionTimedependent2020,st-jeanMeasuringTopologicalInvariants2021,xiaoObservationTopologicalPhase2021}.
The MCD is the signed momentum distribution of the chiral symmetric system and defined as
\begin{align}
    C_j(t) &= 4\Tr[\hat{\rho}_0 \hat{\mathcal{U}}_j^\dagger(t) (\hat{n}\otimes\hat{\Gamma}) \hat{\mathcal{U}}_j(t)], \label{eq:MCD}
\end{align}
where $\hat{n}$ is the momentum operator defined by $\hat{n}\ket{n}=n\ket{n}$ with $\ket{n}$ being the momentum lattice basis defined below Eq.~(\ref{eq:2_ORSDKR_Floquet_op}), and $\hat{\mathcal{U}}_j(t)$ ($j=1,2$) is the time evolution operator corresponding to the Floquet operator $\hat{U}_j(\theta)$ defined in the symmetric time frame.
For example, when $t=mT$ ($m=1,2,\cdots$), we have $\hat{\mathcal{U}}_j(t)=[\hat{\mathcal{U}}_j(T)]^m$ with $\hat{\mathcal{U}}_j(T)=\int_{-\pi}^{\pi} d\theta \ket{\theta}\bra{\theta} \otimes \hat{U}_j(\theta)$.
The initial density matrix $\hat{\rho}_0$ is here assumed to be
\begin{align}
    \hat{\rho}_0 &= \ket{n=0}\bra{n=0} \otimes \frac{\hat{\tau}_0 \otimes \hat{\sigma}_0}{4}. \label{eq:initial_density_matrix}
\end{align}
Equation~(\ref{eq:MCD}) is experimentally evaluated as follows:
Prepare the initial state in a pure state $\ket{\psi_{\tau \sigma}} = \ket{n=0} \otimes \ket{\tau}\otimes\ket{\sigma}$, ($\tau=\mathrm{A,B}$ and $\sigma=\uparrow,\downarrow$), and observe the time evolution of the expectation value of $\hat{n}\otimes\hat{\Gamma}$.
Repeating this procedure for the initial states with all the combinations of $\tau = \mathrm{A, B}$ and $\sigma = \uparrow, \downarrow$, we obtain the expectation value for each initial state and then take the summation for them, which becomes the MCD.
We note that $\ket{\psi_{\tau \sigma}}$ is the eigenstate of $\hat{\Gamma} = \hat{\tau}_z \otimes \hat{\sigma}_0$ and expressed as $\ket{\psi_{\mathrm{A} \sigma}} = \ket{l=0} \otimes \ket{\sigma}$ and $\ket{\psi_{\mathrm{B} \sigma}} = \ket{l=1} \otimes \ket{\sigma}$, where $\ket{l=0}$ and $\ket{l=1}$ are the eigenstates of $\hat{p}$ with eigenvalues $0$ and $h/a$, respectively.
Thus, from the measurement of the momentum distribution for each spin state after a certain period, we can obtain the MCD.

On the other hand, in the quasiposition basis the MCD at $t=mT$ is written as
\begin{align}
    C_j(mT) &= \int_{-\pi}^{\pi} \frac{d\theta}{2\pi} \Tr\left[[\hat{U}_j^\dagger(\theta)]^m \hat{\Gamma} i \frac{\partial}{\partial \theta} [\hat{U}_j(\theta)]^m\right], \label{eq:MCD_trace_position}
\end{align}
which converges to the winding number $w_j$ as $m\to\infty$~\cite{maffeiTopologicalCharacterizationChiral2018}.
Thus, from the MCDs $C_1(mT)$ and $C_2(mT)$ defined with the Floquet operators in the symmetric time frames $\hat{U}_1(\theta)$ and $\hat{U}_2(\theta)$, we obtain the winding numbers $w_0$ and $w_\pi$ through
\begin{align}
    C_0(mT) &= \frac{C_1(mT) + C_2(mT)}{2},\\
    C_\pi(mT) &= \frac{C_1(mT) - C_2(mT)}{2}, \label{eq:MCD_0_pi}
\end{align}
at a sufficiently long time.
The convergence becomes faster when we take the time average of the MCDs.
Here, we consider the time average of the MCD at $t=mT$ as
\begin{align}
    \overline{C}_j(t=mT) &= \frac{1}{m} \sum_{k=1}^{m} C_j(kT), \label{eq:MCD_average}
\end{align}
which converges to $w_j$ as $m\to\infty$~\cite{zhouFloquetTopologicalPhases2018,zhouNonHermitianFloquetTopological2019}.

Figure~\ref{fig:MCD} shows the time evolution of the MCDs $C_0$ and $C_\pi$ in class CII.
The parameters are the same as those in Fig.~\ref{fig:winding_number} and we choose $\lambda_1=3.0\pi\hbar$ and $\lambda_2=0.7\pi\hbar$.
The corresponding points are depicted with the yellow filled circles in Fig.~\ref{fig:winding_number}, for which a pair of the winding numbers is given by $(w_0,w_\pi)=(4,-2)$.
The MCDs $C_0$ and $C_\pi$ converge to the winding numbers $w_0$ and $w_\pi$, respectively, as time evolves.
One can clearly see that the time-averaged MCDs $\overline{C}_0$ and $\overline{C}_\pi$ converge to $w_0$ and $w_\pi$ faster than $C_0$ and $C_\pi$.

\section{\label{sec:experiment}Proposal for experimental setup}
\begin{figure*}
    \includegraphics[width=0.9\linewidth]{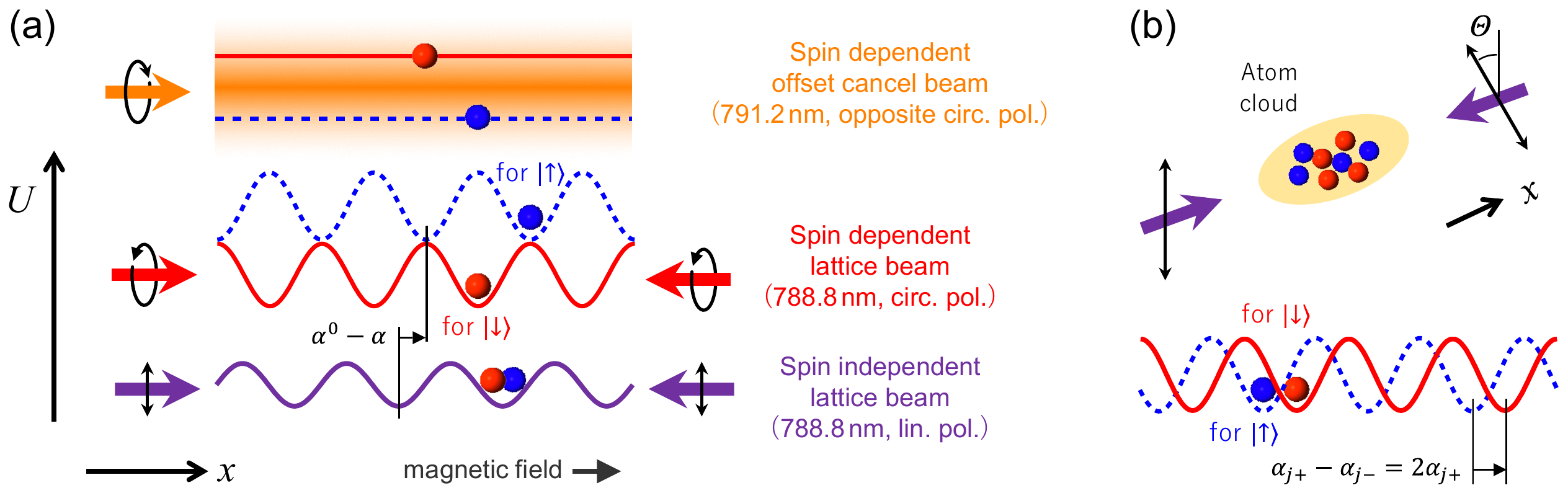}
    \subfloat{\label{fig:setup_lin_circ}}
    \subfloat{\label{fig:setup_lin_enclosing}}
    \caption{\label{fig:setup}(a) Optical lattices created by linearly and circularly polarized lasers.
    A spin-dependent lattice for $\ket{\uparrow}=\ket{F=2,m_F=2}$ and $\ket{\downarrow}=\ket{F=1,m_F=1}$ of $^{87}$Rb can be created by the standing wave of the circularly polarized \add{788.8~nm} laser~\cite{wenExperimentalStudyTuneout2021}.
    A spin-independent lattice for both spin states is created by the standing wave of the linearly polarized \add{788.8~nm} laser.
    To avoid unwanted interference between the spin-dependent and spin-independent lattices, the laser frequency of the circularly polarized laser and the linearly polarized laser is slightly (typically a few hundred MHz) deviated from each other.
    Despite this small frequency difference, the spin-dependent and spin-independent lattices can be regarded as having the same lattice constant in the region of the atom cloud.
    The phase of each lattice can be switched by changing the laser frequency of each lattice beam.
    The spin-dependent global phase shift can be removed by a spin-dependent offset cancel beam created by a running \add{791.2~nm} laser beam with opposite circular polarization.
    This spin-dependent offset cancel beam also cancels unwanted spin-dependent confinement/deconfinement of radial direction of the spin-dependent lattice beam.
    (b) Optical lattices created by linearly polarized lasers with an enclosing angle.
    The phase difference $\alpha_{j +} - \alpha_{j -} = 2\alpha_{j +}$ between two optical lattice potentials for $\ket{\uparrow}=\ket{F=1,m_F=1}$ and $\ket{\downarrow}=\ket{F=1,m_F=-1}$ can be tuned by changing the linear polarization vectors enclosing an angle $\Theta$ which can be switched by using an electro-optical modulator~\cite{mandelCoherentTransportNeutral2003,mandelControlledCollisionsMultiparticle2003}.}
\end{figure*}

In this section, we discuss how to experimentally implement the Hamiltonian~(\ref{eq:SDKR_Hamiltonian}).
After explaining the method to choose the direction of $\bm n_{1,2}$ in Sec.~\ref{subsec:spin_dependence_z}, we propose two methods to simultaneously create spin-dependent and spin-independent lattice potentials in Secs.~\ref{subsec:linear_circular_polarization} and \ref{subsec:inclined_linear_polarization}.
The experimental setup is summarized in Fig.~\ref{fig:setup}.
\add{We discuss the case when the ideal quantum resonance conditions are unsatisfied and study the effects of nonzero quasimomentum component in Sec.~\ref{subsec:quasimomentum}.}

\subsection{\label{subsec:spin_dependence_z}Spin dependence of the kick}
We first show that the direction specifying spin dependence of the kick can be arbitrarily chosen if we can implement the $\hat{\sigma}_z$-dependent optical lattice
\begin{align}
    \hat{V}_z &= V(\hat{x}) \otimes \hat{\sigma}_z, \label{eq:kick_z}
\end{align}
and the spin rotation operator proposed in Refs.~\cite{summyQuantumRandomWalk2016,dadrasQuantumWalkMomentum2018,dadrasExperimentalRealizationMomentumspace2019}
\begin{align}
    \hat{M}(\alpha,\chi) &= e^{-i(\alpha/2)(\sin\chi \hat{\sigma}_x - \cos\chi \hat{\sigma}_y)}. \label{eq:coin_op}
\end{align}
\add{Using Eq.~(\ref{eq:coin_op}), we can arbitrarily rotate the spin direction in Eq.~(\ref{eq:kick_z}) as}
\begin{align}
    \hat{M}(-\alpha,\chi) \hat{V}_z \hat{M}(\alpha,\chi) &= V(\hat{x}) \otimes \bm{n}\cdot\hat{\bm{\sigma}}, \label{eq:kick_potential_arbitrary}
\end{align}
where $\bm{n} = (\sin\alpha \cos\chi, \sin\alpha \sin\chi, \cos\alpha)$.
When the lattice potential in Eq.~(\ref{eq:kick_z}) is applied during an infinitesimal time $\Delta t$, the whole time evolution operator is given by
\begin{align}
    \hat{M}(-\alpha,\chi) e^{-i \hat{V}_z \Delta t / \hbar} \hat{M}(\alpha,\chi) &= e^{-i V(\hat{x}) \otimes \bm{n}\cdot\hat{\bm{\sigma}} \Delta t / \hbar}.
\end{align}
Thus, we can realize the kick potential with arbitrary spin dependence as given in Eq.~(\ref{eq:kick_potential_arbitrary}) by sandwiching the kick $\hat{V}_z$ between $\hat{M}(-\alpha,\chi)$ and $\hat{M}(\alpha,\chi)$.

\subsection{\label{subsec:linear_circular_polarization}Optical lattices created by linearly and circularly polarized lasers}
The spin-independent and spin-dependent kicks can be realized by applying lasers with linear and circular polarizations to a quasi-spin-1/2 BEC.
From the discussion in Sec.~\ref{subsec:spin_dependence_z}, the required spin-independent and spin-dependent potentials are $V^0(\hat{x}) \otimes \hat{\sigma}_0$ and $V(\hat{x}) \otimes \hat{\sigma}_z$, respectively, where
\begin{align}
    V^0(x) &= \lambda^0 \cos\left(\frac{2\pi}{a} \nu^0 x + \alpha^0\right), \label{eq:kick_potential_x_spin_independent} \\
    V(x) &= \lambda \cos\left(\frac{2\pi}{a} \nu x + \alpha\right). \label{eq:kick_potential_x_spin_dependent}
\end{align}
We consider a mixture of $^{87}$Rb BECs in the hyperfine levels $\ket{F=2,m_F=2}$ and $\ket{F=1,m_F=1}$, \add{which we define as $\ket{\uparrow}$ and $\ket{\downarrow}$, respectively, in this setup.}
The spin rotation between the two hyperfine states is realized by appliying microwave pulses~\cite{summyQuantumRandomWalk2016,dadrasQuantumWalkMomentum2018,dadrasExperimentalRealizationMomentumspace2019,mandelCoherentTransportNeutral2003,mandelControlledCollisionsMultiparticle2003}.

These potentials can be realized by two counterpropagating laser beams with linear parallel polarization (lin~$\parallel$~lin) and circular parallel polarization (circ~$\parallel$~circ) [see Fig.~\subref*{fig:setup_lin_circ}].
\add{The ac Stark shift of the atom in the hyperfine state $\ket{F,m_F}$ interacting with the laser field is given by}
\add{%
\begin{align}
    \Delta E_{F,m_F}^{\varepsilon\parallel\varepsilon} &= U_{F,m_F}^{\varepsilon\parallel\varepsilon} (1 + \cos 2kx), \label{eq:energy_shift}
\end{align}
}%
\add{where $\varepsilon=\mathrm{lin}$ (circ) is for linear (circular) polarization} and $k$ is the wavenumber of the laser beam.
\add{(The details of the ac Stark shift is given in Appendix~\ref{appx:optical_lattice}.)}
If the wavelength is sufficiently far from the corresponding optical transitions, the amplitude in the lin~$\parallel$~lin configuration satisfies $\add{U_{F=2,m_F=2}^\mathrm{lin\parallel lin}}=\add{U_{F=1,m_F=1}^\mathrm{lin\parallel lin}}$ (see \add{Appendix~\ref{subsec:lin_para_lin}}), which corresponds to the spin-independent optical lattice.
On the other hand, at the wavelength of \add{788.8~nm}, the amplitude in the circ~$\parallel$~circ configuration satisfies $\add{U_{F=2,m_F=2}^{\mathrm{circ}\parallel\mathrm{circ}}}=-\add{U_{F=1,m_F=1}^{\mathrm{circ}\parallel\mathrm{circ}}}$ (see \add{Appendix~\ref{subsec:circ_para_circ}}), which corresponds to the spin-dependent optical lattice.
Thus, at this wavelength, both spin-independent and spin-dependent optical lattice pulses can be realized.
Note that since Eq.~(\ref{eq:energy_shift}) has a constant offset term \add{$U_{F,m_F}^{\varepsilon\parallel\varepsilon}$}, the resulting potentials are not directly equal to $V^0(\hat{x}) \otimes \hat{\sigma}_0$ for $\varepsilon=\mathrm{lin}$ and $V(\hat{x}) \otimes \hat{\sigma}_z$ for $\varepsilon=\mathrm{circ}$;
\add{$U_{F,m_F}^\mathrm{lin\parallel lin}$} induces a global phase shift, and hence, it is irrelevant;
\add{$U_{F,m_F}^{\mathrm{circ}\parallel\mathrm{circ}}$} creates a spin-dependent phase shift, which should be removed by, e.g., applying a running \add{791.2~nm} laser beam with opposite circular polarization [see Fig.~\subref*{fig:setup_lin_circ} \add{and Appendix~\ref{subsec:circ}}].
It is worth noticing that although the wavelength of the spin-dependent lattice is fixed due to the nature of the spin dependence, we can change the wavelength of the spin-independent lattice in a vast range.
For example, when we use the linearly polarized beams with the wavelength of \add{394.4~nm} or 1578~nm for the spin-independent lattice, we can realize $(\nu_{1,2}^0,\nu_{1,2}) = (2,1),(1,2)$, respectively.
Also, we can even realize $(\nu_{1,2}^0,\nu_{1,2}) = (1, n)$ ($n>2$) by introducing an optical lattice generated by the interference of two laser beams propagating at a relative angle of $\theta_\mathrm{L}$.
For example, by intersecting two $\lambda_\mathrm{L} = 1064~\mathrm{nm}$ laser beams with relative angle $\theta_\mathrm{L}=53.4^\circ$, we can create an optical lattice with a lattice spacing of $d=\lambda_\mathrm{L}/[2\sin(\theta_\mathrm{L}/2)]=1184~\mathrm{nm}$, which corresponds to the $n=3$ case.

\subsection{\label{subsec:inclined_linear_polarization}Optical lattices created by linearly polarized lasers with an enclosing angle}
In this section, we propose a more feasible setup for the class CII case in Table~\ref{tab:class_ORSDKR} than the setup in Sec.~\ref{subsec:linear_circular_polarization}.
More concretely, Eq.~(\ref{eq:SDKR_kick_Hamiltonian}) with $\nu_j^0=\nu_j$ ($j=1,2$) can be realized by using only a pair of linearly polarized counterpropagating beams with an enclosing angle, whereas the setup in Sec.~\ref{subsec:linear_circular_polarization} requires two pairs of counterpropagating beams and an offset cancel beam.
The required kick Hamiltonian is Eq.~(\ref{eq:SDKR_kick_Hamiltonian}) with $\bm{n}_j=(0, 0, 1)$, which is written as
\begin{align}
    \hat{H}_j &= \begin{pmatrix}
        \hat{H}_{j +} & 0 \\
        0 & \hat{H}_{j -}
    \end{pmatrix}, \\
    \hat{H}_{j \pm} &= \lambda_j^0\cos\left(\frac{2\pi}{a}\nu_j^0\hat{x} + \alpha_j^0\right) \pm \lambda_j\cos\left(\frac{2\pi}{a}\nu_j\hat{x} + \alpha_j\right). \label{eq:Hamiltonian_experiment}
\end{align}
Assuming $\nu_j^0=\nu_j$, we can rewrite Eq.~(\ref{eq:Hamiltonian_experiment}) as
\begin{align}
    \hat{H}_{j \pm} &= \lambda_{j \pm} \cos\left(\frac{2\pi}{a}\nu_j\hat{x} + \alpha_{j \pm}\right), \label{eq:Hamiltonian_CII} \\
    \lambda_{j \pm} &= \sqrt{(\lambda_j^0)^2 + (\lambda_j)^2 \pm 2 \lambda_j^0 \lambda_j \cos(\alpha_j^0 - \alpha_j)}, \\
    \tan\alpha_{j \pm} &= \frac{\lambda_j^0 \sin\alpha_j^0 \pm \lambda_j \sin\alpha_j}{\lambda_j^0 \cos\alpha_j^0 \pm \lambda_j \cos\alpha_j}.
\end{align}
Here, we especially focus on the class CII case in Table~\ref{tab:class_ORSDKR}.
Substituting $\alpha_1^0=0$, $\alpha_1=-\pi/2$, $\alpha_2^0=-\pi/2$, $\alpha_2=0$, and $\nu_{1,2}^0=\nu_{1,2}=1$, we obtain
\begin{align}
    \lambda_{1+} &= \lambda_{1-} = \sqrt{(\lambda_1^0)^2 + (\lambda_1)^2}, \label{eq:lambda1p} \\
    \alpha_{1+} &= -\alpha_{1-} = -\tan^{-1}\frac{\lambda_1}{\lambda_1^0}, \label{eq:alpha1p} \\
    \lambda_{2+} &= \lambda_{2-} = \sqrt{(\lambda_2^0)^2 + (\lambda_2)^2}, \label{eq:lambda2p} \\
    \alpha_{2+} &= -\alpha_{2-} = -\tan^{-1}\frac{\lambda_2^0}{\lambda_2}. \label{eq:alpha2p}
\end{align}
Thus, the optical potential that we need is the one that acts on spin-up and down atoms with the same amplitude and the phase difference $\alpha_{j+}-\alpha_{j-}=2\alpha_{j+}$.

To realize the quasi-spin-1/2 system belonging to class CII, we consider a mixture of $^{87}$Rb BECs in the hyperfine levels $\ket{F=1,m_F=1}$ and $\ket{F=1,m_F=-1}$, \add{which we define as $\ket{\uparrow}$ and $\ket{\downarrow}$, respectively, in this setup.}
The spin rotation between the two hyperfine states can be realized by using the Raman transition.
However, these two hyperfine states are unstable against the collision toward the $\ket{F=1,m_F=0}$ state, and the experiment using these two hyperfine states is limited to a short time.
Nevertheless, since this setup has a much simpler laser alignment than the scheme in Sec.~\ref{subsec:linear_circular_polarization}, it is worth considering.
We consider that the atoms are periodically kicked by two counterpropagating pulse lasers with linear polarization vectors enclosing an angle $\Theta$~\cite{mandelCoherentTransportNeutral2003,mandelControlledCollisionsMultiparticle2003} [see Fig.~\subref*{fig:setup_lin_enclosing}].
\add{
At the wavelength of 787.5~nm, the optical lattice potentials for the two internal degrees of freedom are given by
\begin{align}
    V_{F=1,m_F=\pm 1}(x) &= -V [1 + \cos(2kx \pm \Theta)], \label{eq:optical_lattice_lin_theta_lin}
\end{align}
where $V$ is the strength of the optical lattice pulse and $k$ is the wave number of the laser.
(The derivation is shown in Appendix~\ref{subsec:lin_theta_lin}.)
Therefore, we should set $V=\lambda_{j+}$ and $\Theta=\alpha_{j+}$ to realize Eq.~(\ref{eq:Hamiltonian_CII}) with Eqs.~(\ref{eq:lambda1p})--(\ref{eq:alpha2p}).
The relations between experimental parameters $V_j, \Theta_j$ and the model parameters $\lambda_j^0, \lambda_j$ are given by
\begin{align}
    \lambda_1^0 &= V_1 \cos\Theta_1, \\
    \lambda_1 &= -V_1 \sin\Theta_1, \\
    \lambda_2^0 &= V_2 \sin\Theta_2, \\
    \lambda_2 &= -V_2 \cos\Theta_2,
\end{align}
where the additional subscript $j=1,2$ of $V_j$ and $\Theta_j$ corresponds to the potentials for $\hat{H}_j$, and without loss of generality, we have take a possible sign combinations.
}

In a similar way, one can consider the other cases of $\nu_j^0=\nu_j$ and $\alpha_j^0=\alpha_j=0$ or $\pm\pi/2$, e.g., classes BDI and D in Table~\ref{tab:class_ORSDKR}.
However, these classes require the lattice potentials with same phases and different amplitudes for spin-up and down atoms, and thus another quasi-spin mixture is needed.

\add{
\subsection{\label{subsec:quasimomentum}Effects of nonzero quasimomentum component}
One of the major concerns to realize the ORSDKR model in the cold atomic experiments may be the difficulty of achieving the ideal quantum resonance.
In particular, it is difficult to prepare an initial state with $\beta=0$ in Eq.~(\ref{eq:momentum}) due to the residual atomic thermal cloud and the finite size of the system~\cite{dadrasExperimentalRealizationMomentumspace2019,bolikDetectingTopologicalPhase2022}.
Thus, the BEC has a finite width $\Delta\beta$ of the quasimomentum $\beta$, which is typically $\Delta\beta \approx 0.025$~\cite{ryuHighOrderQuantumResonances2006,dadrasExperimentalRealizationMomentumspace2019}.
In Fig.~\ref{fig:MCD_winding_number}, we show the time-averaged MCDs at $t/T=4$ (a) and 10 (b) calculated for Gaussian initial states centered at $p=0$ and $p=h/a$ with a standard deviation $\Delta\beta$ in the momentum space.
The parameters are the same as those in Fig.~\ref{fig:winding_number} and we choose $\lambda_1=0.7\pi\hbar$.
We note that the experiments would be restricted to small kick strengths and short time duration because otherwise the momentum distribution of the BEC extends too broadly.
In the calculation, we restrict the range of the momentum space to $[-50 h/a,50 h/a]$ and $[-120 h/a,120 h/a]$ for the time-averaged MCDs at $t/T=4$ and $t/T=10$, respectively.
We see that even when $\Delta\beta$ is nonzero, the time-averaged MCDs $\overline{C}_0$ and $\overline{C}_\pi$ at $t/T=4$ take the values close to the winding numbers $w_0$ and $w_\pi$, respectively, at the regions far from the topological phase transition points.
However, the deviation from the winding numbers is prominent around the topological phase transition points, and the MCDs for $\Delta\beta=0.025$ do not exhibit convergence behavior as a function of duration.
The reason for the former is that the energy gap is small around the phase transition points so that it takes a long time for the MCDs to converge.
Indeed, the MCDs for $\Delta\beta=0$ at $t/T=10$ approach more to the winding numbers.
On the other hand, the reason for the latter is that the topological protection is violated due to the nonzero $\beta$ component:
If we start from a BEC with a nonzero fixed $\beta$, the Floquet operator is given by Eq.~(\ref{eq:2_ORSDKR_Floquet_op}) with replacing $\hat{l}^2$ with $(\hat{l}+\beta)^2$, which no longer belongs to class CII;
In addition, the modified Floquet operator has different gap-closing points from that for $\beta=0$;
As a result, when the initial state has a finite width $\Delta\beta$ of quasimomentum, the nonzero $\beta$ component blurs the integer limit values of the MCDs in the long-time limit.
Because this effect becomes large as time $t/T$ and the range $(h/a)\Delta l$ of the momentum distribution increase even when $\Delta\beta$ is small, the condition $\Delta l \Delta\beta t/T \ll 1$ should be satisfied so that the additional phase factor does not become dominant.
For example, the range of the momentum distribution is about $[-20h/a,20h/a]$ at $t/T=4$ and $[-40h/a,40h/a]$ at $t/T=10$ for $\lambda_2=0.8\pi\hbar$, while it is $[-40h/a,40h/a]$ at $t/T=4$ and $[-80h/a,80h/a]$ at $t/T=10$ for $\lambda_2=2.0\pi\hbar$.
The above mentioned condition is not satisfied even at $t/T=4$ for $\lambda_2=2.0\pi\hbar$;
The convergence of the MCDs at $t/T=10$ in Fig.~\subref*{fig:MCD_winding_number_b} is worse for $\lambda_2=2.0\pi\hbar$ than for $\lambda_2=0.8\pi\hbar$.
Thus, the winding numbers can be obtained from the time-averaged MCDs when all of the width $\Delta\beta$ of the quasimomentum distribution, the kick strengths, and the time duration $t/T$ are sufficiently small and when the system is not too close to the topological phase transition points.

\begin{figure}
    \add{
    \includegraphics[width=\linewidth]{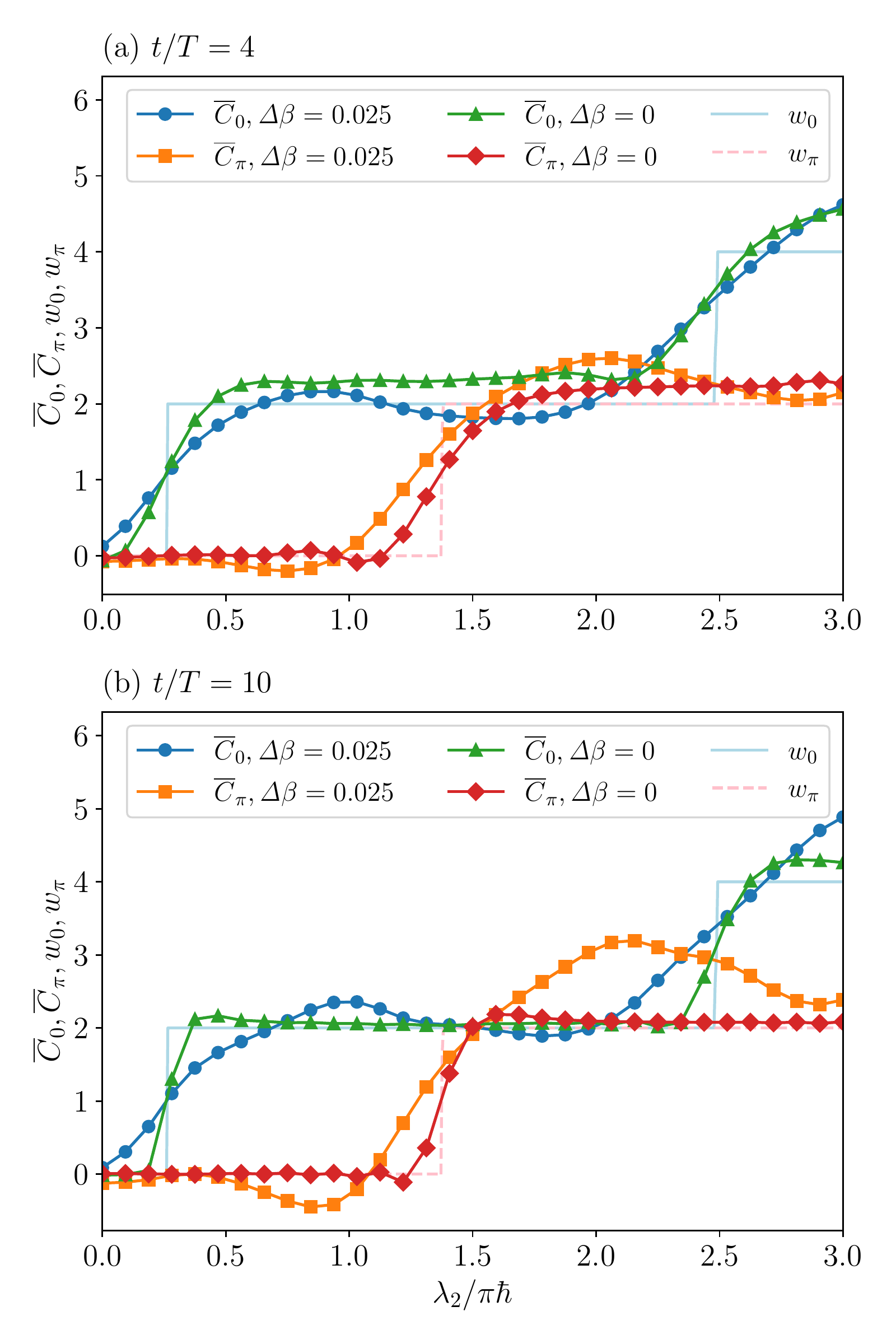}
    \subfloat{\label{fig:MCD_winding_number_a}}
    \subfloat{\label{fig:MCD_winding_number_b}}
    \caption{\label{fig:MCD_winding_number}Time-averaged MCDs $\overline{C}_0$ and $\overline{C}_\pi$ at (a) $t/T=4$ and (b) $t/T=10$ with a quasimomentum deviation $\Delta\beta$ and winding numbers $w_0$ (light blue solid line) and $w_\pi$ (pink dashed line) of the ORSDKR in class CII as functions of the kick strength $\lambda_2$.
    The parameters are the same as those in Fig.~\ref{fig:winding_number} and we choose $\lambda_1=0.7\pi\hbar$ (purple dotted lines in Fig.~\ref{fig:winding_number}).
    In the calculation, we restrict the range of the momentum space to (a) $[-50 h/a,50 h/a]$ and (b) $[-120 h/a,120 h/a]$.}
    }
\end{figure}
}

\section{\label{sec:conclusion}Conclusion}
We have studied Floquet topological phases realized in the ORSDKR model for a one-dimensional quasi-spin-1/2 BEC.
Since class CII requires four internal degrees of freedom, we have combined spin and sublattice degrees of freedom, which are achieved by the on-resonance condition.
Using both spin-dependent and spin-independent kicks, we can realize all the AZ classes with nontrivial topology in one dimension as well as class CII.
We have calculated the winding numbers characterizing the topological phases in the ORSDKR model, which take various values depending on the strengths and phases of the kicking lattices.
These values can be experimentally measured via the MCDs, which, in our model, can be obtained from the momentum distributions of the BEC in the long-time limit.

The ORSDKR model would be experimentally implementable using optical lattice pulses.
As discussed in Sec.~\ref{sec:experiment}, this model can be realized by simultaneously applying two kinds of optical lattice pulses with linear and circular parallel polarizations to the quasi-spin-1/2 BEC.
With the precise control of the phase difference and the compensation of the offset terms in the optical lattices, all the topologically nontrivial classes in one dimension can be accessed.
For the class CII case, it can also be realized by applying the two counterpropagating linearly polarized lasers with the polarization vectors inclined at a certain angle.
\add{Although the finite momentum width of the BEC badly affects observing the quantized winding numbers, the effect is minimal if all of the width of the momentum distribution, the kick strengths, and the time duration are sufficiently small.}

It is important to consider how to observe the edge states in momentum space and topological phase transitions.
Finite-size momentum lattice systems can be created by two-photon Bragg diffraction processes~\cite{gadwayAtomopticsApproachStudying2015,meierAtomopticsSimulatorLattice2016} and topological edge states have been observed~\cite{meierObservationTopologicalSoliton2016,anDirectObservationChiral2017}.
Using this method, edge states for the ORSDKR model might also be detected.
We note that our model undergoes topological phase transitions by changing the parameters of the kicking lattices, \add{as investigated in quenched systems~\cite{derricoBulkDetectionTimedependent2020,jangjanFloquetTopologicalPhase2022}}.
Through the MCDs or edge states, the dynamics of the topological phase transitions might be observed.

The ORSDKR model can be extended to study many-body and non-Hermitian physics in multiple dimensions.
One can introduce interatomic interactions neglected in this paper.
These interactions might induce dynamical instability and localization~\cite{zhangTransitionInstabilityKicked2004,monteiroNonlinearResonancesEnsuremath2009,seetohManybodyDynamicalDelocalization2022,caoInteractiondrivenBreakdownDynamical2022}.
It is also possible to investigate the non-Hermitian regime for the ORSDKR model as Ref.~\cite{zhouNonHermitianFloquetTopological2019}.
The non-Hermitian effects can be experimentally introduced by considering spontaneous emission~\cite{liangDynamicSignaturesNonHermitian2022,zhouEngineeringNonHermitianSkin2022}.
Furthermore, the ORSDKR model in higher dimentions could be achieved by applying additional lasers from multiple directions.

\begin{acknowledgments}
This work was supported by JSPS KAKENHI (Grants No. JP19H01824, No. JP19K14628, No. JP21H01009, and No. JP21H05185), and JST ERATO-FS (Grant No. JPMJER2105).
\end{acknowledgments}

\appendix
\section{\label{appx:basis}Derivation of Eq.~(\ref{eq:ORSDKR_Floquet_op_quasiposition})}
We start with a general kicking potential excluding spin degrees of freedom
\begin{align}
    \hat{V} &= \int_{0}^{a} dx\, V(x) \ket{x} \bra{x}
\end{align}
with $V(x + a) = V(x)$.
By using the Fourier transforms,
\begin{align}
    \braket{x|l} &= \frac{1}{\sqrt{a}}e^{i\frac{2\pi}{a}l x}, \label{eq:Fourier_momentum_lattice_1} \\
    \ket{x} &= \sum_{l=-\infty}^{\infty} \ket{l} \frac{1}{\sqrt{a}}e^{-i\frac{2\pi}{a}l x}, \label{eq:Fourier_momentum_lattice_2} \\
    \ket{l} &= \int_{0}^{a} dx \ket{x} \frac{1}{\sqrt{a}}e^{i\frac{2\pi}{a}l x}, \label{eq:Fourier_momentum_lattice_3}
\end{align}
we represent the potential in the momentum lattice basis $\ket{l}$,
\begin{align}
    \hat{V} &= \sum_{l=-\infty}^{\infty} \sum_{k=-\infty}^{\infty} V_{k} \ket{l + k} \bra{l}, \label{eq:potential_momentum_lattice}
\end{align}
where $V_{k}$ is the Fourier coefficient of $V(x)$ given by
\begin{align}
    V_{k} &= \frac{1}{a} \int_{0}^{a} V(x) e^{-i\frac{2\pi}{a}k x} \,dx.
\end{align}

Due to the translational invariance of Eq.~(\ref{eq:2_ORSDKR_Floquet_op}), $\hat{l} \to \hat{l} + 2$, the momentum lattice $\{\ket{l}\}$ is decomposed into the even and odd lattice sites corresponding to the sublattices A and B.
We define Pauli matrices on the sublattice as
\begin{align}
    \hat{\tau}_0 &= \ket{\mathrm{A}}\bra{\mathrm{A}} + \ket{\mathrm{B}}\bra{\mathrm{B}}, \\
    \hat{\tau}_x &= \ket{\mathrm{A}}\bra{\mathrm{B}} + \ket{\mathrm{B}}\bra{\mathrm{A}}, \\
    \hat{\tau}_y &= -i\ket{\mathrm{A}}\bra{\mathrm{B}} + i \ket{\mathrm{B}}\bra{\mathrm{A}}, \\
    \hat{\tau}_z &= \ket{\mathrm{A}}\bra{\mathrm{A}} - \ket{\mathrm{B}}\bra{\mathrm{B}}.
\end{align}
Here, we choose the new unit cell such that new $n$th unit cell contains old $(2n)$th and $(2n+1)$th momentum lattice.
By using the relations $\ket{l=2n}=\ket{n}\otimes\ket{\mathrm{A}},\ket{l=2n+1}=\ket{n}\otimes\ket{\mathrm{B}}$, we rewrite the potential as
\begin{align}
    \hat{V} &= \sum_{n=-\infty}^{\infty} \sum_{m=-\infty}^{\infty} \ket{n + m} \bra{n} \notag \\
    &\qquad\qquad \otimes (V_{2m} \hat{\tau}_0 + V_{2m + 1} \hat{\tau}_- + V_{2m - 1} \hat{\tau}_+).
    \label{eq:potential_momentum_sublattice}
\end{align}

We further Furier transform the momentum lattice basis $\{\ket{n}\}$ to the quasiposition basis $\{\ket{\theta}\}$:
\begin{align}
    \braket{n|\theta} &= \frac{1}{\sqrt{2\pi}}e^{i\theta n}, \label{eq:Fourier_quasiposition_1} \\
    \ket{\theta} &= \sum_{n=-\infty}^{\infty} \ket{n}\frac{1}{\sqrt{2\pi}}e^{i\theta n}, \label{eq:Fourier_quasiposition_2} \\
    \ket{n} &= \int_{-\pi}^{\pi} d\theta \ket{\theta}\frac{1}{\sqrt{2\pi}}e^{-i\theta n}, \label{eq:Fourier_quasiposition_3}
\end{align}
The resulting potential is given by
\begin{align}
    \hat{V} &= \int_{-\pi}^{\pi} d\theta \ket{\theta} \bra{\theta} \otimes \hat{V}(\theta), \label{eq:kick_potential_quasiposition}
\end{align}
with
\begin{align}
    \hat{V}(\theta) &= \sum_{m=-\infty}^{\infty} (V_{2m} \hat{\tau}_0 + V_{2m + 1} \hat{\tau}_- + V_{2m - 1} \hat{\tau}_+) e^{-i m\theta}.
\end{align}
By using $V_m^*=V_{-m}$ due to the reality of $V(x)$, we can rewrite $\hat{V}(\theta)$ as
\begin{align}
    \hat{V}(\theta) &= \sum_{m=-\infty}^{\infty} (V_{2m} e^{-i m\theta} \hat{\tau}_0 + V_{2m + 1} e^{-i m\theta} \hat{\tau}_- \notag \\
    &\qquad\qquad + V_{2m + 1}^* e^{i m\theta} \hat{\tau}_+) \notag \\
    &= \sum_{m=-\infty}^{\infty} [V_{2m} e^{-im\theta} \hat{\tau}_0 + 2\Real(V_{2m+1}e^{-im\theta}) \hat{\tau}_x \notag \\
    &\qquad\qquad + 2\Imag(V_{2m+1}e^{-im\theta}) \hat{\tau}_y].
    \label{eq:potential_quasiposition}
\end{align}

For the case of ORSDKR model, the $x$ dependence of the kicking potential is given by Eqs.~(\ref{eq:kick_potential_x_spin_independent}) and (\ref{eq:kick_potential_x_spin_dependent}), whose Fourier coefficient is given by
\begin{align}
    V_{k} &= \frac{\lambda}{2} (e^{i\alpha} \delta_{k, \nu} + e^{-i\alpha} \delta_{k, -\nu}).
\end{align}
It follows that $\hat{V}(\theta)$ includes only the $\hat{\tau}_0$ term (the $\hat{\tau}_x$ and $\hat{\tau}_y$ terms) for even (odd) $\nu$, resulting in
\begin{align}
    \hat{V}(\theta) &= \lambda \cos\left(\frac{\nu}{2}\theta - \alpha\right) \hat{h}_{\tau 1}(\theta, \nu),
\end{align}
where $\hat{h}_{\tau 1}(\theta,\nu)$ is defined in Eq.~(\ref{eq:factor_sublattice}).
Reintroducing the spin dependence, we can rewrite the kick terms $e^{-\frac{i}{\hbar}\hat{H}_{1,2}}$ in Eq.~(\ref{eq:2_ORSDKR_Floquet_op}) as
\begin{align}
    e^{-\frac{i}{\hbar}\hat{H}_1} &= \int_{-\pi}^{\pi} d\theta \ket{\theta}\bra{\theta} \otimes e^{-i\hat{h}_1(\theta)}, \\
    e^{-\frac{i}{\hbar}\hat{H}_2} &= \int_{-\pi}^{\pi} d\theta \ket{\theta}\bra{\theta} \otimes e^{-i\hat{\tilde{h}}_2(\theta)},
\end{align}
where $\hat{\tilde{h}}_2(\theta)$ is defined by
\begin{align}
    \hat{\tilde{h}}_2(\theta) &= \Lambda_2^0(\theta) \hat{h}_{\tau 1}(\theta, \nu_2^0) \otimes \hat{\sigma}_0 + \Lambda_2(\theta) \hat{h}_{\tau 1}(\theta, \nu_2) \otimes \bm{n}_2\cdot\hat{\bm{\sigma}}.
\end{align}

The remaining operators in Eq.~(\ref{eq:Fourier_quasiposition}), i.e., the free time evolution operators $e^{\pm i \frac{\pi}{2}\hat{l}^2\otimes\hat{\sigma}_0}$, are also diagonal in the quasiposition basis and given by
\begin{align}
    e^{\pm i \frac{\pi}{2}\hat{l}^2\otimes\hat{\sigma}_0} &= e^{\pm i\frac{\pi}{4}} \sum_{n} \ket{n}\bra{n} \otimes e^{\mp i\frac{\pi}{4}\hat{\tau}_z\otimes\hat{\sigma}_0} \notag \\
    &= e^{\pm i\frac{\pi}{4}} \int_{-\pi}^{\pi} d\theta \ket{\theta}\bra{\theta} \otimes e^{\mp i\frac{\pi}{4}\hat{\tau}_z\otimes\hat{\sigma}_0}.
\end{align}
Acting on $e^{-i\hat{\tilde{h}}_2(\theta)}$ from the both sides, they work as a $\pi/2$ rotation in the sublattice space:
\begin{align}
    e^{-i\frac{\pi}{4}\hat{\tau}_z\otimes\hat{\sigma}_0} e^{-i\hat{\tilde{h}}_2(\theta)} e^{i\frac{\pi}{4}\hat{\tau}_z\otimes\hat{\sigma}_0} &= e^{-i\hat{h}_2(\theta)}.
\end{align}
Getting all factors together, we finally obtain Eq.~(\ref{eq:ORSDKR_Floquet_op_quasiposition}).

\section{\label{appx:symmetry}Symmetry properties of Eqs.~(\ref{eq:symmetric_time_frame_1}) and (\ref{eq:symmetric_time_frame_2})}
\begin{table}
    \caption{\label{tab:kick}Sign change of the coefficient $\Lambda_j(\theta)=\lambda_j \cos(\nu_j\theta/2 - \alpha_j) / \hbar$. Because $\Lambda_j(\theta)$ should be an odd or even function of $\theta$, $\alpha_j$ can take only $0, \pi$ or $\pm\pi/2$.}
    \begin{ruledtabular}
        \begin{tabular}{ccc}
            Operation & $\alpha_j=0, \pi$ & $\alpha_j=\pm\pi/2$ \\
            \hline
            $\theta \to -\theta$ & $+$ & $-$
        \end{tabular}
    \end{ruledtabular}
\end{table}

\begin{table}
    \caption{\label{tab:sublattice}Sign change of sublattice factor $\hat{h}_{\tau j}(\theta,\nu)=\hat{\tau}_0$, $\bm{m}_1(\theta)\cdot\hat{\bm{\tau}}$, or $\bm{m}_2(\theta)\cdot\hat{\bm{\tau}}$ under unitary and antiunitary operations. Here, operator $\hat{S}$ acts on $\hat{h}_{\tau j}(\theta,\nu)$ as $\hat{S} \hat{h}_{\tau j}(\theta,\nu) \hat{S}^{-1}$.}
    \begin{ruledtabular}
        \begin{tabular}{cccc}
            Operation & $\hat{\tau}_0$ & $\bm{m}_1(\theta)\cdot\hat{\bm{\tau}}$ & $\bm{m}_2(\theta)\cdot\hat{\bm{\tau}}$ \\
            \hline
            Unitary: \\
            $\hat{\tau}_0$ & $+$ & $+$ & $+$ \\
            $\hat{\tau}_z$ & $+$ & $-$ & $-$ \\
            Antiunitary: \\
            $\hat{K},\theta \to -\theta$ & $+$ & $+$ & $-$ \\
            $\hat{\tau}_z\hat{K},\theta \to -\theta$ & $+$ & $-$ & $+$
        \end{tabular}
    \end{ruledtabular}
\end{table}

\begin{table}
    \caption{\label{tab:spin}Sign change of spin factor $\hat{h}_{\sigma j}=\hat{\sigma}_0,\bm{n}_j\cdot\hat{\bm{\sigma}}$ under unitary and antiunitary operations. Here, operator $\hat{S}$ acts on $\hat{h}_{\sigma j}$ as $\hat{S} \hat{h}_{\sigma j} \hat{S}^{-1}$. $\bm{n}_{j\perp}$ represents an arbitrary unit vector perpendicular to $\bm{n}_j$.}
    \begin{ruledtabular}
        \begin{tabular}{ccc}
            Operation & $\hat{\sigma}_0$ & $\bm{n}_j\cdot\hat{\bm{\sigma}}$ \\
            \hline
            Unitary: \\
            $\hat{\sigma}_0$ & $+$ & $+$ \\
            $\bm{n}_j\cdot\hat{\bm{\sigma}}$ & $+$ & $+$ \\
            $\bm{n}_{j\perp}\cdot\hat{\bm{\sigma}}$ & $+$ & $-$ \\
            Antiunitary: \\
            $\hat{\sigma}_y \hat{K}, \theta \to -\theta$ & $+$ & $-$ \\
            $\bm{n}_j\cdot\hat{\bm{\sigma}} \hat{\sigma}_y\hat{K}, \theta \to -\theta$ & $+$ & $-$ \\
            $\bm{n}_{j\perp}\cdot\hat{\bm{\sigma}}\hat{\sigma}_y \hat{K}, \theta \to -\theta$ & $+$ & $+$
        \end{tabular}
    \end{ruledtabular}
\end{table}

\begin{table*}
    \caption{\label{tab:factor}Sign change of $\hat{h}_{\tau j}(\theta,\nu) \otimes \hat{h}_{\sigma j}$ under unitary and antiunitary operations. Here, operator $\hat{S}$ acts on $\hat{h}_{\tau j}(\theta,\nu) \otimes \hat{h}_{\sigma j}$ as $\hat{S} \hat{h}_{\tau j}(\theta,\nu) \otimes \hat{h}_{\sigma j} \hat{S}^{-1}$. $\bm{n}_\perp$ represents an unit vector perpendicular to both $\bm{n}_1$ and $\bm{n}_2$. Here, we assume $\bm{n}_1\nparallel\bm{n}_2$ and, for simplicity, $\bm{n}_1\not\perp\bm{n}_2$.}
    \begin{ruledtabular}
        \begin{tabular}{cccccccc}
            Operation & $\hat{\tau}_0 \otimes \hat{\sigma}_0$ & $\bm{m}_1\cdot\hat{\bm{\tau}} \otimes \hat{\sigma}_0$ & $\bm{m}_2\cdot\hat{\bm{\tau}} \otimes \hat{\sigma}_0$ & $\hat{\tau}_0 \otimes \bm{n}_1\cdot\hat{\bm{\sigma}}$ & $\bm{m}_1\cdot\hat{\bm{\tau}} \otimes \bm{n}_1\cdot\hat{\bm{\sigma}}$ & $\hat{\tau}_0 \otimes \bm{n}_2\cdot\hat{\bm{\sigma}}$ & $\bm{m}_2\cdot\hat{\bm{\tau}} \otimes \bm{n}_2\cdot\hat{\bm{\sigma}}$ \\
            \hline
            Unitary: & & & & & & & \\
            $\hat{\tau}_z \otimes \hat{\sigma}_0$ & $+$ & $-$ & $-$ & $+$ & $-$ & $+$ & $-$ \\
            $\hat{\tau}_0 \otimes \bm{n}_\perp\cdot\hat{\bm{\sigma}}$ & $+$ & $+$ & $+$ & $-$ & $-$ & $-$ & $-$ \\
            $\hat{\tau}_z \otimes \bm{n}_\perp\cdot\hat{\bm{\sigma}}$ & $+$ & $-$ & $-$ & $-$ & $+$ & $-$ & $+$ \\
            Antiunitary: & & & & & & & \\
            $\hat{\tau}_0 \otimes \hat{\sigma}_y \hat{K}, \theta \to -\theta$ & $+$ & $+$ & $-$ & $-$ & $-$ & $-$ & $+$ \\
            $\hat{\tau}_z \otimes \hat{\sigma}_y \hat{K}, \theta \to -\theta$ & $+$ & $-$ & $+$ & $-$ & $+$ & $-$ & $-$ \\
            $\hat{\tau}_0 \otimes \bm{n}_\perp\cdot\hat{\bm{\sigma}}\hat{\sigma}_y \hat{K}, \theta \to -\theta$ & $+$ & $+$ & $-$ & $+$ & $+$ & $+$ & $-$ \\
            $\hat{\tau}_z \otimes \bm{n}_\perp\cdot\hat{\bm{\sigma}}\hat{\sigma}_y \hat{K}, \theta \to -\theta$ & $+$ & $-$ & $+$ & $+$ & $-$ & $+$ & $+$
        \end{tabular}
    \end{ruledtabular}
\end{table*}

We discuss the symmetry properties of the Floquet operators $\hat{U}_{1,2}(\theta)$ in the symmetric time frames [Eqs.~(\ref{eq:symmetric_time_frame_1}) and (\ref{eq:symmetric_time_frame_2})].
The result is summarized in Table~\ref{tab:class_ORSDKR}.
We note that $\hat{U}_{1,2}(\theta)$ satisfy Eq.~(\ref{eq:symmetries_Floquet}) when both $e^{-i \hat{h}_1(\theta)}$ and $e^{-i \hat{h}_2(\theta)}$ satisfy Eq.~(\ref{eq:symmetries_Floquet}), which is rewritten as Eq.~(\ref{eq:symmetries_Hamiltonian}).
Since each term appearing in $\hat{h}_j(\theta)$ can be factorized as $\Lambda_j(\theta) \hat{h}_{\tau j}(\theta,\nu) \otimes \hat{h}_{\sigma j}$, where $\Lambda_j(\theta)=\lambda_j \cos(\nu_j\theta/2 - \alpha_j)/\hbar$, $\hat{h}_{\tau j}(\theta,\nu)=\hat{\tau}_0$ or $\bm{m}_j(\theta)\cdot\hat{\bm{\tau}}$, and $\hat{h}_{\sigma j}=\hat{\sigma}_0$ or $\bm{n}_j\cdot\hat{\bm{\sigma}}$ ($j=1,2$), these factors should be invariant up to sign under the unitary or antiunitary operations corresponding to the required symmetry.
We list up the operations that make $\Lambda_j(\theta)$, $\hat{h}_{\tau j}(\theta,\nu)$, and $\hat{h}_{\sigma j}$ invariant in Tables~\ref{tab:kick}, \ref{tab:sublattice}, and \ref{tab:spin}, respectively.
Table~\ref{tab:factor} is the list for the $4\times 4$ matrix $\hat{h}_{\tau j}(\theta,\nu) \otimes \hat{h}_{\sigma j}$ that is obtained by combining the operations for the $2\times 2$ matrices $\hat{h}_{\tau j}(\theta,\nu)$ (Table~\ref{tab:sublattice}) and $\hat{h}_{\sigma j}$ (Table~\ref{tab:spin}).

In the following discussions, we assume $\bm{n}_1\nparallel\bm{n}_2$ and $\bm{n}_1\not\perp\bm{n}_2$.
The former is required for the Floquet operator not to be block-diagonalized.
Otherwise, $\hat{\tau}_0 \otimes \bm{n}_1\cdot\hat{\bm{\sigma}}$ commutes with both $\hat{h}_1(\theta)$ and $\hat{h}_2(\theta)$.
On the other hand, the latter is assumed for simplicity: If $\bm{n}_1\perp\bm{n}_2$, additional symmetry operations may exist for each class.

We can find a chiral operator as follows.
A possible chiral operator is one of the unitary operators, $\hat{\tau}_z \otimes \hat{\sigma}_0$, $\hat{\tau}_0 \otimes\bm{n}_\perp\cdot\hat{\bm{\sigma}}$, and $\hat{\tau}_z \otimes\bm{n}_\perp\cdot\hat{\bm{\sigma}}$, in Table~\ref{tab:factor}.
To satisfy Eq.~(\ref{eq:chiral_Hamiltonian}), we must choose either of $\hat{\tau}_0$ or $\bm{m}_j\cdot\hat{\bm{\tau}}$ as $\hat{h}_{\tau j}(\theta,\nu)$ so that $\hat{h}_{\tau j}(\theta,\nu) \otimes \hat{h}_{\sigma j}$ acquires a minus sign under that unitary operation.
Namely, only the terms $\hat{h}_{\tau j}(\theta,\nu) \otimes \hat{h}_{\sigma j}$ with the entry ``$-$'' in Table~\ref{tab:factor} are allowed.
This determines $\nu_j^0,\nu_j$ to be odd or even.

We can also find a time-reversal operator and a particle-hole operator in the almost same way.
The difference is whether or not to consider the sign change of $\Lambda_j(\theta)$ under $\theta \to -\theta$.
A possible time-reversal operator or particle-hole operator is one of the antiunitary operators, $\hat{\tau}_0 \otimes \hat{\sigma}_y \hat{K}$, $\hat{\tau}_z \otimes \hat{\sigma}_y \hat{K}$, $\hat{\tau}_0 \otimes\bm{n}_\perp\cdot\hat{\bm{\sigma}}\hat{\sigma}_y \hat{K}$, and $\hat{\tau}_z \otimes \bm{n}_\perp\cdot\hat{\bm{\sigma}}\hat{\sigma}_y \hat{K}$, whose squares are $-1$, $-1$, $+1$, and $+1$, respectively, in Table~\ref{tab:factor}.
For a given $\hat{h}_{\tau j}(\theta,\nu) \otimes \hat{h}_{\sigma j}$, we choose $\alpha_j^0,\alpha_j$ so that the whole sign change of $\Lambda_j(\theta) \hat{h}_{\tau j}(\theta,\nu) \otimes \hat{h}_{\sigma j}$ is plus or minus under that time-reversal or particle-hole operation, respectively.
Namely, if the sign change of $\hat{h}_{\tau j}(\theta,\nu) \otimes \hat{h}_{\sigma j}$ is plus (minus) for a time-reversal operator, we choose $\alpha_j^0,\alpha_j=0$ ($\pm\pi/2$), and for a particle-hole operator, $\alpha_j^0,\alpha_j=\pm\pi/2$ ($0$) (see Tables~\ref{tab:kick}).
For the case of class CII discussed in Sec.~\ref{sec:symmetry}, the time-reversal and particle-hole operators are uniquely determined as $\hat{T}=\hat{\tau}_0 \otimes \hat{\sigma}_y \hat{K}$ and $\hat{C}=\hat{\tau}_z \otimes \hat{\sigma}_y \hat{K}$, respectively, by choosing $\alpha_1^0=0$.

Now, we consider the other classes BDI, DIII, D, and AIII that can be achieved in the ORSDKR.
Without loss of generality, we choose $\alpha_1^0=0$.
In the case of $\lambda_1^0=0$, we choose $\alpha_1=0$.

\paragraph*{\add{Class BDI.}}
A Hamiltonian in class BDI has the symmetries satisfying $\hat{T}^2=+1$, $\hat{C}^2=+1$, and $\hat{\Gamma}^2=1$.
The antiunitary operators that square to $+1$ in Table~\ref{tab:factor} are $\hat{\tau}_0 \otimes \bm{n}_\perp\cdot\hat{\bm{\sigma}} \hat{\sigma}_y \hat{K}$ and $\hat{\tau}_z \otimes \bm{n}_\perp\cdot\hat{\bm{\sigma}} \hat{\sigma}_y \hat{K}$.
Therefore, the chiral operator is $\hat{\Gamma}=\hat{\tau}_z \otimes \hat{\sigma}_0$, which is possible when $\nu_j^0$ and $\nu_j$ are all odd from the first row of Table~\ref{tab:factor}.
From the choice of $\alpha_1^0=0$, the time-reversal and particle-hole operators are determined as $\hat{T}=\hat{\tau}_0 \otimes \bm{n}_\perp\cdot\hat{\bm{\sigma}} \hat{\sigma}_y \hat{K}$ and $\hat{C}=\hat{\tau}_0 \otimes \bm{n}_\perp\cdot\hat{\bm{\sigma}} \hat{\sigma}_y \hat{K}$, respectively.
Finally, according to the signs in Table~\ref{tab:factor}, the remaining phases are determined as $\alpha_1=0,\alpha_2^0=\alpha_2=\pi/2$ modulo $\pi$.

\paragraph*{\add{Class DIII.}}
A Hamiltonian in class DIII has the symmetries satisfying $\hat{T}^2=-1$, $\hat{C}^2=+1$, and $\hat{\Gamma}^2=1$.
In the case of $\lambda_1^0\neq 0$, choosing $\alpha_1^0=0$ determines $\nu_1^0$ to be odd, $\hat{T}=\hat{\tau}_0 \otimes \hat{\sigma}_y \hat{K}$, and $\hat{C}=\hat{\tau}_z \otimes \bm{n}_\perp\cdot\hat{\bm{\sigma}} \hat{\sigma}_y \hat{K}$.
It follows that the chiral operator is given by $\hat{\Gamma}=\hat{\tau}_z \otimes \bm{n}_\perp\cdot\hat{\bm{\sigma}}$, with which the possible terms in $\hat{h}_j(\theta)$ are given by odd $\nu_2^0$ and even $\nu_{1,2}$ with $\alpha_2^0=\alpha_{1,2}=\pi/2$ modulo $\pi$.
On the other hand, in the case of $\lambda_1^0=0$, $\lambda_1$ should be nonzero so that the Floquet operator is not block-diagonalized.
In this case, choosing $\alpha_1=0$ determines $\nu_1$ to be odd, $\hat{T}=\hat{\tau}_z \otimes \hat{\sigma}_y \hat{K}$, and $\hat{C}=\hat{\tau}_z \otimes \bm{n}_\perp\cdot\hat{\bm{\sigma}} \hat{\sigma}_y \hat{K}$.
The chiral operator $\hat{\Gamma}=\hat{\tau}_0 \otimes \bm{n}_\perp\cdot\hat{\bm{\sigma}}$ requires $\lambda_2^0=0$.
From the symmetry properties of $\hat{T}$ and $\hat{C}$, we obtain $\alpha_2=\pi/2$ modulo $\pi$, with which $\nu_2$ can take both even and odd.
We note, however, that for the case of $\lambda_1^0=\lambda_2^0=0$, $\nu_1$ and $\nu_2$ should have different parity so that the Floquet operator is not block-diagonalized.
Thus, $\nu_2$ is determined to be even.

\paragraph*{\add{Class D.}}
A Hamiltonian in class D has the particle-hole symmetry satisfying $\hat{C}^2=+1$.
Choosing $\alpha_1^0=0$ determines $\nu_1^0$ to be odd and $\hat{C}=\hat{\tau}_z \otimes \bm{n}_\perp\cdot\hat{\bm{\sigma}} \hat{\sigma}_y \hat{K}$.
In this case, we must choose $\nu_1$ to be odd (even) with $\alpha_1=0$ ($\pi/2$) modulo $\pi$.
$\nu_2^0$ and $\nu_2$ can take both even and odd as long as $\alpha_2^0=\alpha_2=\pi/2$ modulo $\pi$.
We should choose $\nu_2^0$ and $\nu_2$ not to belong to the other classes.
We can also assume $\lambda_1^0=0, \lambda_1\neq 0$ and $\alpha_1=0$.
However, this does not change the particle-hole operator, and the obtained result is included in the case of odd $\nu_1$ in the above argument.

\paragraph*{\add{Class AIII.}}
A Hamiltonian in class AIII has the chiral symmetry satisfying $\hat{\Gamma}^2=1$.
We can choose $\hat{\tau}_z \otimes \hat{\sigma}_0$, $\hat{\tau}_0 \otimes \bm{n}_\perp\cdot\hat{\bm{\sigma}}$, and $\hat{\tau}_z \otimes \bm{n}_\perp\cdot\hat{\bm{\sigma}}$ as chiral operators.
For $\hat{\Gamma}=\hat{\tau}_z \otimes \hat{\sigma}_0$, we must choose $\nu_{1,2}^0,\nu_{1,2}$ to be odd.
For $\hat{\Gamma}=\hat{\tau}_0 \otimes \bm{n}_\perp\cdot\hat{\bm{\sigma}}$, we must choose $\lambda_{1,2}^0=0$.
For $\hat{\Gamma}=\hat{\tau}_z \otimes \bm{n}_\perp\cdot\hat{\bm{\sigma}}$, we must choose $\nu_{1,2}^0$ to be odd and $\nu_{1,2}$ to be even.
In all cases, we should choose $\alpha_{1,2}^0,\alpha_{1,2}$ not to belong to the other classes.

\add{
\section{\label{appx:bulk_edge_correspondence}Bulk-edge correspondence for chiral symmetric Floquet systems}
We show the relation between the number of edge states and winding numbers in chiral symmetric Floquet systems.
As discussed in Refs.~\cite{asbothBulkboundaryCorrespondenceChiral2013,asbothChiralSymmetryBulkboundary2014,liuChiralFlowOnedimensional2018}, the index theorem in one-dimensional chiral symmetric Floquet systems guarantees
\begin{align}
    n_{\epsilon,+}^\mathrm{L} - n_{\epsilon,-}^\mathrm{L} &= w_\epsilon, \label{eq:index_theorem_left} \\
    n_{\epsilon,+}^\mathrm{R} - n_{\epsilon,-}^\mathrm{R} &= -w_\epsilon, \label{eq:index_theorem_right}
\end{align}
where $n_{\epsilon,\pm}^\mathrm{L/R}$ denotes the number of edge states with positive ($+$) or negative ($-$) chirality localized at a left/right edge with quasienergy $\epsilon=0,\pi$ (in units of $\hbar/T$).
Here, we have assumed that the system size is sufficiently large and adopted the Floquet operator $\hat{\mathcal{U}}_1$ in the symmetric time frame.
The total number of edge states with quasienergy $\epsilon$ in a finite system is $n_\epsilon=n_{\epsilon,+}^\mathrm{L} + n_{\epsilon,-}^\mathrm{L} + n_{\epsilon,+}^\mathrm{R} + n_{\epsilon,-}^\mathrm{R}$.
As shown in Ref.~\cite{xiongAnisotropicMagneticResponses2017}, in a chiral symmetric time-independent lattice model with finite range hopping, the number of edge states localized at one side with either chirality, $n_{0,+}^\mathrm{L/R}$ or $n_{0,-}^\mathrm{L/R}$, always vanishes.
Because this is also applicable to the corresponding Floquet systems by considering effective Hamiltonians, we can deduce
\begin{align}
    n_{\epsilon,+}^\mathrm{L} &= n_{\epsilon,-}^\mathrm{R} = 0 \quad\text{for}\quad w_\epsilon \leq 0, \label{eq:chiral_vanish_n} \\
    n_{\epsilon,-}^\mathrm{L} &= n_{\epsilon,+}^\mathrm{R} = 0 \quad\text{for}\quad w_\epsilon \geq 0, \label{eq:chiral_vanish_p}
\end{align}
which follow from the fact that $n_\epsilon^\mathrm{L/R}$ is a non-negative integer.
Combining Eqs.~(\ref{eq:index_theorem_left})--(\ref{eq:chiral_vanish_p}), we obtain
\begin{align}
    n_\epsilon &= 2|w_\epsilon|.
\end{align}
This relation is consistent with the numerical results shown in Fig.~\ref{fig:spectrum}.
}

\add{
\section{\label{appx:optical_lattice}Spin-dependent and spin-independent optical lattices}
We show how to realize the spin-dependent and spin-independent optical lattices such as Eqs.~(\ref{eq:energy_shift}) and (\ref{eq:optical_lattice_lin_theta_lin}) in Secs.~\ref{subsec:linear_circular_polarization} and \ref{subsec:inclined_linear_polarization}.
We consider the $^{87}$Rb atom under an oscillating electric field,
\begin{align}
    \bm{\mathcal{E}}(\bm{x},t) &= \frac{1}{2} \mathcal{E}_0 \bm{\varepsilon} e^{i(\bm{k}\cdot\bm{x} - \omega t)} + \mathrm{c.c.},
\end{align}
where $\mathcal{E}_0$ is a complex amplitude, $\bm{\varepsilon}$ is a polarization vector, $\bm{k}$ is a wave vector, $\omega$ is a frequency of the electric field, and $\mathrm{c.c.}$ stands for the complex conjugate of the preceding term.
According to the second-order time-dependent perturbation theory, the ac Stark shift for the hyperfine states $\ket{F,m_F}$ is decomposed into the irreducible tensor terms as~\cite{rosenbuschAcStarkShift2009,lekienDynamicalPolarizabilityAtoms2013}
\begin{align}
    \Delta E_{F,m_F} &= -\frac{|\mathcal{E}_0|^2}{4} \left(\alpha^\mathrm{s} + \alpha^\mathrm{v} (i\bm{\varepsilon} \times \bm{\varepsilon}^*) \cdot \bm{e}_B \frac{m_F}{2F}\right. \notag \\
    &\quad \left.+ \alpha^\mathrm{t} \frac{3|\bm{\varepsilon}\cdot\bm{e}_B|^2 - 1}{2} \frac{3m_F^2 - F(F + 1)}{F(2F - 1)}\right),
\end{align}
where $\alpha^\mathrm{s}$, $\alpha^\mathrm{v}$, and $\alpha^\mathrm{t}$ are the scalar, vector, and tensor polarizabilities, respectively, and $\bm{e}_B$ is a unit vector along the external magnetic field.
Here, $\lambda_\mathrm{D_1}=2\pi c/\omega_\mathrm{D_1}=794.98~\mathrm{nm}$ and $\lambda_\mathrm{D_2}=2\pi c/\omega_\mathrm{D_2}=780.24~\mathrm{nm}$ are the wavelengths of the $\mathrm{D}_1$ line (the $5 ^2\mathrm{S}_{1/2} \to 5 ^2\mathrm{P}_{1/2}$ transition) and $\mathrm{D}_2$ line (the $5 ^2\mathrm{S}_{1/2} \to 5 ^2\mathrm{P}_{3/2}$ transition) of the $^{87}$Rb atom, respectively.
Within the rotating-wave approximation and the nonrelativistic approximation, the polarizabilities for the ground state $5 ^2\mathrm{S}_{1/2}$ are given by~\cite{lekienDynamicalPolarizabilityAtoms2013,Note1}
\begin{align}
    \alpha^\mathrm{s} &\approx -\frac{1}{6\hbar} |\braket{5 ^2\mathrm{P}_{1/2}\|\hat{\bm{d}}\|5 ^2\mathrm{S}_{1/2}}|^2 \left(\frac{2}{\delta_\mathrm{D_2}} + \frac{1}{\delta_\mathrm{D_1}}\right), \notag \\
    \alpha^\mathrm{v} &\approx \frac{1}{3\hbar} g_F F |\braket{5 ^2\mathrm{P}_{1/2}\|\hat{\bm{d}}\|5 ^2\mathrm{S}_{1/2}}|^2 \left(\frac{1}{\delta_\mathrm{D_2}} - \frac{1}{\delta_\mathrm{D_1}}\right), \notag \\
    \alpha^\mathrm{t} &\approx 0,
    \label{eq:polarizability_approx}
\end{align}
where $\delta_{\mathrm{D}_i}=\omega-\omega_{\mathrm{D}_i}$ ($i=1,2$) is the detuning from the $\mathrm{D}_i$ line, $\hat{\bm{d}}$ is the dipole operator of the atom, and $\braket{5 ^2\mathrm{P}_{1/2}\|\hat{\bm{d}}\|5 ^2\mathrm{S}_{1/2}}$ is the reduced matrix element.
\footnotetext{We note that the sign of the vector polarizability $\alpha^\mathrm{v}$ is different from Eq.~(2) in Ref.~\cite{wenExperimentalStudyTuneout2021} but our choice of the sign is consistent with the result of the reference.
Whereas the expressions of these polarizabilities only include the contributions from the $\mathrm{D_1}$ and $\mathrm{D_2}$ lines, the calculated tune-out wavelengths of the optical lattices coincide with the measured values in Ref.~\cite{wenExperimentalStudyTuneout2021} up to four significant digits.}
$g_F$ is the hyperfine $g$-factor defined by
\begin{align}
    g_F &= g_J \frac{F(F + 1) + J(J + 1) - I(I + 1)}{2F(F + 1)}, \\
    g_J &= 1 + \frac{J(J + 1) + S(S + 1) - L(L + 1)}{2J(J + 1)},
\end{align}
where $g_J$ is the Land\'{e} $g$-factor, $L$ is the electron orbital angular momentum, $S$ is the electron spin angular momentum, $J$ is the total angular momentum of the electrons, and $I$ is the nuclear spin angular momentum.
For the ground state $5 ^2\mathrm{S}_{1/2}$ of the $^{87}$Rb atom, we obtain $g_J=2$, $g_{F=2}=1/2$, and $g_{F=1}=-1/2$.
In the following discussions, we choose the propagation direction, $\bm{k}$, parallel or antiparallel to $\bm{e}_B$.

\subsection{\label{subsec:lin_para_lin}Counterpropagating linearly polarized lasers}
We first consider two counterpropagating electric fields with the same amplitude and linear parallel polarization [see the configuration for the spin-independent lattice beam in Fig.~\subref*{fig:setup_lin_circ}]~\cite{wenExperimentalStudyTuneout2021}.
The total electric field is given by
\begin{align}
    \bm{\mathcal{E}}(\bm{x},t) &= \frac{1}{2} \mathcal{E}_0 \bm{\varepsilon} e^{i(\bm{k}\cdot\bm{x} - \omega t)} + \frac{1}{2} \mathcal{E}_0 \bm{\varepsilon} e^{i(-\bm{k}\cdot\bm{x} - \omega t)} + \mathrm{c.c.} \notag \\
    &= \mathcal{E}_0 \cos(\bm{k}\cdot\bm{x}) \bm{\varepsilon} e^{-i\omega t} + \mathrm{c.c.}
\end{align}
Since the polarization vector $\bm{\varepsilon}$ for the linear polarization can be chosen to be real, the ac Stark shift is
\begin{align}
    \Delta E_{F,m_F}^{\mathrm{lin}\parallel\mathrm{lin}} &= -|\mathcal{E}_0|^2 \cos^2(\bm{k}\cdot\bm{x}) \alpha^\mathrm{s}. \label{eq:ac_stark_shift_lin_para_lin}
\end{align}
Since this energy shift does not depend on $F$ or $m_F$, the resulting optical lattice potential (\ref{eq:energy_shift}) is spin-independent.

\subsection{\label{subsec:circ_para_circ}Counterpropagating circularly polarized lasers}
Next, we consider two counterpropagating electric fields with the same amplitude and circular parallel polarization [see the configuration for the spin-dependent lattice beam in Fig.~\subref*{fig:setup_lin_circ}]~\cite{wenExperimentalStudyTuneout2021}.
The total electric field is given by
\begin{align}
    \bm{\mathcal{E}}(\bm{x},t) &= \frac{1}{2} \mathcal{E}_0 \sqrt{2} \bm{\varepsilon}_\mathrm{L(R)} e^{i(\bm{k}\cdot\bm{x} - \omega t)} \notag \\
    &\quad + \frac{1}{2} \mathcal{E}_0 \sqrt{2} \bm{\varepsilon}_\mathrm{L(R)} e^{i(-\bm{k}\cdot\bm{x} - \omega t)} + \mathrm{c.c.} \notag \\
    &= \mathcal{E}_0 \cos(\bm{k}\cdot\bm{x}) \bm{\varepsilon}_\mathrm{L(R)} e^{-i\omega t} + \mathrm{c.c.},
\end{align}
where $\bm{\varepsilon}_\mathrm{L(R)}=(\bm{e}_1 \pm i \bm{e}_2)/\sqrt{2}$ is a polarization vector for the left (right) circular polarization defined for the electric field propagating along $\bm{k}$, i.e., $\bm{e}_1$ and $\bm{e}_2$ are two orthonormal vectors in the plane perpendicular to the wave vector $\bm{k}$ such that $\bm{e}_1\times\bm{e}_2=\bm{e}_k(\equiv\bm{k}/|\bm{k}|)$.
Here, we choose $\bm{e}_k=\bm{e}_B$.
The resulting ac Stark shift is given by
\begin{align}
    \Delta E_{F,m_F}^{\mathrm{circ}\parallel\mathrm{circ}} &= -|\mathcal{E}_0|^2 \cos^2(\bm{k}\cdot\bm{x}) \left(\alpha^\mathrm{s} \pm \alpha^\mathrm{v} \frac{m_F}{2F}\right), \label{eq:ac_stark_shift_circ_para_circ}
\end{align}
where the upper (lower) sign is for the left (right) circular polarization.
Here, we consider the left circular polarization and two hyperfine levels $\ket{F=2,m_F=2}$ and $\ket{F=1,m_F=1}$.
Taking into account the $F$ dependence of $\alpha^\mathrm{v}$, the amplitudes of the energy shifts have the same modulus but opposite sign between the two hyperfine states, i.e., $\Delta E_{F=2,m_F=2}^{\mathrm{circ}\parallel\mathrm{circ}}=-\Delta E_{F=1,m_F=1}^{\mathrm{circ}\parallel\mathrm{circ}}$, when the wavelength is
\begin{align}
    \lambda &= \frac{12\lambda_\mathrm{D_1} \lambda_\mathrm{D_2}}{5\lambda_\mathrm{D_1} + 7\lambda_\mathrm{D_2}} = 788.8~\mathrm{nm}.
\end{align}
At this wavelength, the resulting optical lattice potential (\ref{eq:energy_shift}) gives opposite sign amplitudes for the two hyperfine states.

\subsection{\label{subsec:circ}One-way propagating circularly polarized laser}
We consider a one-way propagating circularly polarized electric field [see the configuration for the spin-dependent offset cancel beam in Fig.~\subref*{fig:setup_lin_circ}].
The electric field is given by
\begin{align}
    \bm{\mathcal{E}}(\bm{x},t) &= \frac{1}{2} \mathcal{E}_0 \sqrt{2} \bm{\varepsilon}_\mathrm{L(R)} e^{i(\bm{k}\cdot\bm{x} - \omega t)} + \mathrm{c.c.},
\end{align}
which induces the uniform ac Stark shift
\begin{align}
    \Delta E_{F,m_F}^\mathrm{circ} &= -\frac{|\mathcal{E}_0|^2}{2} \left(\alpha^\mathrm{s} \pm \alpha^\mathrm{v} \frac{m_F}{2F}\right), \label{eq:ac_stark_shift_circ}
\end{align}
where the upper (lower) sign is for the left (right) circular polarization.
We consider the right circular polarization (opposite to the ones considered in Sec.~\ref{subsec:circ_para_circ}) and two hyperfine levels $\ket{F=2,m_F=2}$ and $\ket{F=1,m_F=1}$.
The amplitudes of the energy shifts have the same modulus but opposite sign between the two hyperfine states, i.e., $\Delta E_{F=2,m_F=2}^\mathrm{circ}=-\Delta E_{F=1,m_F=1}^\mathrm{circ}$, when the wavelength is
\begin{align}
    \lambda &= \frac{4\lambda_\mathrm{D_1} \lambda_\mathrm{D_2}}{\lambda_\mathrm{D_1} + 3\lambda_\mathrm{D_2}} = 791.2~\mathrm{nm}.
\end{align}
Because the resulting potential has the opposite sign against the one in the circ~$\parallel$~circ configuration for each hyperfine state, it acts the offset cancel beam.

\subsection{\label{subsec:lin_theta_lin}Counterpropagating linearly polarized lasers with an enclosing angle}
We consider two counterpropagating linearly polarized electric fields with same amplitude and an enclosing angle $\Theta$ [see the configuration in Fig.~\subref*{fig:setup_lin_enclosing}]~\cite{mandelCoherentTransportNeutral2003,mandelControlledCollisionsMultiparticle2003}.
The total electric field is given by
\begin{align}
    \bm{\mathcal{E}}(\bm{x},t) &= \frac{1}{2} \mathcal{E}_0 \bm{e}_1 e^{i(\bm{k}\cdot\bm{x} - \omega t)} \notag \\
    &\quad + \frac{1}{2} \mathcal{E}_0 (\cos\Theta \bm{e}_1 - \sin\Theta \bm{e}_2) e^{i(-\bm{k}\cdot\bm{x} - \omega t)} + \mathrm{c.c.} \notag \\
    &= \frac{1}{2} \sqrt{2} \mathcal{E}_0 \left[e^{i\Theta/2} \cos\left(\bm{k}\cdot\bm{x} - \frac{\Theta}{2}\right) \bm{\varepsilon}_\mathrm{L}\right. \notag \\
    &\quad \left.+ e^{-i\Theta/2} \cos\left(\bm{k}\cdot\bm{x} + \frac{\Theta}{2}\right) \bm{\varepsilon}_\mathrm{R}\right] e^{-i\omega t} + \mathrm{c.c.},
\end{align}
and thus the ac Stark shift reads
\begin{align}
    \Delta E_{F,m_F} &= -\frac{|\mathcal{E}_0|^2}{2} \Bigl(\alpha^\mathrm{s} [1 + \cos\Theta \cos(2\bm{k}\cdot\bm{x})] \notag \\
    &\quad + \alpha^\mathrm{v} \sin\Theta \sin(2\bm{k}\cdot\bm{x}) \frac{m_F}{2F}\Bigr). \label{eq:ac_stark_shift_lin_theta_lin}
\end{align}
Substituting Eq.~(\ref{eq:polarizability_approx}) to Eq.~(\ref{eq:ac_stark_shift_lin_theta_lin}), we obtain
\begin{align}
    \Delta E_{F,m_F} &= \frac{|\mathcal{E}_0|^2}{2} \frac{1}{6\hbar} |\braket{5 ^2\mathrm{P}_{1/2}\|\hat{\bm{d}}\|5 ^2\mathrm{S}_{1/2}}|^2 \notag \\
    &\quad \times \left[\frac{2}{\delta_\mathrm{D_2}} + \frac{1}{\delta_\mathrm{D_1}} + A \cos(2\bm{k}\cdot\bm{x} + \alpha)\right],
\end{align}
with
\begin{align}
    A &= \left[\left(\frac{2}{\delta_\mathrm{D_2}} + \frac{1}{\delta_\mathrm{D_1}}\right)^2 \cos^2\Theta\right. \notag \\
    &\quad \left.+ g_F^2 m_F^2 \left(\frac{1}{\delta_\mathrm{D_2}} - \frac{1}{\delta_\mathrm{D_1}}\right)^2 \sin^2\Theta\right]^{1/2}, \\
    \tan\alpha &= g_F m_F \left(\frac{1}{\delta_\mathrm{D_2}} - \frac{1}{\delta_\mathrm{D_1}}\right) \left(\frac{2}{\delta_\mathrm{D_2}} + \frac{1}{\delta_\mathrm{D_1}}\right)^{-1} \tan\Theta.
\end{align}
For simplicity, we choose the wavelength such that
\begin{align}
    g_F \left(\frac{1}{\delta_\mathrm{D_2}} - \frac{1}{\delta_\mathrm{D_1}}\right) \left(\frac{2}{\delta_\mathrm{D_2}} + \frac{1}{\delta_\mathrm{D_1}}\right)^{-1} &= -1
\end{align}
is satisfied, which yields
\begin{align}
    \lambda &= \frac{3\lambda_\mathrm{D_1} \lambda_\mathrm{D_2}}{(1 - g_F) \lambda_\mathrm{D_1} + (2 + g_F) \lambda_\mathrm{D_2}}, \label{eq:wavelength_lin_theta_lin} \\
    A &= \left|\frac{2}{\delta_\mathrm{D_2}} + \frac{1}{\delta_\mathrm{D_1}}\right| \sqrt{\cos^2\Theta + m_F^2 \sin^2\Theta}, \\
    \tan\alpha &= m_F \tan\Theta.
\end{align}
Considering the two hyperfine levels $\ket{F=1,m_F=1}$ and $\ket{F=1,m_F=-1}$ of the $^{87}$Rb atom, the wavelength of Eq.~(\ref{eq:wavelength_lin_theta_lin}) is determined to be $\lambda=787.5~\mathrm{nm}$, at which the ac Stark shifts for the two hyperfine states are given by
\begin{align}
    \Delta E_{F=1,m_F=\pm 1} &= \frac{|\mathcal{E}_0|^2}{2} \frac{1}{6\hbar} |\braket{5 ^2\mathrm{P}_{1/2}\|\hat{\bm{d}}\|5 ^2\mathrm{S}_{1/2}}|^2 \notag \\
    &\quad \times \left(\frac{2}{\delta_\mathrm{D_2}} + \frac{1}{\delta_\mathrm{D_1}}\right) \left[1 + \cos(2\bm{k}\cdot\bm{x} \pm \Theta)\right].
\end{align}
The resulting optical lattice potential is summarized as Eq.~(\ref{eq:optical_lattice_lin_theta_lin}), which has the same amplitude and opposite phase for the two hyperfine states.
[We add the minus sign in the right-hand-side of Eq.~(\ref{eq:optical_lattice_lin_theta_lin}) because of $2/\delta_\mathrm{D_2}+1/\delta_\mathrm{D_1} < 0$ for this wavelength.]
}

\bibliography{reference}

\begin{thebibliography}{79}%
\makeatletter
\providecommand \@ifxundefined [1]{%
 \@ifx{#1\undefined}
}%
\providecommand \@ifnum [1]{%
 \ifnum #1\expandafter \@firstoftwo
 \else \expandafter \@secondoftwo
 \fi
}%
\providecommand \@ifx [1]{%
 \ifx #1\expandafter \@firstoftwo
 \else \expandafter \@secondoftwo
 \fi
}%
\providecommand \natexlab [1]{#1}%
\providecommand \enquote  [1]{``#1''}%
\providecommand \bibnamefont  [1]{#1}%
\providecommand \bibfnamefont [1]{#1}%
\providecommand \citenamefont [1]{#1}%
\providecommand \href@noop [0]{\@secondoftwo}%
\providecommand \href [0]{\begingroup \@sanitize@url \@href}%
\providecommand \@href[1]{\@@startlink{#1}\@@href}%
\providecommand \@@href[1]{\endgroup#1\@@endlink}%
\providecommand \@sanitize@url [0]{\catcode `\\12\catcode `\$12\catcode `\&12\catcode `\#12\catcode `\^12\catcode `\_12\catcode `\%12\relax}%
\providecommand \@@startlink[1]{}%
\providecommand \@@endlink[0]{}%
\providecommand \url  [0]{\begingroup\@sanitize@url \@url }%
\providecommand \@url [1]{\endgroup\@href {#1}{\urlprefix }}%
\providecommand \urlprefix  [0]{URL }%
\providecommand \Eprint [0]{\href }%
\providecommand \doibase [0]{https://doi.org/}%
\providecommand \selectlanguage [0]{\@gobble}%
\providecommand \bibinfo  [0]{\@secondoftwo}%
\providecommand \bibfield  [0]{\@secondoftwo}%
\providecommand \translation [1]{[#1]}%
\providecommand \BibitemOpen [0]{}%
\providecommand \bibitemStop [0]{}%
\providecommand \bibitemNoStop [0]{.\EOS\space}%
\providecommand \EOS [0]{\spacefactor3000\relax}%
\providecommand \BibitemShut  [1]{\csname bibitem#1\endcsname}%
\let\auto@bib@innerbib\@empty
\bibitem [{\citenamefont {Oka}\ and\ \citenamefont {Aoki}(2009)}]{okaPhotovoltaicHallEffect2009}%
  \BibitemOpen
  \bibfield  {author} {\bibinfo {author} {\bibfnamefont {T.}~\bibnamefont {Oka}}\ and\ \bibinfo {author} {\bibfnamefont {H.}~\bibnamefont {Aoki}},\ }\bibfield  {title} {\bibinfo {title} {Photovoltaic {{Hall}} effect in graphene},\ }\href {https://doi.org/10.1103/PhysRevB.79.081406} {\bibfield  {journal} {\bibinfo  {journal} {Phys. Rev. B}\ }\textbf {\bibinfo {volume} {79}},\ \bibinfo {pages} {081406(R)} (\bibinfo {year} {2009})}\BibitemShut {NoStop}%
\bibitem [{\citenamefont {Lindner}\ \emph {et~al.}(2011)\citenamefont {Lindner}, \citenamefont {Refael},\ and\ \citenamefont {Galitski}}]{lindnerFloquetTopologicalInsulator2011}%
  \BibitemOpen
  \bibfield  {author} {\bibinfo {author} {\bibfnamefont {N.~H.}\ \bibnamefont {Lindner}}, \bibinfo {author} {\bibfnamefont {G.}~\bibnamefont {Refael}},\ and\ \bibinfo {author} {\bibfnamefont {V.}~\bibnamefont {Galitski}},\ }\bibfield  {title} {\bibinfo {title} {Floquet topological insulator in semiconductor quantum wells},\ }\href {https://doi.org/10.1038/nphys1926} {\bibfield  {journal} {\bibinfo  {journal} {Nat. Phys.}\ }\textbf {\bibinfo {volume} {7}},\ \bibinfo {pages} {490} (\bibinfo {year} {2011})}\BibitemShut {NoStop}%
\bibitem [{\citenamefont {Goldman}\ \emph {et~al.}(2014)\citenamefont {Goldman}, \citenamefont {Juzeli{\=u}nas}, \citenamefont {{\"O}hberg},\ and\ \citenamefont {Spielman}}]{goldmanLightinducedGaugeFields2014}%
  \BibitemOpen
  \bibfield  {author} {\bibinfo {author} {\bibfnamefont {N.}~\bibnamefont {Goldman}}, \bibinfo {author} {\bibfnamefont {G.}~\bibnamefont {Juzeli{\=u}nas}}, \bibinfo {author} {\bibfnamefont {P.}~\bibnamefont {{\"O}hberg}},\ and\ \bibinfo {author} {\bibfnamefont {I.~B.}\ \bibnamefont {Spielman}},\ }\bibfield  {title} {\bibinfo {title} {Light-induced gauge fields for ultracold atoms},\ }\href {https://doi.org/10.1088/0034-4885/77/12/126401} {\bibfield  {journal} {\bibinfo  {journal} {Rep. Prog. Phys.}\ }\textbf {\bibinfo {volume} {77}},\ \bibinfo {pages} {126401} (\bibinfo {year} {2014})}\BibitemShut {NoStop}%
\bibitem [{\citenamefont {Goldman}\ \emph {et~al.}(2016)\citenamefont {Goldman}, \citenamefont {Budich},\ and\ \citenamefont {Zoller}}]{goldmanTopologicalQuantumMatter2016}%
  \BibitemOpen
  \bibfield  {author} {\bibinfo {author} {\bibfnamefont {N.}~\bibnamefont {Goldman}}, \bibinfo {author} {\bibfnamefont {J.~C.}\ \bibnamefont {Budich}},\ and\ \bibinfo {author} {\bibfnamefont {P.}~\bibnamefont {Zoller}},\ }\bibfield  {title} {\bibinfo {title} {Topological quantum matter with ultracold gases in optical lattices},\ }\href {https://doi.org/10.1038/nphys3803} {\bibfield  {journal} {\bibinfo  {journal} {Nat. Phys.}\ }\textbf {\bibinfo {volume} {12}},\ \bibinfo {pages} {639} (\bibinfo {year} {2016})}\BibitemShut {NoStop}%
\bibitem [{\citenamefont {Eckardt}(2017)}]{eckardtColloquiumAtomicQuantum2017}%
  \BibitemOpen
  \bibfield  {author} {\bibinfo {author} {\bibfnamefont {A.}~\bibnamefont {Eckardt}},\ }\bibfield  {title} {\bibinfo {title} {Colloquium: {{Atomic}} quantum gases in periodically driven optical lattices},\ }\href {https://doi.org/10.1103/RevModPhys.89.011004} {\bibfield  {journal} {\bibinfo  {journal} {Rev. Mod. Phys.}\ }\textbf {\bibinfo {volume} {89}},\ \bibinfo {pages} {011004} (\bibinfo {year} {2017})}\BibitemShut {NoStop}%
\bibitem [{\citenamefont {Zhang}\ \emph {et~al.}(2018)\citenamefont {Zhang}, \citenamefont {Zhu}, \citenamefont {Zhao}, \citenamefont {Yan},\ and\ \citenamefont {Zhu}}]{zhangTopologicalQuantumMatter2018}%
  \BibitemOpen
  \bibfield  {author} {\bibinfo {author} {\bibfnamefont {D.-W.}\ \bibnamefont {Zhang}}, \bibinfo {author} {\bibfnamefont {Y.-Q.}\ \bibnamefont {Zhu}}, \bibinfo {author} {\bibfnamefont {Y.~X.}\ \bibnamefont {Zhao}}, \bibinfo {author} {\bibfnamefont {H.}~\bibnamefont {Yan}},\ and\ \bibinfo {author} {\bibfnamefont {S.-L.}\ \bibnamefont {Zhu}},\ }\bibfield  {title} {\bibinfo {title} {Topological quantum matter with cold atoms},\ }\href {https://doi.org/10.1080/00018732.2019.1594094} {\bibfield  {journal} {\bibinfo  {journal} {Adv. Phys.}\ }\textbf {\bibinfo {volume} {67}},\ \bibinfo {pages} {253} (\bibinfo {year} {2018})}\BibitemShut {NoStop}%
\bibitem [{\citenamefont {Kitagawa}\ \emph {et~al.}(2010{\natexlab{a}})\citenamefont {Kitagawa}, \citenamefont {Rudner}, \citenamefont {Berg},\ and\ \citenamefont {Demler}}]{kitagawaExploringTopologicalPhases2010}%
  \BibitemOpen
  \bibfield  {author} {\bibinfo {author} {\bibfnamefont {T.}~\bibnamefont {Kitagawa}}, \bibinfo {author} {\bibfnamefont {M.~S.}\ \bibnamefont {Rudner}}, \bibinfo {author} {\bibfnamefont {E.}~\bibnamefont {Berg}},\ and\ \bibinfo {author} {\bibfnamefont {E.}~\bibnamefont {Demler}},\ }\bibfield  {title} {\bibinfo {title} {Exploring topological phases with quantum walks},\ }\href {https://doi.org/10.1103/PhysRevA.82.033429} {\bibfield  {journal} {\bibinfo  {journal} {Phys. Rev. A}\ }\textbf {\bibinfo {volume} {82}},\ \bibinfo {pages} {033429} (\bibinfo {year} {2010}{\natexlab{a}})}\BibitemShut {NoStop}%
\bibitem [{\citenamefont {Kitagawa}\ \emph {et~al.}(2012)\citenamefont {Kitagawa}, \citenamefont {Broome}, \citenamefont {Fedrizzi}, \citenamefont {Rudner}, \citenamefont {Berg}, \citenamefont {Kassal}, \citenamefont {{Aspuru-Guzik}}, \citenamefont {Demler},\ and\ \citenamefont {White}}]{kitagawaObservationTopologicallyProtected2012}%
  \BibitemOpen
  \bibfield  {author} {\bibinfo {author} {\bibfnamefont {T.}~\bibnamefont {Kitagawa}}, \bibinfo {author} {\bibfnamefont {M.~A.}\ \bibnamefont {Broome}}, \bibinfo {author} {\bibfnamefont {A.}~\bibnamefont {Fedrizzi}}, \bibinfo {author} {\bibfnamefont {M.~S.}\ \bibnamefont {Rudner}}, \bibinfo {author} {\bibfnamefont {E.}~\bibnamefont {Berg}}, \bibinfo {author} {\bibfnamefont {I.}~\bibnamefont {Kassal}}, \bibinfo {author} {\bibfnamefont {A.}~\bibnamefont {{Aspuru-Guzik}}}, \bibinfo {author} {\bibfnamefont {E.}~\bibnamefont {Demler}},\ and\ \bibinfo {author} {\bibfnamefont {A.~G.}\ \bibnamefont {White}},\ }\bibfield  {title} {\bibinfo {title} {Observation of topologically protected bound states in photonic quantum walks},\ }\href {https://doi.org/10.1038/ncomms1872} {\bibfield  {journal} {\bibinfo  {journal} {Nat. Commun.}\ }\textbf {\bibinfo {volume} {3}},\ \bibinfo {pages} {882} (\bibinfo {year} {2012})}\BibitemShut {NoStop}%
\bibitem [{\citenamefont {Kitagawa}(2012)}]{kitagawaTopologicalPhenomenaQuantum2012}%
  \BibitemOpen
  \bibfield  {author} {\bibinfo {author} {\bibfnamefont {T.}~\bibnamefont {Kitagawa}},\ }\bibfield  {title} {\bibinfo {title} {Topological phenomena in quantum walks: elementary introduction to the physics of topological phases},\ }\href {https://doi.org/10.1007/s11128-012-0425-4} {\bibfield  {journal} {\bibinfo  {journal} {Quantum Inf. Process.}\ }\textbf {\bibinfo {volume} {11}},\ \bibinfo {pages} {1107} (\bibinfo {year} {2012})}\BibitemShut {NoStop}%
\bibitem [{\citenamefont {Kitagawa}\ \emph {et~al.}(2010{\natexlab{b}})\citenamefont {Kitagawa}, \citenamefont {Berg}, \citenamefont {Rudner},\ and\ \citenamefont {Demler}}]{kitagawaTopologicalCharacterizationPeriodically2010}%
  \BibitemOpen
  \bibfield  {author} {\bibinfo {author} {\bibfnamefont {T.}~\bibnamefont {Kitagawa}}, \bibinfo {author} {\bibfnamefont {E.}~\bibnamefont {Berg}}, \bibinfo {author} {\bibfnamefont {M.}~\bibnamefont {Rudner}},\ and\ \bibinfo {author} {\bibfnamefont {E.}~\bibnamefont {Demler}},\ }\bibfield  {title} {\bibinfo {title} {Topological characterization of periodically driven quantum systems},\ }\href {https://doi.org/10.1103/PhysRevB.82.235114} {\bibfield  {journal} {\bibinfo  {journal} {Phys. Rev. B}\ }\textbf {\bibinfo {volume} {82}},\ \bibinfo {pages} {235114} (\bibinfo {year} {2010}{\natexlab{b}})}\BibitemShut {NoStop}%
\bibitem [{\citenamefont {Rudner}\ \emph {et~al.}(2013)\citenamefont {Rudner}, \citenamefont {Lindner}, \citenamefont {Berg},\ and\ \citenamefont {Levin}}]{rudnerAnomalousEdgeStates2013}%
  \BibitemOpen
  \bibfield  {author} {\bibinfo {author} {\bibfnamefont {M.~S.}\ \bibnamefont {Rudner}}, \bibinfo {author} {\bibfnamefont {N.~H.}\ \bibnamefont {Lindner}}, \bibinfo {author} {\bibfnamefont {E.}~\bibnamefont {Berg}},\ and\ \bibinfo {author} {\bibfnamefont {M.}~\bibnamefont {Levin}},\ }\bibfield  {title} {\bibinfo {title} {Anomalous {{Edge States}} and the {{Bulk-Edge Correspondence}} for {{Periodically Driven Two-Dimensional Systems}}},\ }\href {https://doi.org/10.1103/PhysRevX.3.031005} {\bibfield  {journal} {\bibinfo  {journal} {Phys. Rev. X}\ }\textbf {\bibinfo {volume} {3}},\ \bibinfo {pages} {031005} (\bibinfo {year} {2013})}\BibitemShut {NoStop}%
\bibitem [{\citenamefont {Carpentier}\ \emph {et~al.}(2015)\citenamefont {Carpentier}, \citenamefont {Delplace}, \citenamefont {Fruchart},\ and\ \citenamefont {Gaw{\k{e}}dzki}}]{carpentierTopologicalIndexPeriodically2015}%
  \BibitemOpen
  \bibfield  {author} {\bibinfo {author} {\bibfnamefont {D.}~\bibnamefont {Carpentier}}, \bibinfo {author} {\bibfnamefont {P.}~\bibnamefont {Delplace}}, \bibinfo {author} {\bibfnamefont {M.}~\bibnamefont {Fruchart}},\ and\ \bibinfo {author} {\bibfnamefont {K.}~\bibnamefont {Gaw{\k{e}}dzki}},\ }\bibfield  {title} {\bibinfo {title} {Topological {{Index}} for {{Periodically Driven Time-Reversal Invariant 2D Systems}}},\ }\href {https://doi.org/10.1103/PhysRevLett.114.106806} {\bibfield  {journal} {\bibinfo  {journal} {Phys. Rev. Lett.}\ }\textbf {\bibinfo {volume} {114}},\ \bibinfo {pages} {106806} (\bibinfo {year} {2015})}\BibitemShut {NoStop}%
\bibitem [{\citenamefont {Nathan}\ and\ \citenamefont {Rudner}(2015)}]{nathanTopologicalSingularitiesGeneral2015}%
  \BibitemOpen
  \bibfield  {author} {\bibinfo {author} {\bibfnamefont {F.}~\bibnamefont {Nathan}}\ and\ \bibinfo {author} {\bibfnamefont {M.~S.}\ \bibnamefont {Rudner}},\ }\bibfield  {title} {\bibinfo {title} {Topological singularities and the general classification of {{Floquet}}\textendash{{Bloch}} systems},\ }\href {https://doi.org/10.1088/1367-2630/17/12/125014} {\bibfield  {journal} {\bibinfo  {journal} {New J. Phys.}\ }\textbf {\bibinfo {volume} {17}},\ \bibinfo {pages} {125014} (\bibinfo {year} {2015})}\BibitemShut {NoStop}%
\bibitem [{\citenamefont {Cooper}\ \emph {et~al.}(2019)\citenamefont {Cooper}, \citenamefont {Dalibard},\ and\ \citenamefont {Spielman}}]{cooperTopologicalBandsUltracold2019}%
  \BibitemOpen
  \bibfield  {author} {\bibinfo {author} {\bibfnamefont {N.~R.}\ \bibnamefont {Cooper}}, \bibinfo {author} {\bibfnamefont {J.}~\bibnamefont {Dalibard}},\ and\ \bibinfo {author} {\bibfnamefont {I.~B.}\ \bibnamefont {Spielman}},\ }\bibfield  {title} {\bibinfo {title} {Topological bands for ultracold atoms},\ }\href {https://doi.org/10.1103/RevModPhys.91.015005} {\bibfield  {journal} {\bibinfo  {journal} {Rev. Mod. Phys.}\ }\textbf {\bibinfo {volume} {91}},\ \bibinfo {pages} {015005} (\bibinfo {year} {2019})}\BibitemShut {NoStop}%
\bibitem [{\citenamefont {Roy}\ and\ \citenamefont {Harper}(2017)}]{royPeriodicTableFloquet2017}%
  \BibitemOpen
  \bibfield  {author} {\bibinfo {author} {\bibfnamefont {R.}~\bibnamefont {Roy}}\ and\ \bibinfo {author} {\bibfnamefont {F.}~\bibnamefont {Harper}},\ }\bibfield  {title} {\bibinfo {title} {Periodic table for {{Floquet}} topological insulators},\ }\href {https://doi.org/10.1103/PhysRevB.96.155118} {\bibfield  {journal} {\bibinfo  {journal} {Phys. Rev. B}\ }\textbf {\bibinfo {volume} {96}},\ \bibinfo {pages} {155118} (\bibinfo {year} {2017})}\BibitemShut {NoStop}%
\bibitem [{\citenamefont {Yao}\ \emph {et~al.}(2017)\citenamefont {Yao}, \citenamefont {Yan},\ and\ \citenamefont {Wang}}]{yaoTopologicalInvariantsFloquet2017}%
  \BibitemOpen
  \bibfield  {author} {\bibinfo {author} {\bibfnamefont {S.}~\bibnamefont {Yao}}, \bibinfo {author} {\bibfnamefont {Z.}~\bibnamefont {Yan}},\ and\ \bibinfo {author} {\bibfnamefont {Z.}~\bibnamefont {Wang}},\ }\bibfield  {title} {\bibinfo {title} {Topological invariants of {{Floquet}} systems: {{General}} formulation, special properties, and {{Floquet}} topological defects},\ }\href {https://doi.org/10.1103/PhysRevB.96.195303} {\bibfield  {journal} {\bibinfo  {journal} {Phys. Rev. B}\ }\textbf {\bibinfo {volume} {96}},\ \bibinfo {pages} {195303} (\bibinfo {year} {2017})}\BibitemShut {NoStop}%
\bibitem [{\citenamefont {Ozawa}\ \emph {et~al.}(2019)\citenamefont {Ozawa}, \citenamefont {Price}, \citenamefont {Amo}, \citenamefont {Goldman}, \citenamefont {Hafezi}, \citenamefont {Lu}, \citenamefont {Rechtsman}, \citenamefont {Schuster}, \citenamefont {Simon}, \citenamefont {Zilberberg},\ and\ \citenamefont {Carusotto}}]{ozawaTopologicalPhotonics2019}%
  \BibitemOpen
  \bibfield  {author} {\bibinfo {author} {\bibfnamefont {T.}~\bibnamefont {Ozawa}}, \bibinfo {author} {\bibfnamefont {H.~M.}\ \bibnamefont {Price}}, \bibinfo {author} {\bibfnamefont {A.}~\bibnamefont {Amo}}, \bibinfo {author} {\bibfnamefont {N.}~\bibnamefont {Goldman}}, \bibinfo {author} {\bibfnamefont {M.}~\bibnamefont {Hafezi}}, \bibinfo {author} {\bibfnamefont {L.}~\bibnamefont {Lu}}, \bibinfo {author} {\bibfnamefont {M.~C.}\ \bibnamefont {Rechtsman}}, \bibinfo {author} {\bibfnamefont {D.}~\bibnamefont {Schuster}}, \bibinfo {author} {\bibfnamefont {J.}~\bibnamefont {Simon}}, \bibinfo {author} {\bibfnamefont {O.}~\bibnamefont {Zilberberg}},\ and\ \bibinfo {author} {\bibfnamefont {I.}~\bibnamefont {Carusotto}},\ }\bibfield  {title} {\bibinfo {title} {Topological photonics},\ }\href {https://doi.org/10.1103/RevModPhys.91.015006} {\bibfield  {journal} {\bibinfo  {journal} {Rev. Mod. Phys.}\ }\textbf {\bibinfo {volume} {91}},\ \bibinfo {pages} {015006} (\bibinfo {year} {2019})}\BibitemShut {NoStop}%
\bibitem [{\citenamefont {Harper}\ \emph {et~al.}(2020)\citenamefont {Harper}, \citenamefont {Roy}, \citenamefont {Rudner},\ and\ \citenamefont {Sondhi}}]{harperTopologyBrokenSymmetry2020}%
  \BibitemOpen
  \bibfield  {author} {\bibinfo {author} {\bibfnamefont {F.}~\bibnamefont {Harper}}, \bibinfo {author} {\bibfnamefont {R.}~\bibnamefont {Roy}}, \bibinfo {author} {\bibfnamefont {M.~S.}\ \bibnamefont {Rudner}},\ and\ \bibinfo {author} {\bibfnamefont {S.}~\bibnamefont {Sondhi}},\ }\bibfield  {title} {\bibinfo {title} {Topology and {{Broken Symmetry}} in {{Floquet Systems}}},\ }\href {https://doi.org/10.1146/annurev-conmatphys-031218-013721} {\bibfield  {journal} {\bibinfo  {journal} {Annu. Rev. Cond. Mat. Phys.}\ }\textbf {\bibinfo {volume} {11}},\ \bibinfo {pages} {345} (\bibinfo {year} {2020})}\BibitemShut {NoStop}%
\bibitem [{\citenamefont {Rudner}\ and\ \citenamefont {Lindner}(2020)}]{rudnerBandStructureEngineering2020}%
  \BibitemOpen
  \bibfield  {author} {\bibinfo {author} {\bibfnamefont {M.~S.}\ \bibnamefont {Rudner}}\ and\ \bibinfo {author} {\bibfnamefont {N.~H.}\ \bibnamefont {Lindner}},\ }\bibfield  {title} {\bibinfo {title} {Band structure engineering and non-equilibrium dynamics in {{Floquet}} topological insulators},\ }\href {https://doi.org/10.1038/s42254-020-0170-z} {\bibfield  {journal} {\bibinfo  {journal} {Nat Rev Phys}\ }\textbf {\bibinfo {volume} {2}},\ \bibinfo {pages} {229} (\bibinfo {year} {2020})}\BibitemShut {NoStop}%
\bibitem [{\citenamefont {Xie}\ \emph {et~al.}(2020)\citenamefont {Xie}, \citenamefont {Deng}, \citenamefont {Xiao}, \citenamefont {Gou}, \citenamefont {Chen}, \citenamefont {Yi},\ and\ \citenamefont {Yan}}]{xieTopologicalQuantumWalks2020}%
  \BibitemOpen
  \bibfield  {author} {\bibinfo {author} {\bibfnamefont {D.}~\bibnamefont {Xie}}, \bibinfo {author} {\bibfnamefont {T.-S.}\ \bibnamefont {Deng}}, \bibinfo {author} {\bibfnamefont {T.}~\bibnamefont {Xiao}}, \bibinfo {author} {\bibfnamefont {W.}~\bibnamefont {Gou}}, \bibinfo {author} {\bibfnamefont {T.}~\bibnamefont {Chen}}, \bibinfo {author} {\bibfnamefont {W.}~\bibnamefont {Yi}},\ and\ \bibinfo {author} {\bibfnamefont {B.}~\bibnamefont {Yan}},\ }\bibfield  {title} {\bibinfo {title} {Topological {{Quantum Walks}} in {{Momentum Space}} with a {{Bose-Einstein Condensate}}},\ }\href {https://doi.org/10.1103/PhysRevLett.124.050502} {\bibfield  {journal} {\bibinfo  {journal} {Phys. Rev. Lett.}\ }\textbf {\bibinfo {volume} {124}},\ \bibinfo {pages} {050502} (\bibinfo {year} {2020})}\BibitemShut {NoStop}%
\bibitem [{\citenamefont {Wintersperger}\ \emph {et~al.}(2020)\citenamefont {Wintersperger}, \citenamefont {Braun}, \citenamefont {{\"U}nal}, \citenamefont {Eckardt}, \citenamefont {Liberto}, \citenamefont {Goldman}, \citenamefont {Bloch},\ and\ \citenamefont {Aidelsburger}}]{winterspergerRealizationAnomalousFloquet2020}%
  \BibitemOpen
  \bibfield  {author} {\bibinfo {author} {\bibfnamefont {K.}~\bibnamefont {Wintersperger}}, \bibinfo {author} {\bibfnamefont {C.}~\bibnamefont {Braun}}, \bibinfo {author} {\bibfnamefont {F.~N.}\ \bibnamefont {{\"U}nal}}, \bibinfo {author} {\bibfnamefont {A.}~\bibnamefont {Eckardt}}, \bibinfo {author} {\bibfnamefont {M.~D.}\ \bibnamefont {Liberto}}, \bibinfo {author} {\bibfnamefont {N.}~\bibnamefont {Goldman}}, \bibinfo {author} {\bibfnamefont {I.}~\bibnamefont {Bloch}},\ and\ \bibinfo {author} {\bibfnamefont {M.}~\bibnamefont {Aidelsburger}},\ }\bibfield  {title} {\bibinfo {title} {Realization of an anomalous {{Floquet}} topological system with ultracold atoms},\ }\href {https://doi.org/10.1038/s41567-020-0949-y} {\bibfield  {journal} {\bibinfo  {journal} {Nat. Phys.}\ }\textbf {\bibinfo {volume} {16}},\ \bibinfo {pages} {1058} (\bibinfo {year} {2020})}\BibitemShut {NoStop}%
\bibitem [{\citenamefont {Gao}\ \emph {et~al.}(2016)\citenamefont {Gao}, \citenamefont {Gao}, \citenamefont {Shi}, \citenamefont {Yang}, \citenamefont {Lin}, \citenamefont {Xu}, \citenamefont {Joannopoulos}, \citenamefont {Solja{\v c}i{\'c}}, \citenamefont {Chen}, \citenamefont {Lu}, \citenamefont {Chong},\ and\ \citenamefont {Zhang}}]{gaoProbingTopologicalProtection2016}%
  \BibitemOpen
  \bibfield  {author} {\bibinfo {author} {\bibfnamefont {F.}~\bibnamefont {Gao}}, \bibinfo {author} {\bibfnamefont {Z.}~\bibnamefont {Gao}}, \bibinfo {author} {\bibfnamefont {X.}~\bibnamefont {Shi}}, \bibinfo {author} {\bibfnamefont {Z.}~\bibnamefont {Yang}}, \bibinfo {author} {\bibfnamefont {X.}~\bibnamefont {Lin}}, \bibinfo {author} {\bibfnamefont {H.}~\bibnamefont {Xu}}, \bibinfo {author} {\bibfnamefont {J.~D.}\ \bibnamefont {Joannopoulos}}, \bibinfo {author} {\bibfnamefont {M.}~\bibnamefont {Solja{\v c}i{\'c}}}, \bibinfo {author} {\bibfnamefont {H.}~\bibnamefont {Chen}}, \bibinfo {author} {\bibfnamefont {L.}~\bibnamefont {Lu}}, \bibinfo {author} {\bibfnamefont {Y.}~\bibnamefont {Chong}},\ and\ \bibinfo {author} {\bibfnamefont {B.}~\bibnamefont {Zhang}},\ }\bibfield  {title} {\bibinfo {title} {Probing topological protection using a designer surface plasmon structure},\ }\href {https://doi.org/10.1038/ncomms11619} {\bibfield  {journal} {\bibinfo  {journal} {Nat. Commun.}\ }\textbf {\bibinfo {volume} {7}},\ \bibinfo {pages} {11619} (\bibinfo {year} {2016})}\BibitemShut {NoStop}%
\bibitem [{\citenamefont {Mukherjee}\ \emph {et~al.}(2017)\citenamefont {Mukherjee}, \citenamefont {Spracklen}, \citenamefont {Valiente}, \citenamefont {Andersson}, \citenamefont {{\"O}hberg}, \citenamefont {Goldman},\ and\ \citenamefont {Thomson}}]{mukherjeeExperimentalObservationAnomalous2017}%
  \BibitemOpen
  \bibfield  {author} {\bibinfo {author} {\bibfnamefont {S.}~\bibnamefont {Mukherjee}}, \bibinfo {author} {\bibfnamefont {A.}~\bibnamefont {Spracklen}}, \bibinfo {author} {\bibfnamefont {M.}~\bibnamefont {Valiente}}, \bibinfo {author} {\bibfnamefont {E.}~\bibnamefont {Andersson}}, \bibinfo {author} {\bibfnamefont {P.}~\bibnamefont {{\"O}hberg}}, \bibinfo {author} {\bibfnamefont {N.}~\bibnamefont {Goldman}},\ and\ \bibinfo {author} {\bibfnamefont {R.~R.}\ \bibnamefont {Thomson}},\ }\bibfield  {title} {\bibinfo {title} {Experimental observation of anomalous topological edge modes in a slowly driven photonic lattice},\ }\href {https://doi.org/10.1038/ncomms13918} {\bibfield  {journal} {\bibinfo  {journal} {Nat. Commun.}\ }\textbf {\bibinfo {volume} {8}},\ \bibinfo {pages} {13918} (\bibinfo {year} {2017})}\BibitemShut {NoStop}%
\bibitem [{\citenamefont {Maczewsky}\ \emph {et~al.}(2017)\citenamefont {Maczewsky}, \citenamefont {Zeuner}, \citenamefont {Nolte},\ and\ \citenamefont {Szameit}}]{maczewskyObservationPhotonicAnomalous2017}%
  \BibitemOpen
  \bibfield  {author} {\bibinfo {author} {\bibfnamefont {L.~J.}\ \bibnamefont {Maczewsky}}, \bibinfo {author} {\bibfnamefont {J.~M.}\ \bibnamefont {Zeuner}}, \bibinfo {author} {\bibfnamefont {S.}~\bibnamefont {Nolte}},\ and\ \bibinfo {author} {\bibfnamefont {A.}~\bibnamefont {Szameit}},\ }\bibfield  {title} {\bibinfo {title} {Observation of photonic anomalous {{Floquet}} topological insulators},\ }\href {https://doi.org/10.1038/ncomms13756} {\bibfield  {journal} {\bibinfo  {journal} {Nat. Commun.}\ }\textbf {\bibinfo {volume} {8}},\ \bibinfo {pages} {13756} (\bibinfo {year} {2017})}\BibitemShut {NoStop}%
\bibitem [{\citenamefont {Cheng}\ \emph {et~al.}(2019)\citenamefont {Cheng}, \citenamefont {Pan}, \citenamefont {Wang}, \citenamefont {Zhang}, \citenamefont {Yu}, \citenamefont {Gover}, \citenamefont {Zhang}, \citenamefont {Li}, \citenamefont {Zhou},\ and\ \citenamefont {Zhu}}]{chengObservationAnomalousEnsuremath2019}%
  \BibitemOpen
  \bibfield  {author} {\bibinfo {author} {\bibfnamefont {Q.}~\bibnamefont {Cheng}}, \bibinfo {author} {\bibfnamefont {Y.}~\bibnamefont {Pan}}, \bibinfo {author} {\bibfnamefont {H.}~\bibnamefont {Wang}}, \bibinfo {author} {\bibfnamefont {C.}~\bibnamefont {Zhang}}, \bibinfo {author} {\bibfnamefont {D.}~\bibnamefont {Yu}}, \bibinfo {author} {\bibfnamefont {A.}~\bibnamefont {Gover}}, \bibinfo {author} {\bibfnamefont {H.}~\bibnamefont {Zhang}}, \bibinfo {author} {\bibfnamefont {T.}~\bibnamefont {Li}}, \bibinfo {author} {\bibfnamefont {L.}~\bibnamefont {Zhou}},\ and\ \bibinfo {author} {\bibfnamefont {S.}~\bibnamefont {Zhu}},\ }\bibfield  {title} {\bibinfo {title} {Observation of {{Anomalous}} $\ensuremath{\pi}$ {{Modes}} in {{Photonic Floquet Engineering}}},\ }\href {https://doi.org/10.1103/PhysRevLett.122.173901} {\bibfield  {journal} {\bibinfo  {journal} {Phys. Rev. Lett.}\ }\textbf {\bibinfo {volume} {122}},\ \bibinfo {pages} {173901} (\bibinfo {year} {2019})}\BibitemShut {NoStop}%
\bibitem [{\citenamefont {Casati}\ \emph {et~al.}(1979)\citenamefont {Casati}, \citenamefont {Chirikov}, \citenamefont {Izraelev},\ and\ \citenamefont {Ford}}]{casatiStochasticBehaviorQuantum1979}%
  \BibitemOpen
  \bibfield  {author} {\bibinfo {author} {\bibfnamefont {G.}~\bibnamefont {Casati}}, \bibinfo {author} {\bibfnamefont {B.~V.}\ \bibnamefont {Chirikov}}, \bibinfo {author} {\bibfnamefont {F.~M.}\ \bibnamefont {Izraelev}},\ and\ \bibinfo {author} {\bibfnamefont {J.}~\bibnamefont {Ford}},\ }\bibfield  {title} {\bibinfo {title} {Stochastic behavior of a quantum pendulum under a periodic perturbation},\ }in\ \href {https://doi.org/10.1007/BFb0021757} {\emph {\bibinfo {booktitle} {Stoch. {{Behav}}. {{Class}}. {{Quantum Hamiltonian Syst}}.}}},\ \bibinfo {series and number} {Lecture {{Notes}} in {{Physics}}},\ \bibinfo {editor} {edited by\ \bibinfo {editor} {\bibfnamefont {G.}~\bibnamefont {Casati}}\ and\ \bibinfo {editor} {\bibfnamefont {J.}~\bibnamefont {Ford}}}\ (\bibinfo  {publisher} {{Springer}},\ \bibinfo {address} {{Berlin, Heidelberg}},\ \bibinfo {year} {1979})\ pp.\ \bibinfo {pages} {334--352}\BibitemShut {NoStop}%
\bibitem [{\citenamefont {Izrailev}(1990)}]{izrailevSimpleModelsQuantum1990}%
  \BibitemOpen
  \bibfield  {author} {\bibinfo {author} {\bibfnamefont {F.~M.}\ \bibnamefont {Izrailev}},\ }\bibfield  {title} {\bibinfo {title} {Simple models of quantum chaos: {{Spectrum}} and eigenfunctions},\ }\href {https://doi.org/10.1016/0370-1573(90)90067-C} {\bibfield  {journal} {\bibinfo  {journal} {Phys. Rep.}\ }\textbf {\bibinfo {volume} {196}},\ \bibinfo {pages} {299} (\bibinfo {year} {1990})}\BibitemShut {NoStop}%
\bibitem [{\citenamefont {Raizen}(1999)}]{raizenQuantumChaosCold1999}%
  \BibitemOpen
  \bibfield  {author} {\bibinfo {author} {\bibfnamefont {M.~G.}\ \bibnamefont {Raizen}},\ }\bibfield  {title} {\bibinfo {title} {Quantum {{Chaos}} with {{Cold Atoms}}},\ }\href {https://doi.org/10.1016/S1049-250X(08)60218-9} {\bibfield  {journal} {\bibinfo  {journal} {Adv. At. Mol. Opt. Phys.}\ }\textbf {\bibinfo {volume} {41}},\ \bibinfo {pages} {43} (\bibinfo {year} {1999})}\BibitemShut {NoStop}%
\bibitem [{\citenamefont {Sadgrove}\ and\ \citenamefont {Wimberger}(2011)}]{sadgrovePseudoclassicalMethodAtomOptics2011}%
  \BibitemOpen
  \bibfield  {author} {\bibinfo {author} {\bibfnamefont {M.}~\bibnamefont {Sadgrove}}\ and\ \bibinfo {author} {\bibfnamefont {S.}~\bibnamefont {Wimberger}},\ }\bibfield  {title} {\bibinfo {title} {A {{Pseudoclassical Method}} for the {{Atom-Optics Kicked Rotor}}: from {{Theory}} to {{Experiment}} and {{Back}}},\ }\href {https://doi.org/10.1016/B978-0-12-385508-4.00007-3} {\bibfield  {journal} {\bibinfo  {journal} {Adv. At. Mol. Opt. Phys.}\ }\textbf {\bibinfo {volume} {60}},\ \bibinfo {pages} {315} (\bibinfo {year} {2011})}\BibitemShut {NoStop}%
\bibitem [{\citenamefont {Chab{\'e}}\ \emph {et~al.}(2008)\citenamefont {Chab{\'e}}, \citenamefont {Lemari{\'e}}, \citenamefont {Gr{\'e}maud}, \citenamefont {Delande}, \citenamefont {Szriftgiser},\ and\ \citenamefont {Garreau}}]{chabeExperimentalObservationAnderson2008}%
  \BibitemOpen
  \bibfield  {author} {\bibinfo {author} {\bibfnamefont {J.}~\bibnamefont {Chab{\'e}}}, \bibinfo {author} {\bibfnamefont {G.}~\bibnamefont {Lemari{\'e}}}, \bibinfo {author} {\bibfnamefont {B.}~\bibnamefont {Gr{\'e}maud}}, \bibinfo {author} {\bibfnamefont {D.}~\bibnamefont {Delande}}, \bibinfo {author} {\bibfnamefont {P.}~\bibnamefont {Szriftgiser}},\ and\ \bibinfo {author} {\bibfnamefont {J.~C.}\ \bibnamefont {Garreau}},\ }\bibfield  {title} {\bibinfo {title} {Experimental {{Observation}} of the {{Anderson Metal-Insulator Transition}} with {{Atomic Matter Waves}}},\ }\href {https://doi.org/10.1103/PhysRevLett.101.255702} {\bibfield  {journal} {\bibinfo  {journal} {Phys. Rev. Lett.}\ }\textbf {\bibinfo {volume} {101}},\ \bibinfo {pages} {255702} (\bibinfo {year} {2008})}\BibitemShut {NoStop}%
\bibitem [{\citenamefont {Lemari{\'e}}\ \emph {et~al.}(2010)\citenamefont {Lemari{\'e}}, \citenamefont {Lignier}, \citenamefont {Delande}, \citenamefont {Szriftgiser},\ and\ \citenamefont {Garreau}}]{lemarieCriticalStateAnderson2010}%
  \BibitemOpen
  \bibfield  {author} {\bibinfo {author} {\bibfnamefont {G.}~\bibnamefont {Lemari{\'e}}}, \bibinfo {author} {\bibfnamefont {H.}~\bibnamefont {Lignier}}, \bibinfo {author} {\bibfnamefont {D.}~\bibnamefont {Delande}}, \bibinfo {author} {\bibfnamefont {P.}~\bibnamefont {Szriftgiser}},\ and\ \bibinfo {author} {\bibfnamefont {J.~C.}\ \bibnamefont {Garreau}},\ }\bibfield  {title} {\bibinfo {title} {Critical {{State}} of the {{Anderson Transition}}: {{Between}} a {{Metal}} and an {{Insulator}}},\ }\href {https://doi.org/10.1103/PhysRevLett.105.090601} {\bibfield  {journal} {\bibinfo  {journal} {Phys. Rev. Lett.}\ }\textbf {\bibinfo {volume} {105}},\ \bibinfo {pages} {090601} (\bibinfo {year} {2010})}\BibitemShut {NoStop}%
\bibitem [{\citenamefont {Karski}\ \emph {et~al.}(2009)\citenamefont {Karski}, \citenamefont {F{\"o}rster}, \citenamefont {Choi}, \citenamefont {Steffen}, \citenamefont {Alt}, \citenamefont {Meschede},\ and\ \citenamefont {Widera}}]{karskiQuantumWalkPosition2009}%
  \BibitemOpen
  \bibfield  {author} {\bibinfo {author} {\bibfnamefont {M.}~\bibnamefont {Karski}}, \bibinfo {author} {\bibfnamefont {L.}~\bibnamefont {F{\"o}rster}}, \bibinfo {author} {\bibfnamefont {J.-M.}\ \bibnamefont {Choi}}, \bibinfo {author} {\bibfnamefont {A.}~\bibnamefont {Steffen}}, \bibinfo {author} {\bibfnamefont {W.}~\bibnamefont {Alt}}, \bibinfo {author} {\bibfnamefont {D.}~\bibnamefont {Meschede}},\ and\ \bibinfo {author} {\bibfnamefont {A.}~\bibnamefont {Widera}},\ }\bibfield  {title} {\bibinfo {title} {Quantum {{Walk}} in {{Position Space}} with {{Single Optically Trapped Atoms}}},\ }\href {https://doi.org/10.1126/science.1174436} {\bibfield  {journal} {\bibinfo  {journal} {Science}\ }\textbf {\bibinfo {volume} {325}},\ \bibinfo {pages} {174} (\bibinfo {year} {2009})}\BibitemShut {NoStop}%
\bibitem [{\citenamefont {Summy}\ and\ \citenamefont {Wimberger}(2016)}]{summyQuantumRandomWalk2016}%
  \BibitemOpen
  \bibfield  {author} {\bibinfo {author} {\bibfnamefont {G.}~\bibnamefont {Summy}}\ and\ \bibinfo {author} {\bibfnamefont {S.}~\bibnamefont {Wimberger}},\ }\bibfield  {title} {\bibinfo {title} {Quantum random walk of a {{Bose-Einstein}} condensate in momentum space},\ }\href {https://doi.org/10.1103/PhysRevA.93.023638} {\bibfield  {journal} {\bibinfo  {journal} {Phys. Rev. A}\ }\textbf {\bibinfo {volume} {93}},\ \bibinfo {pages} {023638} (\bibinfo {year} {2016})}\BibitemShut {NoStop}%
\bibitem [{\citenamefont {Dadras}\ \emph {et~al.}(2018)\citenamefont {Dadras}, \citenamefont {Gresch}, \citenamefont {Groiseau}, \citenamefont {Wimberger},\ and\ \citenamefont {Summy}}]{dadrasQuantumWalkMomentum2018}%
  \BibitemOpen
  \bibfield  {author} {\bibinfo {author} {\bibfnamefont {S.}~\bibnamefont {Dadras}}, \bibinfo {author} {\bibfnamefont {A.}~\bibnamefont {Gresch}}, \bibinfo {author} {\bibfnamefont {C.}~\bibnamefont {Groiseau}}, \bibinfo {author} {\bibfnamefont {S.}~\bibnamefont {Wimberger}},\ and\ \bibinfo {author} {\bibfnamefont {G.~S.}\ \bibnamefont {Summy}},\ }\bibfield  {title} {\bibinfo {title} {Quantum {{Walk}} in {{Momentum Space}} with a {{Bose-Einstein Condensate}}},\ }\href {https://doi.org/10.1103/PhysRevLett.121.070402} {\bibfield  {journal} {\bibinfo  {journal} {Phys. Rev. Lett.}\ }\textbf {\bibinfo {volume} {121}},\ \bibinfo {pages} {070402} (\bibinfo {year} {2018})}\BibitemShut {NoStop}%
\bibitem [{\citenamefont {Dadras}\ \emph {et~al.}(2019)\citenamefont {Dadras}, \citenamefont {Gresch}, \citenamefont {Groiseau}, \citenamefont {Wimberger},\ and\ \citenamefont {Summy}}]{dadrasExperimentalRealizationMomentumspace2019}%
  \BibitemOpen
  \bibfield  {author} {\bibinfo {author} {\bibfnamefont {S.}~\bibnamefont {Dadras}}, \bibinfo {author} {\bibfnamefont {A.}~\bibnamefont {Gresch}}, \bibinfo {author} {\bibfnamefont {C.}~\bibnamefont {Groiseau}}, \bibinfo {author} {\bibfnamefont {S.}~\bibnamefont {Wimberger}},\ and\ \bibinfo {author} {\bibfnamefont {G.~S.}\ \bibnamefont {Summy}},\ }\bibfield  {title} {\bibinfo {title} {Experimental realization of a momentum-space quantum walk},\ }\href {https://doi.org/10.1103/PhysRevA.99.043617} {\bibfield  {journal} {\bibinfo  {journal} {Phys. Rev. A}\ }\textbf {\bibinfo {volume} {99}},\ \bibinfo {pages} {043617} (\bibinfo {year} {2019})}\BibitemShut {NoStop}%
\bibitem [{\citenamefont {Wang}\ \emph {et~al.}(2013)\citenamefont {Wang}, \citenamefont {Wang}, \citenamefont {Guarneri}, \citenamefont {Casati},\ and\ \citenamefont {Gong}}]{wangExponentialQuantumSpreading2013}%
  \BibitemOpen
  \bibfield  {author} {\bibinfo {author} {\bibfnamefont {H.}~\bibnamefont {Wang}}, \bibinfo {author} {\bibfnamefont {J.}~\bibnamefont {Wang}}, \bibinfo {author} {\bibfnamefont {I.}~\bibnamefont {Guarneri}}, \bibinfo {author} {\bibfnamefont {G.}~\bibnamefont {Casati}},\ and\ \bibinfo {author} {\bibfnamefont {J.}~\bibnamefont {Gong}},\ }\bibfield  {title} {\bibinfo {title} {Exponential quantum spreading in a class of kicked rotor systems near high-order resonances},\ }\href {https://doi.org/10.1103/PhysRevE.88.052919} {\bibfield  {journal} {\bibinfo  {journal} {Phys. Rev. E}\ }\textbf {\bibinfo {volume} {88}},\ \bibinfo {pages} {052919} (\bibinfo {year} {2013})}\BibitemShut {NoStop}%
\bibitem [{\citenamefont {Zhou}\ and\ \citenamefont {Gong}(2018)}]{zhouFloquetTopologicalPhases2018}%
  \BibitemOpen
  \bibfield  {author} {\bibinfo {author} {\bibfnamefont {L.}~\bibnamefont {Zhou}}\ and\ \bibinfo {author} {\bibfnamefont {J.}~\bibnamefont {Gong}},\ }\bibfield  {title} {\bibinfo {title} {Floquet topological phases in a spin-1/2 double kicked rotor},\ }\href {https://doi.org/10.1103/PhysRevA.97.063603} {\bibfield  {journal} {\bibinfo  {journal} {Phys. Rev. A}\ }\textbf {\bibinfo {volume} {97}},\ \bibinfo {pages} {063603} (\bibinfo {year} {2018})}\BibitemShut {NoStop}%
\bibitem [{\citenamefont {Zhou}\ and\ \citenamefont {Pan}(2019)}]{zhouNonHermitianFloquetTopological2019}%
  \BibitemOpen
  \bibfield  {author} {\bibinfo {author} {\bibfnamefont {L.}~\bibnamefont {Zhou}}\ and\ \bibinfo {author} {\bibfnamefont {J.}~\bibnamefont {Pan}},\ }\bibfield  {title} {\bibinfo {title} {Non-{{Hermitian Floquet}} topological phases in the double-kicked rotor},\ }\href {https://doi.org/10.1103/PhysRevA.100.053608} {\bibfield  {journal} {\bibinfo  {journal} {Phys. Rev. A}\ }\textbf {\bibinfo {volume} {100}},\ \bibinfo {pages} {053608} (\bibinfo {year} {2019})}\BibitemShut {NoStop}%
\bibitem [{\citenamefont {Zhou}(2021)}]{zhouFloquetSecondOrderTopological2021}%
  \BibitemOpen
  \bibfield  {author} {\bibinfo {author} {\bibfnamefont {L.}~\bibnamefont {Zhou}},\ }\bibfield  {title} {\bibinfo {title} {Floquet {{Second-Order Topological Phases}} in {{Momentum Space}}},\ }\href {https://doi.org/10.3390/nano11051170} {\bibfield  {journal} {\bibinfo  {journal} {Nanomaterials}\ }\textbf {\bibinfo {volume} {11}},\ \bibinfo {pages} {1170} (\bibinfo {year} {2021})}\BibitemShut {NoStop}%
\bibitem [{\citenamefont {Bolik}\ \emph {et~al.}(2022)\citenamefont {Bolik}, \citenamefont {Groiseau}, \citenamefont {Clark}, \citenamefont {Summy}, \citenamefont {Liu},\ and\ \citenamefont {Wimberger}}]{bolikDetectingTopologicalPhase2022}%
  \BibitemOpen
  \bibfield  {author} {\bibinfo {author} {\bibfnamefont {N.}~\bibnamefont {Bolik}}, \bibinfo {author} {\bibfnamefont {C.}~\bibnamefont {Groiseau}}, \bibinfo {author} {\bibfnamefont {J.~H.}\ \bibnamefont {Clark}}, \bibinfo {author} {\bibfnamefont {G.~S.}\ \bibnamefont {Summy}}, \bibinfo {author} {\bibfnamefont {Y.}~\bibnamefont {Liu}},\ and\ \bibinfo {author} {\bibfnamefont {S.}~\bibnamefont {Wimberger}},\ }\bibfield  {title} {\bibinfo {title} {Detecting topological phase transitions in a double kicked quantum rotor},\ }\href {https://doi.org/10.1103/PhysRevA.106.043318} {\bibfield  {journal} {\bibinfo  {journal} {Phys. Rev. A}\ }\textbf {\bibinfo {volume} {106}},\ \bibinfo {pages} {043318} (\bibinfo {year} {2022})}\BibitemShut {NoStop}%
\bibitem [{\citenamefont {Schnyder}\ \emph {et~al.}(2008)\citenamefont {Schnyder}, \citenamefont {Ryu}, \citenamefont {Furusaki},\ and\ \citenamefont {Ludwig}}]{schnyderClassificationTopologicalInsulators2008}%
  \BibitemOpen
  \bibfield  {author} {\bibinfo {author} {\bibfnamefont {A.~P.}\ \bibnamefont {Schnyder}}, \bibinfo {author} {\bibfnamefont {S.}~\bibnamefont {Ryu}}, \bibinfo {author} {\bibfnamefont {A.}~\bibnamefont {Furusaki}},\ and\ \bibinfo {author} {\bibfnamefont {A.~W.~W.}\ \bibnamefont {Ludwig}},\ }\bibfield  {title} {\bibinfo {title} {Classification of topological insulators and superconductors in three spatial dimensions},\ }\href {https://doi.org/10.1103/PhysRevB.78.195125} {\bibfield  {journal} {\bibinfo  {journal} {Phys. Rev. B}\ }\textbf {\bibinfo {volume} {78}},\ \bibinfo {pages} {195125} (\bibinfo {year} {2008})}\BibitemShut {NoStop}%
\bibitem [{\citenamefont {Kitaev}(2009)}]{kitaevPeriodicTableTopological2009}%
  \BibitemOpen
  \bibfield  {author} {\bibinfo {author} {\bibfnamefont {A.}~\bibnamefont {Kitaev}},\ }\bibfield  {title} {\bibinfo {title} {Periodic table for topological insulators and superconductors},\ }\href {https://doi.org/10.1063/1.3149495} {\bibfield  {journal} {\bibinfo  {journal} {AIP Conf. Proc.}\ }\textbf {\bibinfo {volume} {1134}},\ \bibinfo {pages} {22} (\bibinfo {year} {2009})}\BibitemShut {NoStop}%
\bibitem [{\citenamefont {Ryu}\ \emph {et~al.}(2010)\citenamefont {Ryu}, \citenamefont {Schnyder}, \citenamefont {Furusaki},\ and\ \citenamefont {Ludwig}}]{ryuTopologicalInsulatorsSuperconductors2010}%
  \BibitemOpen
  \bibfield  {author} {\bibinfo {author} {\bibfnamefont {S.}~\bibnamefont {Ryu}}, \bibinfo {author} {\bibfnamefont {A.~P.}\ \bibnamefont {Schnyder}}, \bibinfo {author} {\bibfnamefont {A.}~\bibnamefont {Furusaki}},\ and\ \bibinfo {author} {\bibfnamefont {A.~W.~W.}\ \bibnamefont {Ludwig}},\ }\bibfield  {title} {\bibinfo {title} {Topological insulators and superconductors: tenfold way and dimensional hierarchy},\ }\href {https://doi.org/10.1088/1367-2630/12/6/065010} {\bibfield  {journal} {\bibinfo  {journal} {New J. Phys.}\ }\textbf {\bibinfo {volume} {12}},\ \bibinfo {pages} {065010} (\bibinfo {year} {2010})}\BibitemShut {NoStop}%
\bibitem [{\citenamefont {Gentile}\ \emph {et~al.}(2015)\citenamefont {Gentile}, \citenamefont {Cuoco},\ and\ \citenamefont {Ortix}}]{gentileEdgeStatesTopological2015}%
  \BibitemOpen
  \bibfield  {author} {\bibinfo {author} {\bibfnamefont {P.}~\bibnamefont {Gentile}}, \bibinfo {author} {\bibfnamefont {M.}~\bibnamefont {Cuoco}},\ and\ \bibinfo {author} {\bibfnamefont {C.}~\bibnamefont {Ortix}},\ }\bibfield  {title} {\bibinfo {title} {Edge {{States}} and {{Topological Insulating Phases Generated}} by {{Curving}} a {{Nanowire}} with {{Rashba Spin-Orbit Coupling}}},\ }\href {https://doi.org/10.1103/PhysRevLett.115.256801} {\bibfield  {journal} {\bibinfo  {journal} {Phys. Rev. Lett.}\ }\textbf {\bibinfo {volume} {115}},\ \bibinfo {pages} {256801} (\bibinfo {year} {2015})}\BibitemShut {NoStop}%
\bibitem [{\citenamefont {Liu}\ \emph {et~al.}(2019)\citenamefont {Liu}, \citenamefont {Li},\ and\ \citenamefont {Wang}}]{liuFractionalChargedEdge2019}%
  \BibitemOpen
  \bibfield  {author} {\bibinfo {author} {\bibfnamefont {Z.-X.}\ \bibnamefont {Liu}}, \bibinfo {author} {\bibfnamefont {Z.-H.}\ \bibnamefont {Li}},\ and\ \bibinfo {author} {\bibfnamefont {A.-M.}\ \bibnamefont {Wang}},\ }\bibfield  {title} {\bibinfo {title} {Fractional charged edge states in ladder topological insulators},\ }\href {https://doi.org/10.1088/1361-648X/aaff16} {\bibfield  {journal} {\bibinfo  {journal} {J. Phys.: Condens. Matter}\ }\textbf {\bibinfo {volume} {31}},\ \bibinfo {pages} {125402} (\bibinfo {year} {2019})}\BibitemShut {NoStop}%
\bibitem [{\citenamefont {Zhou}\ and\ \citenamefont {Du}(2020)}]{zhouFloquetTopologicalPhases2020}%
  \BibitemOpen
  \bibfield  {author} {\bibinfo {author} {\bibfnamefont {L.}~\bibnamefont {Zhou}}\ and\ \bibinfo {author} {\bibfnamefont {Q.}~\bibnamefont {Du}},\ }\bibfield  {title} {\bibinfo {title} {Floquet topological phases with fourfold-degenerate edge modes in a driven spin-1/2 {{Creutz}} ladder},\ }\href {https://doi.org/10.1103/PhysRevA.101.033607} {\bibfield  {journal} {\bibinfo  {journal} {Phys. Rev. A}\ }\textbf {\bibinfo {volume} {101}},\ \bibinfo {pages} {033607} (\bibinfo {year} {2020})}\BibitemShut {NoStop}%
\bibitem [{\citenamefont {Zhou}(2020)}]{zhouNonHermitianFloquetPhases2020}%
  \BibitemOpen
  \bibfield  {author} {\bibinfo {author} {\bibfnamefont {L.}~\bibnamefont {Zhou}},\ }\bibfield  {title} {\bibinfo {title} {Non-{{Hermitian Floquet Phases}} with {{Even-Integer Topological Invariants}} in a {{Periodically Quenched Two-Leg Ladder}}},\ }\href {https://doi.org/10.3390/e22070746} {\bibfield  {journal} {\bibinfo  {journal} {Entropy}\ }\textbf {\bibinfo {volume} {22}},\ \bibinfo {pages} {746} (\bibinfo {year} {2020})}\BibitemShut {NoStop}%
\bibitem [{\citenamefont {Malard}\ \emph {et~al.}(2020)\citenamefont {Malard}, \citenamefont {Brandao}, \citenamefont {{de Brito}},\ and\ \citenamefont {Johannesson}}]{malardMulticriticalityOnedimensionalTopological2020}%
  \BibitemOpen
  \bibfield  {author} {\bibinfo {author} {\bibfnamefont {M.}~\bibnamefont {Malard}}, \bibinfo {author} {\bibfnamefont {D.}~\bibnamefont {Brandao}}, \bibinfo {author} {\bibfnamefont {P.~E.}\ \bibnamefont {{de Brito}}},\ and\ \bibinfo {author} {\bibfnamefont {H.}~\bibnamefont {Johannesson}},\ }\bibfield  {title} {\bibinfo {title} {Multicriticality in a one-dimensional topological band insulator},\ }\href {https://doi.org/10.1103/PhysRevResearch.2.033246} {\bibfield  {journal} {\bibinfo  {journal} {Phys. Rev. Res.}\ }\textbf {\bibinfo {volume} {2}},\ \bibinfo {pages} {033246} (\bibinfo {year} {2020})}\BibitemShut {NoStop}%
\bibitem [{\citenamefont {Meier}\ \emph {et~al.}(2016{\natexlab{a}})\citenamefont {Meier}, \citenamefont {An},\ and\ \citenamefont {Gadway}}]{meierObservationTopologicalSoliton2016}%
  \BibitemOpen
  \bibfield  {author} {\bibinfo {author} {\bibfnamefont {E.~J.}\ \bibnamefont {Meier}}, \bibinfo {author} {\bibfnamefont {F.~A.}\ \bibnamefont {An}},\ and\ \bibinfo {author} {\bibfnamefont {B.}~\bibnamefont {Gadway}},\ }\bibfield  {title} {\bibinfo {title} {Observation of the topological soliton state in the {{Su}}\textendash{{Schrieffer}}\textendash{{Heeger}} model},\ }\href {https://doi.org/10.1038/ncomms13986} {\bibfield  {journal} {\bibinfo  {journal} {Nat. Commun.}\ }\textbf {\bibinfo {volume} {7}},\ \bibinfo {pages} {13986} (\bibinfo {year} {2016}{\natexlab{a}})}\BibitemShut {NoStop}%
\bibitem [{\citenamefont {Meier}\ \emph {et~al.}(2018)\citenamefont {Meier}, \citenamefont {An}, \citenamefont {Dauphin}, \citenamefont {Maffei}, \citenamefont {Massignan}, \citenamefont {Hughes},\ and\ \citenamefont {Gadway}}]{meierObservationTopologicalAnderson2018}%
  \BibitemOpen
  \bibfield  {author} {\bibinfo {author} {\bibfnamefont {E.~J.}\ \bibnamefont {Meier}}, \bibinfo {author} {\bibfnamefont {F.~A.}\ \bibnamefont {An}}, \bibinfo {author} {\bibfnamefont {A.}~\bibnamefont {Dauphin}}, \bibinfo {author} {\bibfnamefont {M.}~\bibnamefont {Maffei}}, \bibinfo {author} {\bibfnamefont {P.}~\bibnamefont {Massignan}}, \bibinfo {author} {\bibfnamefont {T.~L.}\ \bibnamefont {Hughes}},\ and\ \bibinfo {author} {\bibfnamefont {B.}~\bibnamefont {Gadway}},\ }\bibfield  {title} {\bibinfo {title} {Observation of the topological {{Anderson}} insulator in disordered atomic wires},\ }\href {https://doi.org/10.1126/science.aat3406} {\bibfield  {journal} {\bibinfo  {journal} {Science}\ }\textbf {\bibinfo {volume} {362}},\ \bibinfo {pages} {929} (\bibinfo {year} {2018})}\BibitemShut {NoStop}%
\bibitem [{\citenamefont {Xie}\ \emph {et~al.}(2019)\citenamefont {Xie}, \citenamefont {Gou}, \citenamefont {Xiao}, \citenamefont {Gadway},\ and\ \citenamefont {Yan}}]{xieTopologicalCharacterizationsExtended2019}%
  \BibitemOpen
  \bibfield  {author} {\bibinfo {author} {\bibfnamefont {D.}~\bibnamefont {Xie}}, \bibinfo {author} {\bibfnamefont {W.}~\bibnamefont {Gou}}, \bibinfo {author} {\bibfnamefont {T.}~\bibnamefont {Xiao}}, \bibinfo {author} {\bibfnamefont {B.}~\bibnamefont {Gadway}},\ and\ \bibinfo {author} {\bibfnamefont {B.}~\bibnamefont {Yan}},\ }\bibfield  {title} {\bibinfo {title} {Topological characterizations of an extended {{Su}}\textendash{{Schrieffer}}\textendash{{Heeger}} model},\ }\href {https://doi.org/10.1038/s41534-019-0159-6} {\bibfield  {journal} {\bibinfo  {journal} {npj Quantum Inf.}\ }\textbf {\bibinfo {volume} {5}},\ \bibinfo {pages} {55} (\bibinfo {year} {2019})}\BibitemShut {NoStop}%
\bibitem [{\citenamefont {Cardano}\ \emph {et~al.}(2017)\citenamefont {Cardano}, \citenamefont {D'Errico}, \citenamefont {Dauphin}, \citenamefont {Maffei}, \citenamefont {Piccirillo}, \citenamefont {{de Lisio}}, \citenamefont {De~Filippis}, \citenamefont {Cataudella}, \citenamefont {Santamato}, \citenamefont {Marrucci}, \citenamefont {Lewenstein},\ and\ \citenamefont {Massignan}}]{cardanoDetectionZakPhases2017}%
  \BibitemOpen
  \bibfield  {author} {\bibinfo {author} {\bibfnamefont {F.}~\bibnamefont {Cardano}}, \bibinfo {author} {\bibfnamefont {A.}~\bibnamefont {D'Errico}}, \bibinfo {author} {\bibfnamefont {A.}~\bibnamefont {Dauphin}}, \bibinfo {author} {\bibfnamefont {M.}~\bibnamefont {Maffei}}, \bibinfo {author} {\bibfnamefont {B.}~\bibnamefont {Piccirillo}}, \bibinfo {author} {\bibfnamefont {C.}~\bibnamefont {{de Lisio}}}, \bibinfo {author} {\bibfnamefont {G.}~\bibnamefont {De~Filippis}}, \bibinfo {author} {\bibfnamefont {V.}~\bibnamefont {Cataudella}}, \bibinfo {author} {\bibfnamefont {E.}~\bibnamefont {Santamato}}, \bibinfo {author} {\bibfnamefont {L.}~\bibnamefont {Marrucci}}, \bibinfo {author} {\bibfnamefont {M.}~\bibnamefont {Lewenstein}},\ and\ \bibinfo {author} {\bibfnamefont {P.}~\bibnamefont {Massignan}},\ }\bibfield  {title} {\bibinfo {title} {Detection of {{Zak}} phases and topological invariants in a chiral quantum walk of twisted photons},\ }\href {https://doi.org/10.1038/ncomms15516} {\bibfield  {journal} {\bibinfo  {journal} {Nat. Commun.}\ }\textbf {\bibinfo {volume} {8}},\ \bibinfo {pages} {15516} (\bibinfo {year} {2017})}\BibitemShut {NoStop}%
\bibitem [{\citenamefont {Maffei}\ \emph {et~al.}(2018)\citenamefont {Maffei}, \citenamefont {Dauphin}, \citenamefont {Cardano}, \citenamefont {Lewenstein},\ and\ \citenamefont {Massignan}}]{maffeiTopologicalCharacterizationChiral2018}%
  \BibitemOpen
  \bibfield  {author} {\bibinfo {author} {\bibfnamefont {M.}~\bibnamefont {Maffei}}, \bibinfo {author} {\bibfnamefont {A.}~\bibnamefont {Dauphin}}, \bibinfo {author} {\bibfnamefont {F.}~\bibnamefont {Cardano}}, \bibinfo {author} {\bibfnamefont {M.}~\bibnamefont {Lewenstein}},\ and\ \bibinfo {author} {\bibfnamefont {P.}~\bibnamefont {Massignan}},\ }\bibfield  {title} {\bibinfo {title} {Topological characterization of chiral models through their long time dynamics},\ }\href {https://doi.org/10.1088/1367-2630/aa9d4c} {\bibfield  {journal} {\bibinfo  {journal} {New J. Phys.}\ }\textbf {\bibinfo {volume} {20}},\ \bibinfo {pages} {013023} (\bibinfo {year} {2018})}\BibitemShut {NoStop}%
\bibitem [{\citenamefont {D'Errico}\ \emph {et~al.}(2020)\citenamefont {D'Errico}, \citenamefont {Di~Colandrea}, \citenamefont {Barboza}, \citenamefont {Dauphin}, \citenamefont {Lewenstein}, \citenamefont {Massignan}, \citenamefont {Marrucci},\ and\ \citenamefont {Cardano}}]{derricoBulkDetectionTimedependent2020}%
  \BibitemOpen
  \bibfield  {author} {\bibinfo {author} {\bibfnamefont {A.}~\bibnamefont {D'Errico}}, \bibinfo {author} {\bibfnamefont {F.}~\bibnamefont {Di~Colandrea}}, \bibinfo {author} {\bibfnamefont {R.}~\bibnamefont {Barboza}}, \bibinfo {author} {\bibfnamefont {A.}~\bibnamefont {Dauphin}}, \bibinfo {author} {\bibfnamefont {M.}~\bibnamefont {Lewenstein}}, \bibinfo {author} {\bibfnamefont {P.}~\bibnamefont {Massignan}}, \bibinfo {author} {\bibfnamefont {L.}~\bibnamefont {Marrucci}},\ and\ \bibinfo {author} {\bibfnamefont {F.}~\bibnamefont {Cardano}},\ }\bibfield  {title} {\bibinfo {title} {Bulk detection of time-dependent topological transitions in quenched chiral models},\ }\href {https://doi.org/10.1103/PhysRevResearch.2.023119} {\bibfield  {journal} {\bibinfo  {journal} {Phys. Rev. Res.}\ }\textbf {\bibinfo {volume} {2}},\ \bibinfo {pages} {023119} (\bibinfo {year} {2020})}\BibitemShut {NoStop}%
\bibitem [{\citenamefont {{St-Jean}}\ \emph {et~al.}(2021)\citenamefont {{St-Jean}}, \citenamefont {Dauphin}, \citenamefont {Massignan}, \citenamefont {Real}, \citenamefont {Jamadi}, \citenamefont {Milicevic}, \citenamefont {Lema{\^i}tre}, \citenamefont {Harouri}, \citenamefont {Le~Gratiet}, \citenamefont {Sagnes}, \citenamefont {Ravets}, \citenamefont {Bloch},\ and\ \citenamefont {Amo}}]{st-jeanMeasuringTopologicalInvariants2021}%
  \BibitemOpen
  \bibfield  {author} {\bibinfo {author} {\bibfnamefont {P.}~\bibnamefont {{St-Jean}}}, \bibinfo {author} {\bibfnamefont {A.}~\bibnamefont {Dauphin}}, \bibinfo {author} {\bibfnamefont {P.}~\bibnamefont {Massignan}}, \bibinfo {author} {\bibfnamefont {B.}~\bibnamefont {Real}}, \bibinfo {author} {\bibfnamefont {O.}~\bibnamefont {Jamadi}}, \bibinfo {author} {\bibfnamefont {M.}~\bibnamefont {Milicevic}}, \bibinfo {author} {\bibfnamefont {A.}~\bibnamefont {Lema{\^i}tre}}, \bibinfo {author} {\bibfnamefont {A.}~\bibnamefont {Harouri}}, \bibinfo {author} {\bibfnamefont {L.}~\bibnamefont {Le~Gratiet}}, \bibinfo {author} {\bibfnamefont {I.}~\bibnamefont {Sagnes}}, \bibinfo {author} {\bibfnamefont {S.}~\bibnamefont {Ravets}}, \bibinfo {author} {\bibfnamefont {J.}~\bibnamefont {Bloch}},\ and\ \bibinfo {author} {\bibfnamefont {A.}~\bibnamefont {Amo}},\ }\bibfield  {title} {\bibinfo {title} {Measuring {{Topological Invariants}} in a {{Polaritonic Analog}} of {{Graphene}}},\ }\href {https://doi.org/10.1103/PhysRevLett.126.127403} {\bibfield  {journal} {\bibinfo  {journal} {Phys. Rev. Lett.}\ }\textbf {\bibinfo {volume} {126}},\ \bibinfo {pages} {127403} (\bibinfo {year} {2021})}\BibitemShut {NoStop}%
\bibitem [{\citenamefont {Xiao}\ \emph {et~al.}(2021)\citenamefont {Xiao}, \citenamefont {Xie}, \citenamefont {Dong}, \citenamefont {Chen}, \citenamefont {Yi},\ and\ \citenamefont {Yan}}]{xiaoObservationTopologicalPhase2021}%
  \BibitemOpen
  \bibfield  {author} {\bibinfo {author} {\bibfnamefont {T.}~\bibnamefont {Xiao}}, \bibinfo {author} {\bibfnamefont {D.}~\bibnamefont {Xie}}, \bibinfo {author} {\bibfnamefont {Z.}~\bibnamefont {Dong}}, \bibinfo {author} {\bibfnamefont {T.}~\bibnamefont {Chen}}, \bibinfo {author} {\bibfnamefont {W.}~\bibnamefont {Yi}},\ and\ \bibinfo {author} {\bibfnamefont {B.}~\bibnamefont {Yan}},\ }\bibfield  {title} {\bibinfo {title} {Observation of topological phase with critical localization in a quasi-periodic lattice},\ }\href {https://doi.org/10.1016/j.scib.2021.07.025} {\bibfield  {journal} {\bibinfo  {journal} {Sci. Bull.}\ }\textbf {\bibinfo {volume} {66}},\ \bibinfo {pages} {2175} (\bibinfo {year} {2021})}\BibitemShut {NoStop}%
\bibitem [{\citenamefont {Moore}\ \emph {et~al.}(1995)\citenamefont {Moore}, \citenamefont {Robinson}, \citenamefont {Bharucha}, \citenamefont {Sundaram},\ and\ \citenamefont {Raizen}}]{mooreAtomOpticsRealization1995}%
  \BibitemOpen
  \bibfield  {author} {\bibinfo {author} {\bibfnamefont {F.~L.}\ \bibnamefont {Moore}}, \bibinfo {author} {\bibfnamefont {J.~C.}\ \bibnamefont {Robinson}}, \bibinfo {author} {\bibfnamefont {C.~F.}\ \bibnamefont {Bharucha}}, \bibinfo {author} {\bibfnamefont {B.}~\bibnamefont {Sundaram}},\ and\ \bibinfo {author} {\bibfnamefont {M.~G.}\ \bibnamefont {Raizen}},\ }\bibfield  {title} {\bibinfo {title} {Atom {{Optics Realization}} of the {{Quantum}} $\ensuremath{\delta}$-{{Kicked Rotor}}},\ }\href {https://doi.org/10.1103/PhysRevLett.75.4598} {\bibfield  {journal} {\bibinfo  {journal} {Phys. Rev. Lett.}\ }\textbf {\bibinfo {volume} {75}},\ \bibinfo {pages} {4598} (\bibinfo {year} {1995})}\BibitemShut {NoStop}%
\bibitem [{\citenamefont {Wang}\ and\ \citenamefont {Gong}(2008)}]{wangProposalColdatomRealization2008}%
  \BibitemOpen
  \bibfield  {author} {\bibinfo {author} {\bibfnamefont {J.}~\bibnamefont {Wang}}\ and\ \bibinfo {author} {\bibfnamefont {J.}~\bibnamefont {Gong}},\ }\bibfield  {title} {\bibinfo {title} {Proposal of a cold-atom realization of quantum maps with {{Hofstadter}}'s butterfly spectrum},\ }\href {https://doi.org/10.1103/PhysRevA.77.031405} {\bibfield  {journal} {\bibinfo  {journal} {Phys. Rev. A}\ }\textbf {\bibinfo {volume} {77}},\ \bibinfo {pages} {031405(R)} (\bibinfo {year} {2008})}\BibitemShut {NoStop}%
\bibitem [{\citenamefont {Asb{\'o}th}\ and\ \citenamefont {Obuse}(2013)}]{asbothBulkboundaryCorrespondenceChiral2013}%
  \BibitemOpen
  \bibfield  {author} {\bibinfo {author} {\bibfnamefont {J.~K.}\ \bibnamefont {Asb{\'o}th}}\ and\ \bibinfo {author} {\bibfnamefont {H.}~\bibnamefont {Obuse}},\ }\bibfield  {title} {\bibinfo {title} {Bulk-boundary correspondence for chiral symmetric quantum walks},\ }\href {https://doi.org/10.1103/PhysRevB.88.121406} {\bibfield  {journal} {\bibinfo  {journal} {Phys. Rev. B}\ }\textbf {\bibinfo {volume} {88}},\ \bibinfo {pages} {121406(R)} (\bibinfo {year} {2013})}\BibitemShut {NoStop}%
\bibitem [{\citenamefont {Xiong}\ \emph {et~al.}(2017)\citenamefont {Xiong}, \citenamefont {Yamakage}, \citenamefont {Kobayashi}, \citenamefont {Sato},\ and\ \citenamefont {Tanaka}}]{xiongAnisotropicMagneticResponses2017}%
  \BibitemOpen
  \bibfield  {author} {\bibinfo {author} {\bibfnamefont {Y.}~\bibnamefont {Xiong}}, \bibinfo {author} {\bibfnamefont {A.}~\bibnamefont {Yamakage}}, \bibinfo {author} {\bibfnamefont {S.}~\bibnamefont {Kobayashi}}, \bibinfo {author} {\bibfnamefont {M.}~\bibnamefont {Sato}},\ and\ \bibinfo {author} {\bibfnamefont {Y.}~\bibnamefont {Tanaka}},\ }\bibfield  {title} {\bibinfo {title} {Anisotropic {{Magnetic Responses}} of {{Topological Crystalline Superconductors}}},\ }\href {https://doi.org/10.3390/cryst7020058} {\bibfield  {journal} {\bibinfo  {journal} {Crystals}\ }\textbf {\bibinfo {volume} {7}},\ \bibinfo {pages} {58} (\bibinfo {year} {2017})}\BibitemShut {NoStop}%
\bibitem [{\citenamefont {Wen}\ \emph {et~al.}(2021)\citenamefont {Wen}, \citenamefont {Meng}, \citenamefont {Wang}, \citenamefont {Chen}, \citenamefont {Huang}, \citenamefont {Wang},\ and\ \citenamefont {Zhang}}]{wenExperimentalStudyTuneout2021}%
  \BibitemOpen
  \bibfield  {author} {\bibinfo {author} {\bibfnamefont {K.}~\bibnamefont {Wen}}, \bibinfo {author} {\bibfnamefont {Z.}~\bibnamefont {Meng}}, \bibinfo {author} {\bibfnamefont {L.}~\bibnamefont {Wang}}, \bibinfo {author} {\bibfnamefont {L.}~\bibnamefont {Chen}}, \bibinfo {author} {\bibfnamefont {L.}~\bibnamefont {Huang}}, \bibinfo {author} {\bibfnamefont {P.}~\bibnamefont {Wang}},\ and\ \bibinfo {author} {\bibfnamefont {J.}~\bibnamefont {Zhang}},\ }\bibfield  {title} {\bibinfo {title} {Experimental study of tune-out wavelengths for spin-dependent optical lattice in {\textsuperscript{87}}{{Rb Bose}}\textendash{{Einstein}} condensation},\ }\href {https://doi.org/10.1364/JOSAB.432448} {\bibfield  {journal} {\bibinfo  {journal} {J. Opt. Soc. Am. B}\ }\textbf {\bibinfo {volume} {38}},\ \bibinfo {pages} {3269} (\bibinfo {year} {2021})}\BibitemShut {NoStop}%
\bibitem [{\citenamefont {Mandel}\ \emph {et~al.}(2003{\natexlab{a}})\citenamefont {Mandel}, \citenamefont {Greiner}, \citenamefont {Widera}, \citenamefont {Rom}, \citenamefont {H{\"a}nsch},\ and\ \citenamefont {Bloch}}]{mandelCoherentTransportNeutral2003}%
  \BibitemOpen
  \bibfield  {author} {\bibinfo {author} {\bibfnamefont {O.}~\bibnamefont {Mandel}}, \bibinfo {author} {\bibfnamefont {M.}~\bibnamefont {Greiner}}, \bibinfo {author} {\bibfnamefont {A.}~\bibnamefont {Widera}}, \bibinfo {author} {\bibfnamefont {T.}~\bibnamefont {Rom}}, \bibinfo {author} {\bibfnamefont {T.~W.}\ \bibnamefont {H{\"a}nsch}},\ and\ \bibinfo {author} {\bibfnamefont {I.}~\bibnamefont {Bloch}},\ }\bibfield  {title} {\bibinfo {title} {Coherent {{Transport}} of {{Neutral Atoms}} in {{Spin-Dependent Optical Lattice Potentials}}},\ }\href {https://doi.org/10.1103/PhysRevLett.91.010407} {\bibfield  {journal} {\bibinfo  {journal} {Phys. Rev. Lett.}\ }\textbf {\bibinfo {volume} {91}},\ \bibinfo {pages} {010407} (\bibinfo {year} {2003}{\natexlab{a}})}\BibitemShut {NoStop}%
\bibitem [{\citenamefont {Mandel}\ \emph {et~al.}(2003{\natexlab{b}})\citenamefont {Mandel}, \citenamefont {Greiner}, \citenamefont {Widera}, \citenamefont {Rom}, \citenamefont {H{\"a}nsch},\ and\ \citenamefont {Bloch}}]{mandelControlledCollisionsMultiparticle2003}%
  \BibitemOpen
  \bibfield  {author} {\bibinfo {author} {\bibfnamefont {O.}~\bibnamefont {Mandel}}, \bibinfo {author} {\bibfnamefont {M.}~\bibnamefont {Greiner}}, \bibinfo {author} {\bibfnamefont {A.}~\bibnamefont {Widera}}, \bibinfo {author} {\bibfnamefont {T.}~\bibnamefont {Rom}}, \bibinfo {author} {\bibfnamefont {T.~W.}\ \bibnamefont {H{\"a}nsch}},\ and\ \bibinfo {author} {\bibfnamefont {I.}~\bibnamefont {Bloch}},\ }\bibfield  {title} {\bibinfo {title} {Controlled collisions for multi-particle entanglement of optically trapped atoms},\ }\href {https://doi.org/10.1038/nature02008} {\bibfield  {journal} {\bibinfo  {journal} {Nature}\ }\textbf {\bibinfo {volume} {425}},\ \bibinfo {pages} {937} (\bibinfo {year} {2003}{\natexlab{b}})}\BibitemShut {NoStop}%
\bibitem [{\citenamefont {Ryu}\ \emph {et~al.}(2006)\citenamefont {Ryu}, \citenamefont {Andersen}, \citenamefont {Vaziri}, \citenamefont {{d'Arcy}}, \citenamefont {Grossman}, \citenamefont {Helmerson},\ and\ \citenamefont {Phillips}}]{ryuHighOrderQuantumResonances2006}%
  \BibitemOpen
  \bibfield  {author} {\bibinfo {author} {\bibfnamefont {C.}~\bibnamefont {Ryu}}, \bibinfo {author} {\bibfnamefont {M.~F.}\ \bibnamefont {Andersen}}, \bibinfo {author} {\bibfnamefont {A.}~\bibnamefont {Vaziri}}, \bibinfo {author} {\bibfnamefont {M.~B.}\ \bibnamefont {{d'Arcy}}}, \bibinfo {author} {\bibfnamefont {J.~M.}\ \bibnamefont {Grossman}}, \bibinfo {author} {\bibfnamefont {K.}~\bibnamefont {Helmerson}},\ and\ \bibinfo {author} {\bibfnamefont {W.~D.}\ \bibnamefont {Phillips}},\ }\bibfield  {title} {\bibinfo {title} {High-{{Order Quantum Resonances Observed}} in a {{Periodically Kicked Bose-Einstein Condensate}}},\ }\href {https://doi.org/10.1103/PhysRevLett.96.160403} {\bibfield  {journal} {\bibinfo  {journal} {Phys. Rev. Lett.}\ }\textbf {\bibinfo {volume} {96}},\ \bibinfo {pages} {160403} (\bibinfo {year} {2006})}\BibitemShut {NoStop}%
\bibitem [{\citenamefont {Gadway}(2015)}]{gadwayAtomopticsApproachStudying2015}%
  \BibitemOpen
  \bibfield  {author} {\bibinfo {author} {\bibfnamefont {B.}~\bibnamefont {Gadway}},\ }\bibfield  {title} {\bibinfo {title} {Atom-optics approach to studying transport phenomena},\ }\href {https://doi.org/10.1103/PhysRevA.92.043606} {\bibfield  {journal} {\bibinfo  {journal} {Phys. Rev. A}\ }\textbf {\bibinfo {volume} {92}},\ \bibinfo {pages} {043606} (\bibinfo {year} {2015})}\BibitemShut {NoStop}%
\bibitem [{\citenamefont {Meier}\ \emph {et~al.}(2016{\natexlab{b}})\citenamefont {Meier}, \citenamefont {An},\ and\ \citenamefont {Gadway}}]{meierAtomopticsSimulatorLattice2016}%
  \BibitemOpen
  \bibfield  {author} {\bibinfo {author} {\bibfnamefont {E.~J.}\ \bibnamefont {Meier}}, \bibinfo {author} {\bibfnamefont {F.~A.}\ \bibnamefont {An}},\ and\ \bibinfo {author} {\bibfnamefont {B.}~\bibnamefont {Gadway}},\ }\bibfield  {title} {\bibinfo {title} {Atom-optics simulator of lattice transport phenomena},\ }\href {https://doi.org/10.1103/PhysRevA.93.051602} {\bibfield  {journal} {\bibinfo  {journal} {Phys. Rev. A}\ }\textbf {\bibinfo {volume} {93}},\ \bibinfo {pages} {051602(R)} (\bibinfo {year} {2016}{\natexlab{b}})}\BibitemShut {NoStop}%
\bibitem [{\citenamefont {An}\ \emph {et~al.}(2017)\citenamefont {An}, \citenamefont {Meier},\ and\ \citenamefont {Gadway}}]{anDirectObservationChiral2017}%
  \BibitemOpen
  \bibfield  {author} {\bibinfo {author} {\bibfnamefont {F.~A.}\ \bibnamefont {An}}, \bibinfo {author} {\bibfnamefont {E.~J.}\ \bibnamefont {Meier}},\ and\ \bibinfo {author} {\bibfnamefont {B.}~\bibnamefont {Gadway}},\ }\bibfield  {title} {\bibinfo {title} {Direct observation of chiral currents and magnetic reflection in atomic flux lattices},\ }\href {https://doi.org/10.1126/sciadv.1602685} {\bibfield  {journal} {\bibinfo  {journal} {Sci. Adv.}\ }\textbf {\bibinfo {volume} {3}},\ \bibinfo {pages} {e1602685} (\bibinfo {year} {2017})}\BibitemShut {NoStop}%
\bibitem [{\citenamefont {Jangjan}\ \emph {et~al.}(2022)\citenamefont {Jangjan}, \citenamefont {Foa~Torres},\ and\ \citenamefont {Hosseini}}]{jangjanFloquetTopologicalPhase2022}%
  \BibitemOpen
  \bibfield  {author} {\bibinfo {author} {\bibfnamefont {M.}~\bibnamefont {Jangjan}}, \bibinfo {author} {\bibfnamefont {L.~E.~F.}\ \bibnamefont {Foa~Torres}},\ and\ \bibinfo {author} {\bibfnamefont {M.~V.}\ \bibnamefont {Hosseini}},\ }\bibfield  {title} {\bibinfo {title} {Floquet topological phase transitions in a periodically quenched dimer},\ }\href {https://doi.org/10.1103/PhysRevB.106.224306} {\bibfield  {journal} {\bibinfo  {journal} {Phys. Rev. B}\ }\textbf {\bibinfo {volume} {106}},\ \bibinfo {pages} {224306} (\bibinfo {year} {2022})}\BibitemShut {NoStop}%
\bibitem [{\citenamefont {Zhang}\ \emph {et~al.}(2004)\citenamefont {Zhang}, \citenamefont {Liu}, \citenamefont {Raizen},\ and\ \citenamefont {Niu}}]{zhangTransitionInstabilityKicked2004}%
  \BibitemOpen
  \bibfield  {author} {\bibinfo {author} {\bibfnamefont {C.}~\bibnamefont {Zhang}}, \bibinfo {author} {\bibfnamefont {J.}~\bibnamefont {Liu}}, \bibinfo {author} {\bibfnamefont {M.~G.}\ \bibnamefont {Raizen}},\ and\ \bibinfo {author} {\bibfnamefont {Q.}~\bibnamefont {Niu}},\ }\bibfield  {title} {\bibinfo {title} {Transition to {{Instability}} in a {{Kicked Bose-Einstein Condensate}}},\ }\href {https://doi.org/10.1103/PhysRevLett.92.054101} {\bibfield  {journal} {\bibinfo  {journal} {Phys. Rev. Lett.}\ }\textbf {\bibinfo {volume} {92}},\ \bibinfo {pages} {054101} (\bibinfo {year} {2004})}\BibitemShut {NoStop}%
\bibitem [{\citenamefont {Monteiro}\ \emph {et~al.}(2009)\citenamefont {Monteiro}, \citenamefont {Ran{\c c}on},\ and\ \citenamefont {Ruostekoski}}]{monteiroNonlinearResonancesEnsuremath2009}%
  \BibitemOpen
  \bibfield  {author} {\bibinfo {author} {\bibfnamefont {T.~S.}\ \bibnamefont {Monteiro}}, \bibinfo {author} {\bibfnamefont {A.}~\bibnamefont {Ran{\c c}on}},\ and\ \bibinfo {author} {\bibfnamefont {J.}~\bibnamefont {Ruostekoski}},\ }\bibfield  {title} {\bibinfo {title} {Nonlinear {{Resonances}} in $\ensuremath{\delta}$-{{Kicked Bose-Einstein Condensates}}},\ }\href {https://doi.org/10.1103/PhysRevLett.102.014102} {\bibfield  {journal} {\bibinfo  {journal} {Phys. Rev. Lett.}\ }\textbf {\bibinfo {volume} {102}},\ \bibinfo {pages} {014102} (\bibinfo {year} {2009})}\BibitemShut {NoStop}%
\bibitem [{\citenamefont {See~Toh}\ \emph {et~al.}(2022)\citenamefont {See~Toh}, \citenamefont {McCormick}, \citenamefont {Tang}, \citenamefont {Su}, \citenamefont {Luo}, \citenamefont {Zhang},\ and\ \citenamefont {Gupta}}]{seetohManybodyDynamicalDelocalization2022}%
  \BibitemOpen
  \bibfield  {author} {\bibinfo {author} {\bibfnamefont {J.~H.}\ \bibnamefont {See~Toh}}, \bibinfo {author} {\bibfnamefont {K.~C.}\ \bibnamefont {McCormick}}, \bibinfo {author} {\bibfnamefont {X.}~\bibnamefont {Tang}}, \bibinfo {author} {\bibfnamefont {Y.}~\bibnamefont {Su}}, \bibinfo {author} {\bibfnamefont {X.-W.}\ \bibnamefont {Luo}}, \bibinfo {author} {\bibfnamefont {C.}~\bibnamefont {Zhang}},\ and\ \bibinfo {author} {\bibfnamefont {S.}~\bibnamefont {Gupta}},\ }\bibfield  {title} {\bibinfo {title} {Many-body dynamical delocalization in a kicked one-dimensional ultracold gas},\ }\href {https://doi.org/10.1038/s41567-022-01721-w} {\bibfield  {journal} {\bibinfo  {journal} {Nat. Phys.}\ }\textbf {\bibinfo {volume} {18}},\ \bibinfo {pages} {1297} (\bibinfo {year} {2022})}\BibitemShut {NoStop}%
\bibitem [{\citenamefont {Cao}\ \emph {et~al.}(2022)\citenamefont {Cao}, \citenamefont {Sajjad}, \citenamefont {Mas}, \citenamefont {Simmons}, \citenamefont {Tanlimco}, \citenamefont {{Nolasco-Martinez}}, \citenamefont {Shimasaki}, \citenamefont {Kondakci}, \citenamefont {Galitski},\ and\ \citenamefont {Weld}}]{caoInteractiondrivenBreakdownDynamical2022}%
  \BibitemOpen
  \bibfield  {author} {\bibinfo {author} {\bibfnamefont {A.}~\bibnamefont {Cao}}, \bibinfo {author} {\bibfnamefont {R.}~\bibnamefont {Sajjad}}, \bibinfo {author} {\bibfnamefont {H.}~\bibnamefont {Mas}}, \bibinfo {author} {\bibfnamefont {E.~Q.}\ \bibnamefont {Simmons}}, \bibinfo {author} {\bibfnamefont {J.~L.}\ \bibnamefont {Tanlimco}}, \bibinfo {author} {\bibfnamefont {E.}~\bibnamefont {{Nolasco-Martinez}}}, \bibinfo {author} {\bibfnamefont {T.}~\bibnamefont {Shimasaki}}, \bibinfo {author} {\bibfnamefont {H.~E.}\ \bibnamefont {Kondakci}}, \bibinfo {author} {\bibfnamefont {V.}~\bibnamefont {Galitski}},\ and\ \bibinfo {author} {\bibfnamefont {D.~M.}\ \bibnamefont {Weld}},\ }\bibfield  {title} {\bibinfo {title} {Interaction-driven breakdown of dynamical localization in a kicked quantum gas},\ }\href {https://doi.org/10.1038/s41567-022-01724-7} {\bibfield  {journal} {\bibinfo  {journal} {Nat. Phys.}\ }\textbf {\bibinfo {volume} {18}},\ \bibinfo {pages} {1302} (\bibinfo {year} {2022})}\BibitemShut {NoStop}%
\bibitem [{\citenamefont {Liang}\ \emph {et~al.}(2022)\citenamefont {Liang}, \citenamefont {Xie}, \citenamefont {Dong}, \citenamefont {Li}, \citenamefont {Li}, \citenamefont {Gadway}, \citenamefont {Yi},\ and\ \citenamefont {Yan}}]{liangDynamicSignaturesNonHermitian2022}%
  \BibitemOpen
  \bibfield  {author} {\bibinfo {author} {\bibfnamefont {Q.}~\bibnamefont {Liang}}, \bibinfo {author} {\bibfnamefont {D.}~\bibnamefont {Xie}}, \bibinfo {author} {\bibfnamefont {Z.}~\bibnamefont {Dong}}, \bibinfo {author} {\bibfnamefont {H.}~\bibnamefont {Li}}, \bibinfo {author} {\bibfnamefont {H.}~\bibnamefont {Li}}, \bibinfo {author} {\bibfnamefont {B.}~\bibnamefont {Gadway}}, \bibinfo {author} {\bibfnamefont {W.}~\bibnamefont {Yi}},\ and\ \bibinfo {author} {\bibfnamefont {B.}~\bibnamefont {Yan}},\ }\bibfield  {title} {\bibinfo {title} {Dynamic {{Signatures}} of {{Non-Hermitian Skin Effect}} and {{Topology}} in {{Ultracold Atoms}}},\ }\href {https://doi.org/10.1103/PhysRevLett.129.070401} {\bibfield  {journal} {\bibinfo  {journal} {Phys. Rev. Lett.}\ }\textbf {\bibinfo {volume} {129}},\ \bibinfo {pages} {070401} (\bibinfo {year} {2022})}\BibitemShut {NoStop}%
\bibitem [{\citenamefont {Zhou}\ \emph {et~al.}(2022)\citenamefont {Zhou}, \citenamefont {Li}, \citenamefont {Yi},\ and\ \citenamefont {Cui}}]{zhouEngineeringNonHermitianSkin2022}%
  \BibitemOpen
  \bibfield  {author} {\bibinfo {author} {\bibfnamefont {L.}~\bibnamefont {Zhou}}, \bibinfo {author} {\bibfnamefont {H.}~\bibnamefont {Li}}, \bibinfo {author} {\bibfnamefont {W.}~\bibnamefont {Yi}},\ and\ \bibinfo {author} {\bibfnamefont {X.}~\bibnamefont {Cui}},\ }\bibfield  {title} {\bibinfo {title} {Engineering non-{{Hermitian}} skin effect with band topology in ultracold gases},\ }\href {https://doi.org/10.1038/s42005-022-01021-y} {\bibfield  {journal} {\bibinfo  {journal} {Commun. Phys.}\ }\textbf {\bibinfo {volume} {5}},\ \bibinfo {pages} {252} (\bibinfo {year} {2022})}\BibitemShut {NoStop}%
\bibitem [{\citenamefont {Asb{\'o}th}\ \emph {et~al.}(2014)\citenamefont {Asb{\'o}th}, \citenamefont {Tarasinski},\ and\ \citenamefont {Delplace}}]{asbothChiralSymmetryBulkboundary2014}%
  \BibitemOpen
  \bibfield  {author} {\bibinfo {author} {\bibfnamefont {J.~K.}\ \bibnamefont {Asb{\'o}th}}, \bibinfo {author} {\bibfnamefont {B.}~\bibnamefont {Tarasinski}},\ and\ \bibinfo {author} {\bibfnamefont {P.}~\bibnamefont {Delplace}},\ }\bibfield  {title} {\bibinfo {title} {Chiral symmetry and bulk-boundary correspondence in periodically driven one-dimensional systems},\ }\href {https://doi.org/10.1103/PhysRevB.90.125143} {\bibfield  {journal} {\bibinfo  {journal} {Phys. Rev. B}\ }\textbf {\bibinfo {volume} {90}},\ \bibinfo {pages} {125143} (\bibinfo {year} {2014})}\BibitemShut {NoStop}%
\bibitem [{\citenamefont {Liu}\ \emph {et~al.}(2018)\citenamefont {Liu}, \citenamefont {Harper},\ and\ \citenamefont {Roy}}]{liuChiralFlowOnedimensional2018}%
  \BibitemOpen
  \bibfield  {author} {\bibinfo {author} {\bibfnamefont {X.}~\bibnamefont {Liu}}, \bibinfo {author} {\bibfnamefont {F.}~\bibnamefont {Harper}},\ and\ \bibinfo {author} {\bibfnamefont {R.}~\bibnamefont {Roy}},\ }\bibfield  {title} {\bibinfo {title} {Chiral flow in one-dimensional {{Floquet}} topological insulators},\ }\href {https://doi.org/10.1103/PhysRevB.98.165116} {\bibfield  {journal} {\bibinfo  {journal} {Phys. Rev. B}\ }\textbf {\bibinfo {volume} {98}},\ \bibinfo {pages} {165116} (\bibinfo {year} {2018})}\BibitemShut {NoStop}%
\bibitem [{\citenamefont {Rosenbusch}\ \emph {et~al.}(2009)\citenamefont {Rosenbusch}, \citenamefont {Ghezali}, \citenamefont {Dzuba}, \citenamefont {Flambaum}, \citenamefont {Beloy},\ and\ \citenamefont {Derevianko}}]{rosenbuschAcStarkShift2009}%
  \BibitemOpen
  \bibfield  {author} {\bibinfo {author} {\bibfnamefont {P.}~\bibnamefont {Rosenbusch}}, \bibinfo {author} {\bibfnamefont {S.}~\bibnamefont {Ghezali}}, \bibinfo {author} {\bibfnamefont {V.~A.}\ \bibnamefont {Dzuba}}, \bibinfo {author} {\bibfnamefont {V.~V.}\ \bibnamefont {Flambaum}}, \bibinfo {author} {\bibfnamefont {K.}~\bibnamefont {Beloy}},\ and\ \bibinfo {author} {\bibfnamefont {A.}~\bibnamefont {Derevianko}},\ }\bibfield  {title} {\bibinfo {title} {ac {{Stark}} shift of the {{Cs}} microwave atomic clock transitions},\ }\href {https://doi.org/10.1103/PhysRevA.79.013404} {\bibfield  {journal} {\bibinfo  {journal} {Phys. Rev. A}\ }\textbf {\bibinfo {volume} {79}},\ \bibinfo {pages} {013404} (\bibinfo {year} {2009})}\BibitemShut {NoStop}%
\bibitem [{\citenamefont {Le~Kien}\ \emph {et~al.}(2013)\citenamefont {Le~Kien}, \citenamefont {Schneeweiss},\ and\ \citenamefont {Rauschenbeutel}}]{lekienDynamicalPolarizabilityAtoms2013}%
  \BibitemOpen
  \bibfield  {author} {\bibinfo {author} {\bibfnamefont {F.}~\bibnamefont {Le~Kien}}, \bibinfo {author} {\bibfnamefont {P.}~\bibnamefont {Schneeweiss}},\ and\ \bibinfo {author} {\bibfnamefont {A.}~\bibnamefont {Rauschenbeutel}},\ }\bibfield  {title} {\bibinfo {title} {Dynamical polarizability of atoms in arbitrary light fields: general theory and application to cesium},\ }\href {https://doi.org/10.1140/epjd/e2013-30729-x} {\bibfield  {journal} {\bibinfo  {journal} {Eur. Phys. J. D}\ }\textbf {\bibinfo {volume} {67}},\ \bibinfo {pages} {92} (\bibinfo {year} {2013})}\BibitemShut {NoStop}%
\bibitem [{Note1()}]{Note1}%
  \BibitemOpen
  \bibinfo {note} {We note that the sign of the vector polarizability $\alpha ^\protect \mathrm {v}$ is different from Eq.~(2) in Ref.~\cite {wenExperimentalStudyTuneout2021} but our choice of the sign is consistent with the result of the reference. Whereas the expressions of these polarizabilities only include the contributions from the $\protect \mathrm {D_1}$ and $\protect \mathrm {D_2}$ lines, the calculated tune-out wavelengths of the optical lattices coincide with the measured values in Ref.~\cite {wenExperimentalStudyTuneout2021} up to four significant digits.}\BibitemShut {Stop}%
\end{thebibliography}%

\end{document}